\documentclass[10pt,a4paper]{article}
\usepackage{jheppub}
\usepackage{amsmath,amssymb,amsfonts,bbold,mathdots, verbatim}
\usepackage{physics}
\usepackage{tikz}
\usetikzlibrary{decorations.markings}
\usepackage{mathrsfs}

\newenvironment{psmallmatrix}
{\left(\begin{smallmatrix}}
{\end{smallmatrix}\right)}

\providecommand{\Lt}{{\tt L}}
\renewcommand{\Lt}{{\tt L}}
\providecommand{\Mt}{{\tt M}}
\renewcommand{\Mt}{{\tt M}}
\providecommand{\Jt}{{\tt J}}
\renewcommand{\Jt}{{\tt J}}
\providecommand{\Pt}{{\tt P}}
\renewcommand{\Pt}{{\tt P}}

\newcommand{\e}{\epsilon}

\newcommand{\s}{\sigma}

\newcommand{\refb}[1]{(\ref{#1})}

\DeclareMathOperator{\extdm}{d}
\newcommand{\extd}{\extdm \!}

\def\beaa{\begin{eqnarray*}}
\def\eeaa{\end{eqnarray*}}
\def\bea{\begin{eqnarray}}
\def\eea{\end{eqnarray}}
\def\be{\begin{equation}}
\def\ee{\end{equation}}
\newcommand{\bes}{\begin{subequations}}
\newcommand{\ees}{\end{subequations}}
\def\ba{\begin{align}}
\def\ea{\end{align}}

\def\vec{\overrightarrow}

\title{BMS Field Theories with $\mathfrak{u}(1)$ Symmetry}

\author[a, b, c]{Arjun Bagchi,} \author[a]{Ritankar Chatterjee,} \author[a, d]{Rishabh Kaushik,} \author[a]{Sanchari Pal,} \author[e]{Max Riegler} \author[a, d]{and Debmalya Sarkar.} \author{\\}

\affiliation[a]{Indian Institute of Technology Kanpur, Kalyanpur, Kanpur 208016. INDIA.\\}
\affiliation[b]{Erwin Schrödinger International Institute for Mathematics and Physics, University of Vienna. 1090 Vienna, AUSTRIA.\\}
\affiliation[c]{Centre de Physique Théorique, Ecole Polytechnique de Paris, 91128 Palaiseau Cedex, FRANCE.\\}
\affiliation[d]{International Centre for Theoretical Sciences (ICTS-TIFR), Tata Institute of Fundamental Research, Shivakote, Hesaraghatta, Bengaluru 560089. INDIA.\\}

\affiliation[e]{Faculty of Physics, University of Vienna, Boltzmanngasse 5, 1090 Vienna, AUSTRIA.\\}
\emailAdd{(abagchi, ritankar, pals)@iitk.ac.in}
\emailAdd{sarkardebmalya01@gmail.com}
\emailAdd{rishabh.kaushik@icts.res.in}
\emailAdd{max.riegler@univie.ac.at}

\preprint{}
\abstract{We investigate quantum field theories in two dimensions (2d) with an underlying Bondi-van der Burgh-Metzner-Sachs (BMS) symmetry augmented by $\mathfrak{u}(1)$ currents. These field theories are expected to holographically capture features of charged versions of cosmological solutions in asymptotically flat 3d spacetimes called Flat Space Cosmologies (FSCs). We conduct a study of the modular properties of these field theories. The characters for the highest weight representations of the symmetry algebra are constructed, and the partition function of the theory is obtained from them. We derive the density of (primary) states and find the entropy and asymptotic values of the structure constants exploiting the modular properties of the partition function and the torus one-point function. The expression for the asymptotic structure constants shows shifts in the weights and one of the central terms and an extra phase compared to the earlier results in the literature for BMS invariant theories without $\mathfrak{u}(1)$ currents present. We reproduce our field results for the structure constants by a bulk computation involving a scalar probe in the background of a charged FSC.}

\begin{document}
\maketitle

\newpage

\section{Introduction} 

Conformal field theories (CFTs) in $d$ dimensions have SO$(d,2)$ as an underlying symmetry group, and these symmetries have been instrumental in understanding various topics in physics, from phase transition to quantum gravity. In two spacetime dimensions, the conformal algebra becomes infinite-dimensional and can be decomposed into two copies of the Virasoro algebra: 
\bea \label{vir}
 &&[\mathcal{L}_{n},\mathcal{L}_{m}]=(n-m)\mathcal{L}_{n+m}+\frac{c}{12}(n^{3}-n)\delta_{n+m,0}, \crcr 
&&[\mathcal{L}_{n}, \mathcal{\bar{L}}_{m}]= 0, \crcr
&&[\mathcal{\bar{L}}_{n},\mathcal{\bar{L}}_{m}]=(n-m)\mathcal{\bar{L}}_{n+m}+\frac{\bar{c}}{12}(n^{3}-n)\delta_{n+m,0}.
\eea
The asymptotic symmetry algebra of Anti de Sitter (AdS) spacetime in three dimensions \cite{Brown:1986nw} is also given by \eqref{vir} with $c=\bar{c}=\frac{3 \ell}{2G}$ which is seen by many as a precursor to the celebrated AdS/CFT correspondence \cite{Maldacena:1997re}. 

Using the conformal bootstrap program \cite{Poland:2016chs}, one can, in principle, chart out the parameter space of all CFTs. The data required to define a 2d CFT as a point in this parameter space are the central charge, the structure constants (which are the constants of the three-point functions that are undetermined by symmetry), and the spectrum of primary operators. Arbitrary sets of data are not consistent and have to obey constraints imposed by crossing symmetry of four-point functions and the modular properties of the 2d CFT on the torus. One significant consequence of modular invariance is the connection between the high-energy limit of the theory with its low-energy limit. In the seminal paper \cite{Cardy:1986ie}, this connection was used to correctly count the density of states in 2d CFT and derive the associated entropy. It was later shown in \cite{Strominger:1997eq,Carlip:1998qw} that this Cardy counting of entropy matches with the Bekenstein-Hawking entropy of BTZ blackholes \cite{Banados:1992wn, Banados:1992gq} in AdS$_3$, thereby providing a holographic check of the AdS$_3$/CFT$_2$ correspondence without resorting to the details of the boundary theory.   

The Cardy analysis used modular properties of the partition function or, equivalently, the zero-point function on the torus to derive the asymptotic density of states. Recently, using modular properties of torus one-point functions, the asymptotic structure constants were derived in \cite{Kraus:2016nwo}. A holographic computation of a one-point function on a BTZ background reproduces this CFT result. See also \cite{Alkalaev:2016ptm, Alkalaev:2020yvq} for a slightly different holographic interpretation. As an extension of that work in \cite{Das:2017vej} the structure constant for a 2d CFT with additional global $\mathfrak{u}(1)$ symmetry was worked out, and connections to BTZ black holes with $\mathfrak{u}(1)$ charges were established. 

\subsection*{Asymptotically Flat Spacetimes, BMS symmetry, and Holography}

An infinite-dimensional symmetry algebra is not unique to asymptotically AdS$_3$ spacetime. About two decades before the seminal work of Brown and Henneaux, it was shown by Bondi, van der Burgh, Metzner, and Sachs \cite{Bondi:1962px, Sachs:1962zza} that the asymptotic symmetries of 4d flat spacetime at future and past null infinity ($\mathscr{I}^\pm$) go beyond the expected Poincar\'e group and are given by an infinite-dimensional symmetry group nowadays called BMS. The same is true for three-dimensional asymptotically flat spacetime \cite{Barnich:2006av}. This is the case we will be interested in for this paper. 
The asymptotic symmetry algebra defined on $\mathscr{I}^\pm$ in three-dimensional asymptotically flat spacetimes is given by \cite{Barnich:2006av}
\begin{subequations}\label{BMSAL}
\begin{align}
    [L_{n},L_{m}]&=(n-m)L_{n+m}+c_{L}(n^{3}-n)\delta_{n+m,0} \\
    [L_{n},M_{m}]&=(n-m)M_{m+n}+c_{M}(n^{3}-n)\delta_{m+n,0}\\
    [M_{n},M_{m}]&=0.
\end{align}
\end{subequations} 
In the above, $M_n$ are angle-dependent translations along the null direction called supertranslations, $L_n$ are diffeomorphisms of the circle at infinity called superrotations, and $c_L, c_M$ are central terms. Performing a canonical analysis for Einstein gravity, it was found that \cite{Barnich:2006av}
\be\label{bc}
c_L = 0, \quad c_M = \frac{3}{G}.
\ee
In what follows in this paper, we will continue to work with the more general case where both central terms are non-zero. This is, e.g., what happens when we consider Topologically Massive Gravity with asymptotically flat boundary conditions \cite{Bagchi:2012yk}. 

Taking cues from how holography works in asymptotically AdS spacetimes, especially in AdS$_3$, it is natural to assume that the dual theory to asymptotically flat spacetimes would be a quantum field theory that inherits the symmetry \refb{BMSAL} and lives on null infinity. This line of thought \cite{Bagchi:2010eg, Bagchi:2012cy} has led to a formulation of holography in flat space that is nowadays called Carrollian holography, which has seen some exciting successes in the case of three bulk spacetime dimensions \cite{Bagchi:2012xr, Bagchi:2010eg, Bagchi:2012cy, Bagchi:2012yk,Barnich:2012aw,Barnich:2012rz,Bagchi:2014iea,Basu:2015evh,Bagchi:2015wna,Bagchi:2016geg,Jiang:2017ecm,Hijano:2017eii,Grumiller:2019xna,Merbis:2019wgk,Bagchi:2021qfe}. In higher dimensions, a notable new connection has been made between 4d scattering amplitudes and the correlation functions of operators of these 3d field theories residing on $\mathscr{I}^\pm$ \cite{Bagchi:2022emh} {\footnote{See also \cite{Donnay:2022aba} for another connection between the Carrollian and Celestial approaches to flat holography.}}. The name Carrollian is derived from a limiting procedure we go on to describe below. The approach is distinct from the formulation of holography in flat space through the celestial holography program, which has led to many discoveries connecting asymptotic symmetries with scattering amplitudes and memory effects. The reader is pointed to the excellent reviews \cite{Strominger:2017zoo, Pasterski:2021rjz, Raclariu:2021zjz} for more details on this. 

Minkowski spacetime can also be understood as an infinite radius of curvature $(\ell\to\infty)$ limit of AdS spacetime. One way to obtain the symmetry algebra in \eqref{BMSAL} is by taking Inönü-Wigner contractions of linear combinations of the holomorphic and non-holomorphic copies of Virasoro algebra \cite{Bagchi:2012cy}. This limit on the bulk side can be interpreted as an ultra-relativistic (UR) or Carrollian limit on 2d conformal field theory \cite{Bagchi:2012cy}, where the speed of light in the boundary theory is taken to zero. In this limit light cones close up, and many non-intuitive things happen in Carrollian theories, which have shown up in various diverse situations like the tensionless limit of string theory \cite{Bagchi:2013bga, Bagchi:2015nca, Bagchi:2020fpr}, the event horizon of black holes \cite{Donnay:2019jiz}, condensed matter physics in the theory of fractons \cite{Bidussi:2021nmp} and of flat bands \cite{Bagchi:2022eui} and in the formulation of dark energy in cosmology \cite{deBoer:2021jej}.  

\subsection*{The focus of this paper}

Holographic properties of 3d asymptotically flat spacetimes have been studied extensively, and several aspects of the dual 2d BMS invariant field theory (BMSFT) have been looked into. Modular properties of BMSFT were analyzed, and a BMS-Cardy formula was derived in \cite{Bagchi:2012xr}. More recently, this BMS-Cardy formula has been used to compute the entropy of generic higher-dimensional black holes \cite{Carlip:2017xne}. In \cite{Bagchi:2020rwb} the three-point coefficient was calculated, and a host of new results were derived, most notably the torus BMS block. We aim to extend the analysis to a BMSFT with additional $\mathfrak{u}(1)$ symmetries. Some aspects of this have appeared in earlier work \cite{Basu:2017aqn}. In this paper, we will derive many new results, which verify some answers of \cite{Basu:2017aqn}. 

The first part of our paper is purely field-theoretic, but we connect this to holography in the last section. In 3d asymptotically flat spacetimes, there are no black hole solutions. There are, however, analogs of BTZ black holes, which now turn out to be cosmological solutions \cite{Cornalba:2002fi} called Flat Space Cosmologies (FSC). These can be understood intrinsically as (shifted-boost) orbifolds of 3d flat space and the $\ell\to\infty$ limit of non-extremal BTZ black holes. These cosmologies come together with a cosmological horizon, and the BMS-Cardy formula of the dual theory \cite{Bagchi:2012xr, Bagchi:2013qva, Riegler:2014bia, Fareghbal:2014qga} counts the Bekenstein-Hawking entropy of this cosmological horizon. The asymptotic structure constants derived through the torus one-point function can be connected to one-point functions of probes around the cosmological horizon of these FSCs \cite{Bagchi:2020rwb}. In our case, the extra $\mathfrak{u}(1)$ symmetry adds charges to the FSC. The entropy and the asymptotic structure constants that we would compute from the boundary are now that of charged FSCs, as we go on to show in this work. 

\subsection*{Why is this important?}
At the outset, it may not be clear to the reader why we are interested in putting extra $\mathfrak{u}(1)$ symmetries into the analysis. It may seem that at least conceptually this is a rather straightforward exercise with limited motivation. This is, however, not the case as can be seen, e.g., by a careful study of the appendices of this work. Let us clarify the relevance of our work for various problems in the following.

\medskip

 BMS or Conformal Carrollian symmetries have been found on generic null manifolds, making them central to the idea of holography in asymptotically flat spacetimes and also on horizons of generic black holes. These symmetries have also been discovered recently in the context of flat band structures in condensed matter physics. Systems with enhanced symmetries on top of BMS symmetries are hence very important for these applications. Below we explicitly treat each application.

\begin{itemize}

\item \textbf{Generic charged black holes:} Carrollian structures appear on generic black hole event horizons \cite{Donnay:2019jiz}. 2d Conformal Carrollian or BMS$_3$ symmetries have been used to understand the entropy of generic uncharged black holes in any dimension in \cite{Carlip:2017xne, Carlip:2019dbu}, using the BMS-Cardy formula \cite{Bagchi:2012xr}. The methods we have outlined in our paper are thus useful for understanding the entropy of generic charged black holes. The three-point function coefficients would be related to perturbations of generic charged black holes and hence quasi-normal modes associated with them. 

\item \textbf{Enhanced symmetries in flat holography:} In $d=4$, it has been understood by looking at scattering amplitudes that the symmetries associated with flat spacetimes actually enhance beyond BMS \cite{Banerjee:2020zlg, Strominger:2021mtt}. A first step to understanding these symmetries holographically is the in-depth study of BMS$\oplus\mathfrak{u}(1)$ field theories in two dimensions, which we have carried out in this work. This would be a steppingstone for generalized non-Lorentzian Kac-Moody algebras that will be addressed in upcoming work. This would then be taken to three boundary dimensions and connected to the additional symmetries that arise from scattering amplitudes in four-dimensional asymptotically flat spacetimes. Various aspects of the somewhat simpler symmetries that we are dealing with here need to be addressed first if there is any hope of building towards 3d Carrollian CFTs with non-abelian currents. It is important to note here that the $\mathfrak{vir}\oplus\mathfrak{u}(1)$ algebra arises as a chiral algebra of $\omega_{1+\infty}$ algebra. We expect similar features to appear for the BMS versions. Since $\omega_{1+\infty}$ appears in the context of scattering in bulk four dimensions \cite{Strominger:2021mtt}, it is very likely that our construction in this paper would prove very useful in this context from the point of view of the Carrollian interpretation of flat space holography. 

\item \textbf{Carroll Fermions with additional symmetry:} Very recently it has been shown that in the physics of flat bands \cite{Bagchi:2022eui}, which is responsible for, e.g., superconductivity at magic angles in graphene and the fractional quantum hall effect, there are emergent Carrollian structures. When one takes massless or gapless systems, this is enhanced to conformal Carroll or BMS symmetry. It has also been explicitly shown that Carrollian fermions can account for the lattice models of such theories in the continuum limit. Condensed matter systems often come with additional charges associated with them. Our methods outlined in this paper would be of direct relevance in these cases. 

\item \textbf{Null strings with additional $\mathfrak{u}(1)$ currents:} Null or tensionless strings have BMS$_3$ as their worldsheet symmetries \cite{Bagchi:2013bga,Bagchi:2015nca}. It is easy to envision strings carrying extra internal charges and hence being equipped with additional worldsheet currents. Our investigations would be of importance in this case as well. 

\item \textbf{Carrollian hydrodynamics:} Carroll hydrodynamics are useful for various problems, including connections to cosmology \cite{deBoer:2021jej} and applications to quark-gluon-plasma \cite{toappear}. Hydrodynamics often come hand in hand with additional $\mathfrak{u}(1)$ fields. The microscopic description of these flows in the case of conformal fluids, which are of relevance, e.g., to the early universe, would be dictated by the field theories we have studied in this paper. 

\item \textbf{BMS free field theories:} Interestingly, it has been noticed in \cite{Hao:2021urq} that the BMS free scalar model comes along naturally with an additional $\mathfrak{u}(1)$ current. These additional currents may always be present when we consider free theories. Our investigations are thus also relevant to these explicit theories.

\item \textbf{Thermalization in Non-Lorentzian CFTs:} The heavy-heavy-light three-point coefficients play an important role in the investigations of the Eigenstate Thermalisation Hypothesis (ETH) in usual CFTs \cite{Basu:2017kzo}. They are expected to play a similar role in BMS invariant theories. An in-depth analysis of ETH requires the investigation of the Generalised Gibbs Ensemble, which has all conserved charges turned on. Thus our present work is of direct importance to the study of thermalization in non-Lorentzian CFTs. 

\item \textbf{3d Flat holography:} We have already written earlier in the section that one of the initial motivations for this work was to understand holography in 3d asymptotically flat spacetimes better, and our computation provides further evidence that this formulation of flat holography in terms of Carrollian CFTs as the boundary theory works very well. The computation of the averaged three-point function of charged Carrollian CFTs, and the reproduction of this answer from a bulk analysis of probes around charged cosmological solutions in 3d flat spacetimes provides another check of the Carrollian holography program. 

\item \textbf{Flat holography with different boundary conditions:} In \cite{Detournay:2016sfv}, it was shown that with a different set of boundary conditions than the usual Barnich-Compere boundary conditions \cite{Barnich:2006av}, the asymptotic symmetries of flat spacetimes enhance to the symmetries that we have been discussing in this paper. Our paper thus, is important for investigations into flat holography with alternative boundary conditions. 
\end{itemize}
Hence it is clearly of importance to look at these field theories with symmetries over and above BMS symmetries. 

\subsection*{Outline of this paper}
The paper is organized as follows. In Section~\ref{sec:Preliminaries}, we derive the symmetry algebra for BMSFT$\oplus\mathfrak{u}(1)$ and define the highest weight representation for the theory. In Section~\ref{sec:BMSCharacter}, we calculate the character of the algebra for the highest weight representation. We do it first by using the commutation relations and then by calculating the trace of the operator. The general structure of the Gram Matrix, which facilitates the construction of trace, is discussed in the appendix. The intrinsic calculation is followed by a limiting analysis that takes the Carroll and Galilean limit of the $\mathfrak{vir}\oplus u(1)$ character. In this section, we also derive the character for multiplet states. In Section~\ref{sec:PartitionFunction}, we carry out a Cardy-like analysis and derive the expression for the density of primary states in two different ways. One way is an approximation using a saddle point analysis, whereas the other way is an exact calculation. In section 5, we compute three-point coefficients using again a saddle point approximation and an exact method. Section~\ref{sec:BulkComputations} contains holographic bulk computations that are dual to the Cardy-like formula and three-point coefficients in Section~\ref{sec:PartitionFunction} and Section~\ref{sec:ThrePointCoefficients}, respectively. We conclude in Section~\ref{sec:Conclusions} by laying out some future work directions. Several appendices complement the main body of the paper. Appendix A recalls the basic features of BMS invariant field theories in $d=2$. Appendix B is a lightening review of required features of CFTs with additional $\mathfrak{u}(1)$ currents. Some very important details of the Hilbert space BMS-U(1) theories is contained in Appendix C. Appendix D and E contain the novelties associated with non-diagonal boost operator and multiplet states. We have deliberately not included the details of this in the main text to make the paper easily accessible to people who wish to avoid these complications. But for the interested reader, we should stress that these appendices contain a lot of novel methods in dealing with non-diagonal structures, which is new work and not available elsewhere. 

\section{BMS Field Theories with $\mathfrak{u}(1)$ Symmetry: Preliminaries}\label{sec:Preliminaries}
Following \cite{Basu:2017aqn}, we begin our explorations of BMS field theories with additional symmetry first by writing down the algebra of interest by an In{\"o}n{\"u}-Wigner contraction of the corresponding relativistic algebra. We then describe some aspects of the representation theory of these algebras. 

\subsection{Symmetry Algebra}
\label{symalg}
In the infinite radius (vanishing cosmological constant) limit, AdS$_{3}$ reduces to 3d flat space. This limit can also be performed on the asymptotic symmetries by an In{\"o}n{\"u}-Wigner contraction. Identifying $\e=\frac{1}{\ell}$, this contraction is given by
\be\label{lim}
L_{n}= \mathcal{L}_{n}-\mathcal{\bar{L}}_{-n}, \quad M_{n}= \epsilon(\mathcal{L}_{n}+\mathcal{\bar{L}}_{-n}).
\ee
It is straightforward to check that using this, one obtains the BMS$_3$ algebra \refb{BMSAL} from two copies of the Virasoro algebra \refb{vir}. The central charges are identified as 
\be{}
c_L=\frac{c-\bar{c}}{12}, \quad c_M= \frac{\e(c+\bar{c})}{12}. 
\ee
Given the Brown-Henneaux central charges of AdS$_3$: $c=\bar{c}=\frac{3 \ell}{2G}$, one precisely reproduces \refb{bc}. As mentioned before, this limit on the boundary amounts to taking the speed of light to zero and is called a Carrollian limit. Quantum field theories in 2d with BMS$_3$ as the underlying symmetries have been extensively discussed in connection with holography in asymptotically flat spacetimes. In the following, we briefly review these field theories and a collection of the results that are relevant to this work in Appendix~\ref{ApA}.

In this work we augment the BMS algebra with $\mathfrak{u}(1)$ symmetries. In particular, we will be interested in an algebra that arises in the Carrollian limit of two copies of the Vir$\oplus\mathfrak{u}(1)$ algebra: 
\begin{equation}\label{viru1}
[\mathcal{L}_{n},\mathcal{L}_{m}]=(n-m)\mathcal{L}_{n+m}+\frac{c}{12}(n^{3}-n)\delta_{n+m,0},\hspace{3mm}[\mathcal{L}_{n}, \mathcal{J}_{m}]= -m j_{n+m} ,\hspace{3mm}[\mathcal{J}_{n},\mathcal{J}_{m}]=kn\delta_{n+m,0}.
\end{equation}
Again, we devote an appendix (Appendix \ref{ApB}) for relevant details of CFTs with additional $\mathfrak{u}(1)$ symmetry. 
Using the defintions \refb{lim} 
\be{}\label{contraction}
L_{n}= \mathcal{L}_{n}-\mathcal{\bar{L}}_{-n}, \quad M_{n}= \epsilon(\mathcal{L}_{n}+\mathcal{\bar{L}}_{-n}), \quad J_{n}= \mathcal{J}_{n}- {\bar{\mathcal{J}}}_{-n}, \quad P_{n}= \epsilon(\mathcal{J}_{n}+ {\bar{\mathcal{J}}}_{-n}),
\ee
one obtains in the limit $\epsilon\to0$
\begin{subequations}\label{bmsu1}
    \begin{align}
        [L_{n},L_{m}]&=(n-m)L_{n+m}+c_{L}(n^{3}-n)\delta_{n+m,0}, \\
        [L_{n},M_{m}]&=(n-m)M_{n+m}+c_{M}(n^{3}-n)\delta_{n+m,0}, \\
        [L_{n},J_{m}]&=-m J_{n+m}, \quad 
        [L_{n},P_{m}] =-m P_{n+m},\quad
        [M_{n},J_{m}] =-m P_{m+n},  \\
        [J_{n},J_{m}]&=\kappa_J\, n\,\delta_{n+m,0}, \quad        [J_{n},P_{m}] =\kappa_P \,n\,\delta_{n+m,0}.
    \end{align}
\end{subequations}
In the above, we have only displayed the non-zero commutators. One also has the following identifications:  
\be
\kappa_J= k-\bar{k}, \quad \kappa_P = \epsilon (k +\bar{k}),
\ee
where $k$ is defined in \refb{viru1} and $\bar{k}$ is its antiholomorphic counterpart. 

\subsection{Primaries and Highest Weight Representation }
We now want to construct the representations of the algebra \refb{bmsu1}. These would correspond to the quantum states in the Hilbert space of the theory whose symmetries are given by the BMS algebra augmented by $\mathfrak{u}(1)$ charges.

\smallskip

We begin by labeling states. We note that the operators $\{L_{0}, M_{0}, J_{0}, P_{0}\}$ commute with each other, and hence they have common eigenstates. Let us label the simultaneous eigenstates as $\vert\psi\rangle\equiv\vert\Delta, \xi , j, p\rangle $, where $\Delta, \xi, j$ and $p$ are the eigenvalues of $L_{0}, M_{0}, J_{0}$ and $P_{0}$ respectively.
    \begin{equation}\label{eq:Eigenvalues}
        L_{0}|\psi\rangle=\Delta\vert\psi\rangle,\quad M_{0}|\psi\rangle=\xi\vert\psi\rangle,\quad J_{0}|\psi\rangle=j\vert\psi\rangle,\quad P_{0}|\psi\rangle=p\vert\psi\rangle. 
    \end{equation}
The commutation relations suggest that the generators $L_{n}, M_{n}, J_{n}$ and $P_{n}$ acting on the state $\vert\psi\rangle$ lowers the $L_{0}$ eigenvalue, $\Delta$ by $n$. In analogy with 2d CFTs and earlier works on BMS field theories, we define a primary state $\vert\psi_{p}\rangle$ whose $L_{0}$ eigenvalue $\Delta_{p}$ is bounded from below for a given value of $\xi$, $j$ and $p$. This is equivalent to imposing the conditions: 
\begin{equation}
    L_{n}\vert\psi_{p}\rangle= M_n\vert\psi_{p}\rangle= J_n\vert\psi_{p}\rangle= P_n\vert\psi_{p}\rangle=0, \quad\text{for}\quad n>0.
\end{equation}
These states $\vert\psi_{p}\rangle$ define a  spectrum of primary states in the theory, which will be central to our constructions going forward. The vacuum state in the theory is defined as a primary that is further annihilated by the operators $\{L_{0,\pm 1},M_{0,\pm 1},J_{0},P_{0}\}$. 

A highest weight module over any primary state can be constructed by acting with raising operators. Any primary state $\vert\psi_{p}\rangle$ gives rise to a tower of descendant states when acted upon by the generators $L_{n}, M_{n}, J_{n}$ and $P_{n}$ with $n<0$, which increase the $\Delta$ eigenvalue and hence are creation operators. The primary and its descendants form a highest weight module $\mathcal{B}(c_{L},c_{M},\kappa_{J},\kappa_{P},\Delta,\xi,j,p)$. For the sake of brevity, we omit the suffix $p$ for denoting a primary state. The Hilbert space for the theory is a direct sum of all the modules obtained from all the primaries in the theory.
The state at level $N$  is given by
\begin{equation}
  \vert\psi_N\rangle \equiv  L^{n_{1}}_{-1}...L^{n_{a}}_{-a}M^{m_{1}}_{-1}...M^{m_{b}}_{-b}J^{r_{1}}_{-1}...J^{r_{c}}_{-c}P^{s_{1}}_{-1}...P^{s_{d}}_{-d}\vert\psi\rangle=L_{\Vec{n}}M_{\Vec{m}}J_{\Vec{r}}P_{\Vec{s}}\vert\psi\rangle,
\end{equation}
where $\Vec{n}=(n_{1},n_{2},...,n_{a})$, $\Vec{m}=(m_{1},m_{2},...,m_{b}),\Vec{r}=(r_{1},r_{2},...,r_{c}),\Vec{s}=(s_{1},s_{2}...s_{d})$. Acting with $L_{0}$ on such a state gives 
\begin{equation}
    L_0 \vert\psi_N\rangle = L_{0}L_{\Vec{n}}M_{\Vec{m}}J_{\Vec{r}}P_{\Vec{s}}\vert\psi\rangle=(N+\Delta)\vert\psi_N\rangle,
\end{equation}
where $N=\Big(\sum_{i} i n_{i}+\sum_{j} j m_{j}+\sum_{k} k r_{k}+\sum_{l} l s_{l}\Big)$.
$N$ is the level of states, and the states are organized according to their level. Level 0 consists of only the primary state $\vert\psi_{p}\rangle$. Level 1 is a four-dimensional space spanned by \hspace{1mm}$L_{-1}\vert\psi_{p}\rangle$,\hspace{1mm} $M_{-1}\vert\psi_{p}\rangle$, $J_{-1}\vert\psi_{p}\rangle$, and $P_{-1}\vert\psi_{p}\rangle$.\\

\section{The BMS $\oplus$ $\mathfrak{u}(1)$ Character}\label{sec:BMSCharacter}
This section is focused on the development and computation of the character of the BMS$\oplus\mathfrak{u}(1)$ algebra. We will do this in two distinct ways. First, we will only use the symmetries of the algebra and the definition of the highest weight representations we constructed above to compute the character. We will then proceed to rederive our results by looking at the singular limit from 2d CFTs with $\mathfrak{u}(1)$ charge which we have alluded to above. 

\subsection{Intrinsic Analysis}
A twisted torus is obtained by twisting the cylinder by an angle $\Omega$ and gluing together its two ends. For 2d BMSFT, the partition function defined on such a twisted torus is given by: 
\begin{align}
    Z_{\textrm{BMS}}=\Tr\left[e^{2\pi i\sigma(L_{0}-c_{L}/2)}e^{2\pi i \rho(M_{0}-c_{M}/2)}\right]. 
\end{align}
Now, for a BMSFT with $\mathfrak{u}(1)$ currents, the partition function gets modified to
\begin{equation}
    Z=\Tr\left[e^{2\pi i\sigma(L_{0}-c_{L}/2)}e^{2\pi i\rho(M_{0}-c_{M}/2)}e^{2\pi i\alpha J_{0}}e^{2\pi i\gamma P_{0}}\right]. 
\end{equation} 
The modular parameters $\sigma, \rho, \alpha$ and $\gamma$ are related\footnote{In comparison to relativistic CFTs on a torus these relations might seem counter-intuitive. However, for BMSFTs, these are sensible Hamiltonian interpretations of the modular parameters as we will see later on.} to the twist of the torus $\Omega$, inverse temperature $\beta$, fermion number $\mu_J$ and $\mathfrak{u}(1)$ charge $\mu_P$ as
    \begin{equation}\label{eq:ModularParametersHamiltonian}
        2\pi \sigma = i \Omega,\quad 2\pi \rho = \beta,\quad \alpha = i \mu_J,\quad \gamma = \mu_P.
    \end{equation}
Since the generators $L_{0}$, $M_{0}$, $P_{0}$ and $J_{0}$ do not mix states belonging to different modules we can define the trace for a particular module $\mathcal{B}(\Delta,\xi,j,p,\kappa_{P},\kappa_{J}, c_{L},c_{M})$ as   
\begin{equation}
    \chi_{(\Delta,\xi,j,p..)}=\Tr_{(\Delta,\xi,j,p..)} e^{2\pi i\sigma(L_{0}-c_{L}/2)}e^{2\pi i\rho(M_{0}-c_{M}/2)}e^{2\pi i\alpha J_{0}}e^{2\pi i\gamma P_{0}}.
\end{equation}
 For any operator $\hat{O}$, acting on a linear vector space $V$, the action of the operator on the basis $\vert\psi_{i}\rangle$ gives
 \begin{equation}
     \hat{O}\vert\psi_{i}\rangle=\hat{O}_{ij}\vert\psi_{j}\rangle. 
 \end{equation}
 For our analysis, we will be concerned with the operator 
\be\label{O}
\hat{O} = \mathcal{O} = e^{2\pi i\sigma(L_{0}-c_{L}/2)}e^{2\pi i\rho(M_{0}-c_{M}/2)}e^{2\pi i\alpha J_{0}}e^{2\pi i\gamma P_{0}}.
\ee
Below we calculate the expression for the BMS$\oplus\mathfrak{u}(1)$ character. The operator $\mathcal{O}$, defined above in \refb{O}, does not mix states belonging to different levels. We can thus perform this calculation level by level. We will explicitly compute the result for the first level and then based on this derive the result for generic level $N$. The details of the general construction are rather involved and are detailed in Appendix \ref{ApC}.  

\subsubsection*{Trace at Level 1} 
To find the trace at level 1 one has to calculate the action of $\mathcal{O}$ on the states $L_{-1}\vert\psi_{p}\rangle$, $M_{-1}\vert\psi_{p}\rangle$, $J_{-1}\vert\psi_{p}\rangle$ and $P_{-1}\vert\psi_{p}\rangle$, where for the sake of brevity we have denoted $\vert\Delta,\xi,j,p\rangle$ as $\vert\psi_{p}\rangle$. 
$J_{0}$ and $P_{0}$ commute with all other generators. The states are eigenstates of $L_{0}$ with eigenvalue $\Delta+1$. Thus, one is only left with determining the action of $e^{2\pi i\rho(M_{0}-c_{M}/2)}$ on the states. By use of the commutators of the algebra, this is a straightforward exercise and yields: 
\begin{eqnarray}
e^{2\pi i\rho M_{0}}L_{-1}\vert\psi_{p}\rangle=\left[1+2\pi i\rho M_{0}+\frac{(2\pi i\rho)^{2}M_{0}^{2}}{2}+...\right]L_{-1}\vert\psi_{p}\rangle =e^{2\pi i\rho\xi}(L_{-1}+2\pi i\rho M_{-1})\vert\psi_{p}\rangle .
\end{eqnarray}
Likewise, one has
\begin{subequations}

\begin{align}
e^{2\pi i\rho M_{0}}M_{-1}\vert\psi_{p}\rangle=e^{2\pi i\rho \xi}M_{-1}\vert\psi_{p}\rangle, \\ e^{2\pi i\rho M_{0}}P_{-1}\vert\psi_{p}\rangle=e^{2\pi i\rho\xi}P_{-1}\vert\psi_{p}\rangle, \\
e^{2\pi i\rho M_{0}}J_{-1}\vert\psi_{p}\rangle=e^{2\pi i\rho\xi}(J_{-1}+2\pi i\rho P_{-1}\vert\psi_{p})\rangle . 
\end{align}
\end{subequations}
Putting the above equations together, one obtains
\begin{equation}
\mathcal{\hat{O}}\begin{bmatrix}L_{-1}\vert\psi_{p}\rangle\\M_{-1}\vert\psi_{p}\rangle\\J_{-1}\vert\psi_{p}\rangle\\P_{-1}\vert\psi_{p}\rangle\end{bmatrix}=e^{2\pi i\sigma(\Delta+1-c_{L}/2)}e^{2\pi i\rho(\xi-c_{M}/2}e^{2\pi i\alpha j}e^{2\pi i\gamma p}\begin{bmatrix}1 & 2\pi i\rho & 0 & 0\\
    0 & 1 & 0 & 0\\
    0 & 0 & 1 & 2\pi i\rho\\
    0 & 0 & 0 & 1\end{bmatrix}\begin{bmatrix}L_{-1}\vert\psi_{p}\rangle\\M_{-1}\vert\psi_{p}\rangle\\J_{-1}\vert\psi_{p}\rangle\\P_{-1}\vert\psi_{p}\rangle\end{bmatrix}.
\end{equation}
The trace of the operator at level 1 is then given by
\begin{equation}
    \Tr \mathcal{O}_{1}=4 e^{2\pi i\sigma(\Delta+1-c_{L}/2)}e^{2\pi i\rho(\xi-c_{M}/2)}e^{2\pi i\alpha j}e^{2\pi i\gamma p} .
\end{equation}

\subsubsection*{Trace at Generic Level N}
A general state is defined by $L_{\Vec{n}}M_{\Vec{m}}J_{\Vec{r}}P_{\Vec{s}}|\psi_{p}\rangle$. The $L_{0}$ eigenvalue of the state is $(\Delta+N)$. The operators $J_{0}$ and $P_{0}$ commute with all the other operators. Thus,
\begin{eqnarray}
e^{2\pi i\sigma L_{0}} L_{\Vec{n}}M_{\Vec{m}}J_{\Vec{r}}P_{\Vec{s}}|\psi_{p}\rangle&=&e^{2\pi i\sigma(\Delta+N)}L_{\Vec{n}}M_{\Vec{m}}J_{\Vec{r}}P_{\Vec{s}}|\psi_{p}\rangle,\nonumber\\ e^{2\pi i\beta P_{0}}L_{\Vec{n}}M_{\Vec{m}}J_{\Vec{r}}P_{\Vec{s}}|\psi_{p}\rangle&=&e^{2\pi i\gamma p}L_{\Vec{n}}M_{\Vec{m}}J_{\Vec{r}}P_{\Vec{s}}|\psi_{p}\rangle,\nonumber\\
   e^{2\pi i\alpha J_{0}}L_{\Vec{n}}M_{\Vec{m}}J_{\Vec{r}}P_{\Vec{s}}|\psi\rangle&=&e^{2\pi i\alpha j}L_{\Vec{n}}M_{\Vec{m}}J_{\Vec{r}}P_{\Vec{s}}|\psi\rangle.
\end{eqnarray}
$M_{0}$ does not commute with all the other operators. The action of $M_{0}$ on the state can be expressed as 
\begin{equation}
   e^{2\pi i\rho M_{0}}L_{\Vec{n}}M_{\Vec{m}}J_{\Vec{r}}P_{\Vec{s}}|\psi_{p}\rangle=[ e^{2\pi i\rho M_{0}},L_{\Vec{n}}M_{\Vec{m}}J_{\Vec{r}}P_{\Vec{s}}]|\psi_{p}\rangle+e^{2\pi i\rho \xi}L_{\Vec{n}}M_{\Vec{m}}J_{\Vec{r}}P_{\Vec{s}}|\psi_{p}\rangle.
\end{equation}
The first term on the right-hand side changes into different states at the same level as can be verified by commuting $M_{0}$ with the other generators. As we will show in Appendix \ref{ApC}, even in the presence of the first term on RHS, the character contribution will be given by only $dim_{N}$, where $dim_{N}$ is the total number of independent states at level $N$. Thus, ultimately one has: 
\begin{align}\label{tr1}
    \Tr\hat{\mathcal{O}}_{N}=dim_{N}\hspace{1mm}e^{2\pi i\sigma(\Delta+N-c_{L}/2)}e^{2\pi i\rho(\xi-c_{M}/2)}e^{2\pi i\alpha j}e^{2\pi i\gamma p}.
\end{align}
Summing over the levels $N$, the character of a BMS$\times \mathfrak{u}(1)$ primary state is given by
\begin{eqnarray}\label{tr2}
   \chi_{(\Delta,\xi,j,p)}&=&e^{-2\pi i(\sigma c_{L}/2+\rho c_{M}/2)}e^{2\pi i\sigma\Delta}e^{2\pi i\rho\xi}e^{2\pi i\alpha j}e^{2\pi i\gamma p} \sum_{N}dim_{N} e^{2\pi i\sigma N}.
      \end{eqnarray}
      
\subsection{The Gram Matrix}
We now employ a different formulation to re-derive the expression for the BMS$\oplus\mathfrak{u}(1)$ character. Since our chosen basis is neither orthogonal nor normalized in a general module, we use the following expression to calculate the trace of any operator. 
\begin{align}
\Tr\hat{\mathcal{O}} =\sum_{i,j}K^{ij} \langle\psi_{i}\vert\hat{\mathcal{O}}\vert\psi_{j}\rangle, 
\end{align}
where $\langle\psi_{i}\vert\psi_{j}\rangle=K_{ij}$ is the $(ij)$-th element of the Gram matrix formed by the inner products of the basis states, and $K^{ij}$ is the matrix inverse. One can calculate the Gram matrix for each level separately as the inner products of two states belonging to different levels are zero. This yields
\begin{align}
    \chi_{\Delta,\xi,j,p}=\sum_{i,j,N}K^{ij}_{(N)}\langle\psi_{i}^{(N)}\vert\hat{\mathcal{O}}\vert\psi_{j}^{(N)}\rangle .
\end{align}
We now explicitly carry out the calculation for the Gram matrix at level 1 and show that the trace for level 1 matches with the expression derived in \eqref{tr1}. The states are ordered as $\vert\psi_{1}\rangle=L_{-1}\ket{\psi}$ , $\vert\psi_{2}\rangle=
J_{-1}\ket{\psi}$, $\vert\psi_{3}\rangle=M_{-1}\ket{\psi}$, $\vert\psi_{3}\rangle=P_{-1}\ket{\psi}$, where $\ket{\psi}= \vert\Delta,\xi,j,p\rangle$ is the primary state. Using the BMS$\oplus\mathfrak{u}(1)$ commutation relations, the Gram matrix at this order turns out to be  
\begin{equation}
    K^{ij}_{(1)}=\begin{bmatrix} 2\Delta & j & 2\xi & p\\
    j & \kappa_{J} & p & \kappa_{P}\\
    2\xi & p & 0 & 0\\
    p & \kappa_{P} & 0 & 0\end{bmatrix} .
    \end{equation}
And the matrix $\langle\psi_{i}\vert\hat{\mathcal{O}}\vert\psi_{j}\rangle$ is 
\begin{equation} 
\langle\psi_{i}\vert\hat{\mathcal{O}}\vert\psi_{j}\rangle= e^{2\pi i\left[\sigma(\Delta+1-\frac{c_{L}}{2}) + \rho(\xi-\frac{c_{M}}{2}) + \alpha j +\gamma p \right]}
    \small\begin{bmatrix}2\Delta+4\pi i\rho\xi & j+2\pi i\rho p & 2\xi & p\\
    j+2\pi i\rho p & \kappa_{J}+2\pi i\rho\kappa_{P} & p & \kappa_{P}\\
    2\xi & p & 0 & 0\\
    p & \kappa_{P} & 0 & 0\\   
    \end{bmatrix}.\normalsize  
\end{equation}   
It is straightforward to verify from the above two matrices that the trace at level 1 is the same as \eqref{tr1} with $N=1$. In Appendix~\ref{ApC} we give a general proof showing that for level $N$, the trace is given by \eqref{tr1}. Hence the character for the module $\mathcal{B}(\Delta,\xi,p,j,c_{L},c_{M})$ is 
\begin{equation}\label{char1}
    \chi_{(\Delta,\xi,j,p)}=e^{2\pi i\sigma(\Delta-c_{L}/2)}e^{2\pi i\rho(\xi-c_{M}/2)}e^{2\pi i\alpha j} e^{2\pi i\gamma p}\sum_{N} dim_{N} e^{2\pi i\sigma N} .  
\end{equation}
For a generic level $N$ let us assume that the contributions from the $L$, $M$, $P$ and $J$ generators are $n_{L}$, $n_{M}$, $n_{P}$ and $n_{J}$, respectively. For non-vacuum primaries the number of states is given by 
\begin{align}
 dim_{N}=\sum_{\substack{n_{L},n_{M},n_{P},n_{J} \\ n_{L}+n_{M}+n_{P}+n_{J}=N}} p(n_{L}) p(n_{M})  p(n_{P})  p(n_{J}) ,   
\end{align}
where $p(n)$ is the partition of the integer $n$. The vacuum is annihilated by $L_{-1}$ and $M_{-1}$, so the descendants of the vacuum will not have states containing these generators. In that case, the number of states at level $N$ is given by
\begin{align}
dim_{N}^{(vac)} = \sum_{\substack{n_{L},n_{M},n_{P},n_{J} \\ n_{L}+n_{M}+n_{P}+n_{J}=N}} (p(n_{L})-p(n_{L}-1))(p(n_{M})-p(n_{M}-1)) p(n_{J})p(n_{P}).
\end{align}
Putting back the expression for $dim_{N}$ in \eqref{char1} one obtains
\begin{eqnarray} 
\label{chcexp}
    \chi_{(\Delta,\xi,j,p)}&=&e^{2\pi i\left[\sigma(\Delta+1-\frac{c_{L}}{2}) + \rho(\xi-\frac{c_{M}}{2}) + \alpha j +\gamma p \right]} \sum_{N}\sum_{\substack{n_{L}+n_{M}\\+n_{P}+n_{J}=N}} p(n_{L}) p(n_{M})  p(n_{P})  p(n_{J}) e^{2\pi i\sigma N}  \nonumber\\
    &=&e^{2\pi i\sigma(\Delta-c_{L}/2)}e^{2\pi i\rho(\xi-c_{M}/2)}e^{2\pi i\alpha j}e^{2\pi i\gamma p}\nonumber\\
    &&\sum_{n_{L}} p(n_{L})e^{2\pi i\sigma n_{L}} \sum_{n_{M}} p(n_{M})e^{2\pi i\sigma n_{M}} \sum_{n_{J}} p(n_{J})e^{2\pi i\sigma n_{J}} \sum_{n_{P}} p(n_{P})e^{2\pi i\sigma n_{P}} \nonumber\\
    &=& e^{2\pi i\sigma(\Delta-c_{L}/2)}e^{2\pi i\rho(\xi-c_{M}/2)}e^{2\pi i\alpha j}e^{2\pi i\gamma p}\frac{1}{\phi^{4}(\sigma)}.
\end{eqnarray}
In the above, we have used the function $\phi(\sigma)$ which is related to the Dedekind  $\eta$ function in the following way
\begin{equation}\label{phidef}
 \sum_{n=0}^{\infty}p(n)e^{2\pi i\sigma n}=\prod_{n=1}^{\infty}\frac{1}{(1-e^{2\pi i\sigma n})} = \frac{1}{\phi(\sigma)} = \frac{e^{2\pi i \sigma/24}}{\eta(\sigma)}.
\end{equation}
Similarly, for the vacuum, we obtain 
\begin{equation}\label{eq:VacuumChar}
    \chi_{vac}=e^{-\pi i\sigma c_{L}}e^{-\pi i\rho c_{M}}\frac{(1-e^{2\pi i\sigma})^{2}}{\phi^{4}(\sigma)}.
\end{equation}

\subsection{Limiting Analysis from Virasoro $\times$ $\mathfrak{u}(1)$ Character}
In this section, we show how the BMS$\oplus\mathfrak{u}(1)$ character arises from singular limits of the highest weight characters of the relativistic Virasoro $\mathfrak{u}(1)$ Kac-Moody algebra. As we have shown in Section~\ref{symalg}, the BMS$\oplus\mathfrak{u}(1)$ algebra arises as an ultra-relativistic contraction of two copies of a parent $\mathfrak{vir}\oplus\mathfrak{u}(1)$ algebra \eqref{viru1}. Interestingly, another contraction of the parent relativistic abelian Kac-Moody algebra yields the same BMS$\oplus\mathfrak{u}(1)$ algebra. 
This is given by 
\be{}\label{contr1}
L_{n}= \mathcal{L}_{n}+\mathcal{\bar{L}}_{n}, \quad M_{n}= \epsilon(\mathcal{L}_{n}-\mathcal{\bar{L}}_{n}), \quad J_{n}= \mathcal{J}_{n} + {\bar{\mathcal{J}}}_{n}, \quad P_{n}= \epsilon(\mathcal{J}_{n} - {\bar{\mathcal{J}}}_{n}).
\ee
It is straightforward to check that this contraction yields \eqref{bmsu1}, with the identifications 
\be
\kappa_J= k+\bar{k}, \quad \kappa_P = \epsilon (k -\bar{k}).
\ee
This limit can be interpreted as a non-relativistic (NR) or Galilean limit where the speed of light, $c$, goes to infinity, as opposed to the ultra-relativistic or Carrollian limit where $c\to0$. The isomorphism of the algebras obtained in these diametrically opposite limits is a well-known consequence of being in two dimensions first noticed in \cite{Bagchi:2010eg}, as the algebra does not distinguish between spatial and temporal contractions.

The reason why it is important to stress these different limits is that while the two limits yield the same algebra, they have very different actions on the parent representations. The Virasoro-Kac-Moody highest weight representations are given by 
\bes
\bea{}
&&\mathcal{L}_{0} |h, \bar{h}, \mathcal{Q}, \bar{\mathcal{Q}}\rangle = h |h, \bar{h}, \mathcal{Q}, \bar{\mathcal{Q}}\rangle,  \quad \mathcal{J}_{0} |h, \bar{h}, \mathcal{Q}, \bar{\mathcal{Q}}\rangle = \mathcal{Q} |h, \bar{h},\mathcal{Q},\bar{\mathcal{Q}}\rangle, \\
&& \mathcal{L}_{n} |h, \bar{h}, \mathcal{Q}, \bar{\mathcal{Q}}\rangle = 0, \quad \mathcal{J}_{n} |h, \bar{h}, \mathcal{Q}, \bar{\mathcal{Q}}\rangle=0, \quad \forall n>0,
\eea
\ees
and similarly for the anti-holomorphic sector. If one takes the non-relativistic contraction \eqref{contr1} on the relativistic highest weight representation, it is clear that this leads to the BMS$\oplus\mathfrak{u}(1)$ highest weight representation that we have been discussing so far. But the Carroll contraction \eqref{contraction} does not. This yields
\bes
\bea{}
&& L_{0}|\psi\rangle=\Delta\vert\psi\rangle,\quad M_{0}|\psi\rangle=\xi\vert\psi\rangle,\quad J_{0}|\psi\rangle=j\vert\psi\rangle,\quad P_{0}|\psi\rangle=p\vert\psi\rangle. \\
&& M_n |\psi\rangle= 0, \quad P_n |\psi\rangle= 0 \quad \forall n\neq 0.
\eea
\ees
Here, as before, $\vert\psi\rangle\equiv\vert\Delta, \xi , j, p\rangle$. The representations so defined are the induced representations of the BMS$\oplus\mathfrak{u}(1)$ algebra. Interestingly, the characters of both these representations seem to coincide. This is the $\mathfrak{u}(1)$ extended version of what was earlier encountered in the pure BMS case in \cite{Bagchi:2019unf}.  
    
\subsection*{Carroll limit}
We begin by looking at the ultra-relativistic or Carrollian limit on the partition function of CFT with $\mathfrak{u}(1)$ symmetry. We use the generators as derived in \eqref{contraction} by taking ultra-relativistic contractions of generators for CFT with $\mathfrak{u}(1)$ 
\begin{eqnarray}
Z(\tau,\Bar{\tau},\nu,\Bar{\nu})&=& \Tr  e^{2\pi i\tau(\mathcal{L}_{0}-c/2)} e^{-2\pi i\bar{\tau}(\bar{\mathcal{L}}_{0}-\bar{c}/2)} e^{2\pi i\nu j_{0}}e^{-2\pi i\bar{\nu} \bar{j}_{0}}\nonumber\\
&=& \Tr (e^{2\pi i (L_{0}-c_{L})(\tau+\bar{\tau})/2}e^{2\pi i (M_{0}-c_{M})(\tau-\bar{\tau})/2\epsilon}e^{2\pi i J_{0}(\nu+\bar{\nu})/2} e^{2\pi i P_{0}(\nu-\bar{\nu})/2\epsilon}).\nonumber
\end{eqnarray}
If we want the parent partition function to smoothly translate to the BMS$\oplus\mathfrak{u}(1)$ partition function, then we  need to define the modular parameters as follows.
\begin{equation}\label{cntrctn}
   \frac{\tau+\bar{\tau}}{2}=\sigma,\hspace{1mm} \frac{\tau-\bar{\tau}}{2\epsilon}=\rho,
   \hspace{1mm}\frac{\nu+\bar{\nu}}{2}=\alpha,
   \hspace{1mm}\frac{\nu-\bar{\nu}}{2\epsilon}=\gamma .
\end{equation}
Hence the partition function for BMSFT with $\mathfrak{u}(1)$ symmetry is, 
\begin{equation}
Z(\sigma,\rho,\alpha,\gamma)=\Tr e^{2\pi i\sigma (L_{0}-c_{L}/2)}e^{2\pi i\rho (M_{0}-c_{M}/2)}e^{2\pi i\alpha J_{0}} e^{2\pi i\gamma P_{0}}.    
\end{equation}
The eigenstates of $\mathcal{L}_{0}$ in Virasoro+$\mathfrak{u}(1)$ module can be expressed as 
\begin{align}
    (\mathcal{L}_{-1})^{l_{1}} (\mathcal{L}_{-2})^{l_{2}}...(\mathcal{L}_{-k})^{l_{k}}(\mathcal{J}_{-1})^{r_{1}}(\mathcal{J}_{-2})^{r_{2}}...(\mathcal{J}_{-s})^{r_{s}}\vert h,\mathcal{Q}\rangle.
\end{align}
This is the $N^{\textrm{th}}$ descendant of the module where $N=n+m=\sum_{i}il_{i}+\sum_{i^{\prime}}i^{\prime}r_{i^{\prime}}$. It is straightforward to see that the number of states with eigenvalue $h+N$ is given by $\sum_{m+n=N}p(n)p(m)$.   
In this section, we show that the character for BMSFT with $\mathfrak{u}(1)$ symmetry can be reached through a limiting analysis of the character of $\mathfrak{vir}\oplus\mathfrak{u}(1)$. The $\mathfrak{vir}\oplus\mathfrak{u}(1)$ character is
\begin{eqnarray}\label{ChrH}
\chi_{h,\mathcal{Q},c}&=&\sum_{N}\sum_{m, n, m+n=N} p(m)p(n)e^{2\pi i\tau (m+n)} e^{2\pi i\nu \mathcal{Q}}e^{2\pi i\tau(h-\frac{c}{24})}=  \frac{e^{2\pi i\tau(h-\frac{c}{24})} e^{2\pi i\nu \mathcal{Q}}}{\phi^{2}(\tau)}.
\end{eqnarray}
The function $\phi$ is defined as in (\ref{phidef}). Taking the UR contraction after multiplying \eqref{ChrH} with the non-holomorphic part, we get,  
\begin{align}
    \chi_{\Delta,\xi,j,p} &\equiv \lim_{\epsilon\rightarrow 0}\chi\bar{\chi}=\lim_{\epsilon\rightarrow 0}\frac{e^{\frac{\pi i \tau}{6}}e^{2\pi i \tau(h-\frac{c}{24})}e^{2\pi i \nu \mathcal{Q}}e^{\frac{-\pi i \bar{\tau}}{6}}e^{-2\pi i \tau(\bar{h}-\frac{\bar{c}}{24})}e^{-2\pi i \bar{\nu} \bar{\mathcal{Q}}}}{(\eta(\tau))^2(\eta(\bar{\tau}))^2}\\
    &=\lim_{\epsilon\rightarrow 0}\frac{1}{(\eta(\tau))^2(\eta(\bar{\tau}))^2}\bigg(e^{\frac{\pi i}{6}(\sigma+\epsilon\rho-(\sigma-\epsilon\rho))}e^{2\pi i\{(\sigma+\epsilon\rho)\frac{1}{2}(\Delta+\frac{\xi}{\epsilon})-(\sigma-\epsilon\rho)\frac{1}{2}(-\Delta+\frac{\xi}{\epsilon})\}} \times \nonumber\\
    &\, e^{2\pi i\{(\alpha+\epsilon\gamma)\frac{1}{2}(j+\frac{p}{\epsilon})-(\alpha-\epsilon\gamma)\frac{1}{2}(-j+\frac{p}{\epsilon})\}} e^{-2\pi i\{(\sigma+\epsilon\rho)\frac{1}{4}(c_L+\frac{c_M}{\epsilon})-(\sigma-\epsilon\rho)\frac{1}{4}(-c_L+\frac{c_M}{\epsilon})\}}\bigg)\nonumber\\
    &=\lim_{\epsilon\rightarrow 0}\frac{1}{(\eta(\tau))^2(\eta(\bar{\tau}))^2}\bigg(e^{2\pi i (\sigma(\Delta-\frac{c_L}{2})+\rho(\xi-\frac{c_M}{2})+\alpha j+\gamma p)}\bigg).
\end{align}
Now considering the denominator,
\begin{align}
    \lim_{\epsilon\rightarrow 0}(\eta(\tau))^2(\eta(\bar{\tau}))^2&=\lim_{\epsilon\rightarrow 0}e^{\frac{2\pi i(\sigma+\epsilon\rho)}{12}}\bigg(\prod_{n=1}^{\infty}(1-e^{2\pi i n(\sigma+\epsilon\rho)})\bigg)^2e^{-\frac{2\pi i (\sigma-\epsilon\rho)}{12}}\bigg(\prod_{n=1}^{\infty}(1-e^{-2\pi i n(\sigma-\epsilon\rho)})\bigg)^2\nonumber\\
    &=\bigg(e^{\frac{2\pi i\sigma}{24}}\prod_{n=1}^{\infty}(1-e^{2\pi i n\sigma})\bigg)^2\bigg(e^{-\frac{2\pi i\sigma}{24}}\prod_{n=1}^{\infty}(1-e^{-2\pi i n\sigma})\bigg)^2=(\eta(\sigma))^2(\eta(-\sigma))^2 .
\end{align}
If $\sigma$ is purely real then $\eta(-\sigma)=\Bar{(\eta(\sigma))}$, which implies
\begin{align}
    \lim_{\epsilon\rightarrow 0}(\eta(\tau))^2(\eta(\bar{\tau}))^2=|\eta(\sigma)|^4=|\phi(\sigma)|^4 .
\end{align}
So for the character we get, 
\begin{align}
    \chi_{\Delta,\xi,j,p}=\frac{1}{|\phi(\sigma)|^4}\bigg(e^{2\pi i (\sigma(\Delta-\frac{c_L}{2})+\rho(\xi-\frac{c_M}{2})+\alpha j+\gamma p)}\bigg) .
\end{align}
This matches with our intrinsic analysis carried out earlier {\footnote {This is true up to the modulus sign, which we comment on later.}}. We here re-emphasize that the Carroll limit on the highest weight characters of the $\mathfrak{vir}\oplus\mathfrak{u}(1)$ algebra is supposed to give one the characters for the induced representation of the BMS$\oplus\mathfrak{u}(1)$ algebra. It seems intriguing that this matches with the intrinsic analysis of the highest weight character of the BMS$\oplus\mathfrak{u}(1)$. 
    
\subsection*{Galilean limit}    
We now focus on the non-relativistic or Galilean contraction. Since the NR contraction maps highest weights to highest weights, one might expect to recover what we obtained using the intrinsic analysis, and this is indeed what happens. The NR limit involves defining the following new quantities 
\bes
\begin{align}
    &(\tau,\bar{\tau})=(\pm\sigma-\epsilon\rho) \ , \ (\nu,\bar{\nu})=(\pm\alpha-\epsilon \gamma),\\
    &(h,\bar{h})=\frac{1}{2}(\Delta\mp\frac{\xi}{\epsilon}) \ , \ (\mathcal{Q},\bar{\mathcal{Q}})=\frac{1}{2}(j\mp \frac{p}{\epsilon}) \ , \ (c,\bar{c})=6(c_L\mp\frac{c_M}{\epsilon}).
\end{align}
\ees
In this limit, the character looks like
\begin{align}
    \chi_{\Delta,\xi,j,p}&\equiv\lim_{\epsilon\rightarrow 0}\chi\bar{\chi}=\lim_{\epsilon\rightarrow 0}\frac{e^{\frac{\pi i \tau}{6}}e^{2\pi i \tau(h-\frac{c}{24})}e^{2\pi i \nu \mathcal{Q}}e^{\frac{-\pi i \bar{\tau}}{6}}e^{-2\pi i \tau(\bar{h}-\frac{\bar{c}}{24})}e^{-2\pi i \bar{\nu} \bar{\mathcal{Q}}}}{(\eta(\tau))^2(\eta(\bar{\tau}))^2} \nonumber\\
    &=\lim_{\epsilon\rightarrow 0}\frac{e^{\frac{\pi i \sigma}{3}}}{(\eta(\tau))^2(\eta(\bar{\tau}))^2}\bigg(e^{2\pi i (\sigma(\Delta-\frac{c_L}{2})+\rho(\xi-\frac{c_M}{2})+\alpha j+\gamma p)}\bigg).
\end{align}
The denominator gives
\begin{align}
    \lim_{\epsilon\rightarrow 0}(\eta(\tau))^2(\eta(\bar{\tau}))^2  =\bigg(e^{\frac{2\pi i\sigma}{24}}\prod_{n=1}^{\infty}(1-e^{2\pi i n\sigma})\bigg)^4=(\eta(\sigma))^4 .
\end{align}
Hence,
\begin{align}
    \chi_{\Delta,\xi,j,p}&=\frac{e^{\frac{\pi i \sigma}{3}}}{(\eta(\sigma))^4}\bigg(e^{2\pi i (\sigma(\Delta-\frac{c_L}{2})+\rho(\xi-\frac{c_M}{2})+\alpha j+\gamma p)}\bigg)\nonumber\\
    &=\frac{1}{(\phi(\sigma))^4}\bigg(e^{2\pi i (\sigma(\Delta-\frac{c_L}{2})+\rho(\xi-\frac{c_M}{2})+\alpha j+\gamma p)}\bigg),
\end{align}
which is the same as what we obtained from the intrinsic one. Thus, one has $(\phi(\sigma))^2=|\phi(\sigma)|^2$, and then the character can be written as,
\begin{align}
    \chi_{\Delta,\xi,j,p}=\frac{1}{|\phi(\sigma)|^4}\bigg(e^{2\pi i (\sigma(\Delta-\frac{c_L}{2})+\rho(\xi-\frac{c_M}{2})+\alpha j+\gamma p)}\bigg).
\end{align}
This matching of the character for the intrinsic and limit cases, as we said before, is not a surprise and, indeed, is a robust check of our analysis. However, that the characters of the induced representations obtained from the Carroll limit and the highest weight characters constructed intrinsically also seem to match is rather remarkable. This suggests that calculations based on these characters as the basic building blocks, e.g., of the partition function and density of states (of primaries) of the theory, are independent of the underlying representation. However, there is an important point to note. The range of validity for the modular parameters $\rho, \s$ for the two representations, the highest weight and the induced, need not be the same. For the highest weights, $\s$ is purely imaginary, while for the induced $\s$ is purely real. Hence even though the same expression for the characters exists for these two representations, the intricate details are not quite the same as one would expect.     

\subsection{Character for Multiplet States}
In this section we discuss the character of a certain BMS$\oplus\mathfrak{u}(1)$ module is constituted of simultaneously diagonalizable $L_{0}$, $P_{0}$ and $J_{0}$ and a block-diagonal $M_{0}$. However, for the purpose of the next sections, we use the expression of character as derived in \eqref{chcexp}.   
We start from a BMS$\oplus\mathfrak{u}(1)$ highest-weight module on a plane\footnote{For details on highest-weight BMS modules see, e.g., \cite{Bagchi:2019unf}.} with primaries and their descendants and label a primary operator $\mathcal{O}=\mathcal{O}(0,0)$ by eigenvalues $(\Delta,\xi,j, p)$ of ($L_{0},M_{0},J_{0},P_{0}$) 
\begin{equation}\label{chh1}
[L_{0},\mathcal{O}]=\Delta\mathcal{O}\hspace{5mm}[M_{0},\mathcal{O}]=\xi\mathcal{O},\hspace{5mm}[P_{0},\mathcal{O}]=p\mathcal{O},\hspace{5mm}[J_{0},\mathcal{O}]=j\mathcal{O}.
\end{equation}
The highest weight conditions are
\begin{equation}\label{chh2}
    [L_{n},\mathcal{O}]=0,\hspace{5mm}[M_{n},\mathcal{O}]=0,\hspace{5mm}[P_{n},\mathcal{O}]=0,\hspace{5mm}[J_{n},\mathcal{O}]=0,\hspace{5mm}n>0.
\end{equation}
In \cite{Bagchi:2009ca} it was shown that the the descendent operators $[L_{-n},\mathcal{O}]$ are not an eigenstate of $M_{0}$, even if $\mathcal{O}$ is a common eigenstate of $L_0$ and $M_0$.  Hence there is a possibility that $L_{0}$ and $M_{0}$ are not diagonal even if we restrict our attention to the subspace spanned by the primary states. Extending the analysis in \cite{Bagchi:2009ca} for the case of BMS$\oplus\mathfrak{u}(1)$, one can see that $L_{0}$, $P_{0}$ and $J_{0}$ can be simultaneously diagonalized, while $M_0$ and $J_{0}$ cannot be simultaneously diagonalized. {\footnote{See \cite{Saha:2022gjw} for a new take on off-diagonal structures and representations for these BMS-invariant field theory.}}

Using Jacobi identity on $P_{0}$, $J_{-n}$ and $\mathcal{O}$ one can show that
\begin{align}\label{chh3}
    [P_{0},[J_{-n},\mathcal{O}]]-[J_{-n},[P_{0},\mathcal{O}]]+[\mathcal{O},[P_{0},J_{-n}]]=0.
\end{align}
The algebra of BMS$\oplus\mathfrak{u}(1)$ dictates that $[P_{0},J_{-n}]=0$ and hence
\begin{align}{\label{chh4}}
    [P_{0},[J_{-n},\mathcal{O}]]=[J_{-n},[P_{0},\mathcal{O}]]=p[J_{-n},\mathcal{O}].
\end{align}
Similarly,
\begin{align}\label{chh5}
    [L_{0},[J_{-n},\mathcal{O}]]&=(\Delta+n)[J_{-n},\mathcal{O}],\nonumber\\ 
    [L_{0},[P_{-n},\mathcal{O}]]&=(\Delta+n)[P_{-n},\mathcal{O}],\nonumber\\ 
    [M_{0},[P_{-n},\mathcal{O}]]&=\xi[P_{-n},\mathcal{O}].
\end{align}
Hence, when including descendant states, one can simultaneously diagonalize $L_{0}$, $P_{0}$, and $J_{0}$. One can also simultaneously diagonalize $M_0$ and $P_{0}$. However, enforcing the Jacobi identities on $M_0$, $J_{-n}$ and $\mathcal{O}$ gives
\begin{align}\label{chh6}
    [M_0,[J_{-n},\mathcal{O}]]=\xi[J_{-n},\mathcal{O}]+n[P_{-n},\mathcal{O}],
\end{align}
implying that $[J_{-n},\mathcal{O}]$ is not an eigenstate of $M_0$ and hence, $M_0$ and $J_{0}$ are not simultaneously diagonalizable.

Thus, we choose a basis where $L_{0}$, $P_{0}$, and $J_{0}$ are diagonalized but $M_0$ is  block-diagonal. The blocks are Jordan cells. In this case, $r$ number of primary operators form a multiplet of rank $r$. This multiplet and the descendants of those $r$ primaries form a module. In previous sections of this work, we have dealt with rank one multiplets/singlets where $M_0$ becomes diagonal while acting on primaries. The highest weight primary multiplet of rank $r$ for BMSFT with $\mathfrak{u}(1)$ symmetry will be \cite{Hao:2021urq}
\begin{align}\label{chh7}
[L_0,\mathcal{O}_i]=\Delta\mathcal{O}_i,\hspace{5mm}[M_0,\mathcal{O}_i]=(\hat{\xi}\mathcal{O})_{i},\hspace{5mm}[P_{0},\mathcal{O}_i]=p\mathcal{O}_i,\hspace{5mm}[J_{0},\mathcal{O}_i]=j\mathcal{O}_i,
\end{align}
where $i=\{1,2...,r\}$ and highest weight conditions are 
\begin{align}\label{chh8}
    [L_{n},\mathcal{O}_i]=0,\hspace{5mm}[M_{n},\mathcal{O}_i]=0,\hspace{5mm}[P_{n},\mathcal{O}_i]=0,\hspace{5mm}[J_{n},\mathcal{O}_i]=0,\hspace{5mm} n>0.
\end{align}
Here $\hat{\xi}$ is Jordan cell of rank $r$
\begin{align}\label{chh9}
\hat{\xi}_r = \left(\begin{array}{ccccc}
\xi & 0 & \cdots & & 0\\
1 & \xi & 0 & \cdots & \vdots\\
0 & 1 &  \xi &\ddots  & \vdots   \\
\vdots & \vdots& &\ddots  & 0\\
0 & \cdots & & 1 & \xi\\
\end{array}\right)_{r\times r}.
\end{align}
In Appendix~\ref{chmlt} we give the details of calculation for the characters of BMS doublet, BMS$\oplus\mathfrak{u}(1)$ doublet. The character of the BMS doublet is given by
\begin{align}
   \chi^{\{\mathcal{O}_1, \mathcal{O}_2\}}_{A} = 2q_{L}^{\Delta_{A}-\frac{c_{L}}{12}}q_{M}^{\xi_{A}-\frac{c_{M}}{12}}\sum_N q^{N}_{L}\hspace{.5mm} \widetilde{dim_N}.
\end{align}
Note that here $dim_N$ is the number of states at level $N$ in a BMS$\oplus\mathfrak{u}(1)$ module whereas $\widetilde{dim_N}$ is the number of states at level $N$ in a BMS module. The previous expression is twice the character for the BMS singlet primaries as derived in \cite{Bagchi:2019unf}. 
The character for the BMS$\oplus\mathfrak{u}(1)$ doublet is found to be 
\begin{align}
\chi^{\{\mathcal{O}_1, \mathcal{O}_2\}}_{A}=2q_{L}^{\Delta_{A}-\frac{c_{L}}{12}}q_{M}^{\xi_{A}-\frac{c_{M}}{12}}e^{2\pi i \gamma j}e^{2\pi i \alpha p}\sum_{N}q^{N}_{L}\hspace{.5mm}{dim_N}=2\chi_{(\Delta,\xi,j,p)},
\end{align}
where expression of  $\chi_{(\Delta,\xi,j,p)}$ is given in \eqref{chcexp}.

\medskip

The character of BMS$\oplus\mathfrak{u}(1)$ rank $r$ multiplet can also be computed using similar methods. We give details in Appendix \ref{ApE}. The final answer is of the following simple form: 
\be 
\chi_{(r)}^{\{\mathcal{O}_i\}} =  r\bigl(q_L^{(\Delta  -\frac{c_L}{2})}q_M^{(\xi-\frac{c_M}{2})}e^{2\pi i \gamma j}e^{2\pi i \alpha p}\bigr)\sum_{N} q_L^N \hspace{.5mm} {dim_N} = r \chi_{(\Delta,\xi,j,p)},
\ee 
i.e. it is $r$ times the BMS$\oplus\mathfrak{u}(1)$ character for singlet primaries. We should stress that even though the formulae of the characters look straight-forward in these multiplets, the computations are not. The interested reader is pointed to the appendices \ref{chmlt} and \ref{ApE} for details of the computations. 
%
\section{Partition Function and Cardy Formula for Primaries}\label{sec:PartitionFunction}
%
The partition function for the primary states of the BMS$\oplus\mathfrak{u}(1)$ theory is given by the summation of the characters of different modules. 
    \begin{equation}
        Z=\sum D(\Delta,\xi,j,p)\chi_{\Delta,\xi,j,p},   
    \end{equation} 
where $D(\Delta,\xi,j,p)$ is the spectral density of the primary state with weights $(\Delta, \xi, j,p)$. Substituting \eqref{chcexp} and \eqref{eq:VacuumChar} in the above relation, the partition function can be written as
    \begin{equation}\label{eq:PrimaryPartitionFunction}
        Z(\sigma,\rho,\alpha,\gamma)=\frac{e^{-2\pi i\left(\sigma\frac{\tilde{c}_L}{2}+\rho\frac{c_M}{2}\right)}}{\eta(\sigma)^4}\sum D(\Delta,\xi,j,p)e^{2\pi i\left(\sigma\Delta+\rho\xi+\alpha j +\gamma p\right)},
    \end{equation}
where $\tilde{c}_L=c_L-\frac{1}{3}$ and we have defined
    \begin{equation}
        D(0,0,0,0):=\tilde{D}(0,0,0,0)\left(1-e^{2\pi i \sigma}\right)^2.
    \end{equation}
$\tilde{D}(0,0,0,0)$ is the spectral density of the vacuum. For a non-degenerate vacuum one has $\tilde{D}(0,0,0,0)=1$. For $-2\pi i \sigma=\Omega\to\infty$, where $\Omega$ is taken to be real, the partition function is dominated by the vacuum contribution
    \begin{equation}\label{eq:Zlow-T}
        Z_{\textrm{high-}\Omega} = \frac{e^{-2\pi i\left(\sigma\frac{\tilde{c}_L}{2}+\rho\frac{c_M}{2}\right)}}{\eta(\sigma)^4}\tilde{D}(0,0,0,0)\left(1-e^{2\pi i \sigma}\right)^2.
    \end{equation}
%
\subsection{Modular Transformation of Partition Function}
%
For deriving the Cardy-like formula it is helpful to know how the BMS$\oplus\mathfrak{u}(1)$ partition function behaves under BMS modular transformations. To better understand these BMS modular transformations, it is instructive to first recall modular transformations in a relativistic CFT. Under modular transformation, the parameters $\tau$ and $\nu$ of the CFT partition function transform as  
    \begin{equation}
       \tau\to\frac{a\tau+b}{c\tau+d},\hspace{2mm}\nu\to\frac{\nu}{c\tau+d},\hspace{2mm}ad-bc=1.
    \end{equation}
Similar equations exist for $\bar{\tau}$ and $\bar{\nu}$. Using the contraction relations (\ref{cntrctn}), it can be shown that the BMS modular parameters $\sigma$, $\rho$, $\alpha$, $\gamma$ transform as
    \begin{equation}\label{eq:ModularTransformations} 
      \sigma\to\frac{a\sigma+b}{c\sigma+d},\hspace{1mm}\rho\to\frac{\rho}{(c\sigma+d)^2},\hspace{1mm} \alpha\to\frac{\alpha}{c\sigma+d},\hspace{1mm}\gamma\to\frac{\gamma}{c\sigma+d}-\frac{c\rho\alpha}{(c\sigma+d)^{2}}.
    \end{equation}
In what follows, we focus on a set of transformations with $a=0, b=-1, c=1, d=0$ that relate different parts of the spectrum of the torus partition function. Applying these transformations to the BMS$\oplus\mathfrak{u}(1)$ partition function, one finds\footnote{One straightforward way to derive this behavior is by, e.g. taking the UR limit of equation \eqref{cftp}.} the following transformation behavior \cite{Basu:2017aqn}
    \begin{equation}\label{eq:PartitionFunctionModularTransformations} 
        Z\left(-\frac{1}{\sigma},\frac{\rho}{\sigma^{2}},\frac{\alpha}{\sigma},\frac{\gamma}{\sigma}-\frac{\alpha\rho}{\sigma^{2}}\right)=e^{\pi i \left(\frac{\alpha^{2}\kappa_{J}}{\sigma}+\frac{2\alpha\gamma \kappa_{P}}{\sigma}-\frac{\alpha^{2}\rho\kappa_{P}}{\sigma^{2}}\right)}Z(\sigma,\rho,\alpha,\gamma).
    \end{equation}
Using this modular transformation, one can obtain the expansion in the low-twist regime from the high-twist expression \eqref{eq:Zlow-T} as
    \begin{equation}\label{eq:Zhigh-T}
        Z_{\textrm{low-}\Omega} = e^{-\pi i \left(\frac{\alpha^{2}\kappa_{J}}{\sigma}+\frac{2\alpha\gamma \kappa_{P}}{\sigma}-\frac{\alpha^{2}\rho\kappa_{P}}{\sigma^{2}}\right)}e^{-2\pi i\left(-\frac{1}{\sigma}\frac{\tilde{c}_L}{2}+\frac{\rho}{\sigma^2}\frac{c_M}{2}\right)}\frac{\tilde{D}(0,0,0,0)}{\eta\left(-\frac{1}{\sigma}\right)^4}\left(1-e^{-\frac{2\pi i}{ \sigma}}\right)^2.
    \end{equation}
%
\subsection{Cardy-Like Formula for Primary States }
%
With all the tools that have been developed so far for BMS field theories with $\mathfrak{u}(1)$ symmetry, one can now derive the expression for the density of primary states under the assumption that $\Delta$ and $\xi$ are both large\footnote{Since both $\xi$ and $c_M$ are dimensionful quantities what we mean here by large $\xi$ is that the ratio $\frac{\xi}{c_M}\gg1$.}.\\ 
Since the thermal entropy is given by the natural logarithm of the spectral density $D(\Delta,\xi,j,p)$ one first has to invert \eqref{eq:PrimaryPartitionFunction}. After replacing the sum with an integral, one obtains
    \begin{equation}\label{eq:PartitionFunctionIntegral}
                Z(\sigma,\rho,\alpha,\gamma)=\frac{e^{-2\pi i\left(\sigma\frac{\tilde{c}_L}{2}+\rho\frac{c_M}{2}\right)}}{\eta(\sigma)^4}\int \extd \lambda\, D(\Delta,\xi,j,p)e^{2\pi i\left(\sigma\Delta+\rho\xi+\alpha j +\gamma p\right)},
    \end{equation}
where
    \begin{equation}
        \extd\lambda = \extd\Delta\left(\frac{\extd\xi}{c_M}\right)\extd j\left(\frac{\extd p}{\kappa_p}\right).
    \end{equation}
The additional factors of $c_M$ and $\kappa_P$ here are necessary since both $\xi$ and $p$ are dimensionful quantities. Without taking these factors into account, the partition function would acquire a physical dimension instead of just being a pure number. In order to invert \eqref{eq:PartitionFunctionIntegral} one has to essentially compute
    \begin{equation}\label{eq:DOSIntegralInversion}
        D(\Delta,\xi,j,p)=\int\extd^4\Omega\,Z(\sigma,\rho,\alpha,\gamma)\eta\left(\sigma\right)^4e^{2\pi i\left(\sigma\frac{\tilde{c}_L}{2}+\rho\frac{c_M}{2}\right)}e^{-2\pi i\left(\sigma\Delta+\rho\xi+\alpha j +\gamma p\right)},
    \end{equation}
where
    \begin{equation}
    \label{weight}
        \extd^4\Omega=\extd\sigma(c_M\extd\rho)\extd\alpha(\kappa_P\extd\gamma).
    \end{equation}
\subsubsection*{Saddle-Point Approximation}
\label{saddle_approx}
We first present an approximate derivation for the expression of density of states $D(\Delta,\xi,j,p)$ that involves a saddle point approximation and uses modular properties of the partition function. We define $\tilde{Z}$ as
\begin{equation}
    \tilde{Z}(\sigma,\rho,\alpha,\gamma)=\eta(\sigma)^{4}e^{2\pi i\left(\sigma\frac{\tilde{c}_{L}}{2}+\rho\frac{c_{M}}{2}\right)}Z(\sigma,\rho,\alpha,\gamma).
\end{equation}
The modular transformation of $\tilde{Z}$ now follows from \eqref{eq:PartitionFunctionModularTransformations},
\begin{eqnarray}
\label{tildemodulartranfo}
    \tilde{Z}(\sigma,\rho,\alpha,\gamma)&=&\frac{\eta^{4}(\sigma)}{\eta^{4}\left(-\frac{1}{\sigma}\right)}e^{2\pi i\left(\sigma\frac{\tilde{c_{L}}}{2}+\rho\frac{c_{M}}{2}+\frac{\tilde{c}_{L}}{2\sigma}-\frac{\rho c_{M}}{2\sigma^{2}}\right)}e^{-\pi i\left(\frac{\alpha^{2}\kappa_{J}}{\sigma}+\frac{2\alpha\gamma\kappa_{P}}{\sigma}-\frac{\alpha^{2}\rho\kappa_{P}}{\sigma^{2}}\right)}\nonumber\\
    &&\times\tilde{Z}\left(-\frac{1}{\sigma},\frac{\rho}{\sigma^{2}},\frac{\alpha}{\sigma},\frac{\gamma}{\sigma}-\frac{\alpha\rho}{\sigma^{2}}\right).
\end{eqnarray}
The modular parameter $\sigma$ is related to the twist of the torus $\Omega$ by $-2\pi i\sigma=\Omega$. The modular transformation relates the low-twist $(\Omega\to 0)$ partition function $Z(\sigma,\rho,\alpha,\gamma)$ to a high-twist $(\Omega\to\infty)$ one. Since the partition function is dominated by the vacuum contribution in a high-twist regime one can approximate $\tilde{Z}\left(-\frac{1}{\sigma},\frac{\rho}{\sigma^{2}},\frac{\alpha}{\sigma},\frac{\gamma}{\sigma}-\frac{\alpha\rho}{\sigma^{2}}\right)$ by $\tilde{D}(0,0,0,0)\left(1-e^{-\frac{2\pi i}{\sigma}}\right)^{2}$.     
Inserting the modular transformed expression of $\tilde{Z}(\sigma,\rho,\alpha,\gamma)$ from \eqref{tildemodulartranfo} in \eqref{eq:DOSIntegralInversion} one obtains,
\begin{eqnarray}
\label{DOSSaddleInt}
 D(\Delta,\xi,j,p)=\int\extd^4\Omega && \frac{\eta^{4}(\sigma)}{\eta^{4}\left(-\frac{1}{\sigma}\right)} \exp(2\pi i\left(\sigma\frac{\tilde{c_{L}}}{2}+\rho\frac{c_{M}}{2}+\frac{\tilde{c}_{L}}{2\sigma}-\frac{\rho c_{M}}{2\sigma^{2}}-\sigma\Delta-\rho\xi-\alpha j-\gamma p\right)) \nonumber \\
 && \times \exp(-\pi i\left(\frac{\alpha^{2}\kappa_{J}}{\sigma}+\frac{2\alpha\gamma\kappa_{P}}{\sigma}-\frac{\alpha^{2}\rho\kappa_{P}}{\sigma^{2}}\right)) \left(1-\exp(-\frac{2\pi i}{\sigma})\right)^{2}.
\end{eqnarray}
The $\alpha$ integration in \eqref{DOSSaddleInt} ia a Gaussian integral which is followed by another Gaussian integration with respect to $\gamma$. For large $\Delta$ and $\xi$, one can perform a saddle point approximation for the two remaining integrations in $\sigma$ and $\rho$. To also account for the logarithmic corrections, we carry out the saddle point approximation up to second order in $\sigma$ and $\rho$.   
\begin{eqnarray}
\label{2ndordersaddle}
D(\Delta,\xi,j,p)&=&c_{M}\int \extd\sigma\extd\rho \left(-\frac{1}{\sigma}\right) \left(1-e^{-\frac{2\pi i}{\sigma}}\right)^{2} e^{2\pi i\left(\frac{\tilde{c}_{L}}{2\sigma}-\frac{\rho c_{M}}{2\sigma^{2}}+\sigma\left(\frac{\tilde{c}_{L}}{2}-\Delta^{\prime}\right)+\rho\left(\frac{c_{M}}{2}-\xi^{\prime}\right)\right)} \\
&=&c_{M}e^{f(\sigma_{c},\xi_{c})}\left(-\frac{1}{\sigma_{c}}\right)\left(1-e^{-\frac{2\pi i}{\sigma_{c}}}\right)^{2}\int \extd\sigma\extd\rho e^{\frac{1}{2}f_{\sigma\sigma}|_{(\sigma_{c},\rho_{c})}(\sigma-\sigma_{c})^{2}+f_{\sigma\rho}|_{(\sigma_{c},\rho_{c})}(\sigma-\sigma_{c})(\rho-\rho_{c})},\nonumber
\end{eqnarray}
where 
\begin{equation}
f=2\pi i\left(\frac{\tilde{c}_{L}}{2\sigma}-\frac{\rho c_{M}}{2\sigma^{2}}+\sigma\left(\frac{\tilde{c}_{L}}{2}-\Delta^{\prime}\right)+\rho\left(\frac{c_{M}}{2}-\xi^{\prime}\right)\right), \quad f_{\sigma\sigma}=\frac{\partial^{2}f}{\partial^{2}\sigma},\quad f_{\sigma\rho}=\frac{\partial^{2}f}{\partial\sigma\partial\rho},
\end{equation}
\begin{equation}
 \sigma_{c}=i\sqrt{\frac{c_{M}}{2\xi^{\prime}}}, \quad  \rho_{c}=i\frac{\xi^{\prime}c_{L}-\Delta^{\prime}c_{M}}{\sqrt{c_{M}}(2\xi^{\prime})^{3/2}},
 \end{equation}
 are the saddle points that maximize $f$. We have introduced the spectral flow invariant quantities \cite{Basu:2017aqn}
    \begin{equation}\label{eq:SpectralFlowInvariants}
        \Delta'=\Delta-\frac{jp}{\kappa_{P}}+\frac{p^{2}\kappa_{J}}{2\kappa_{P}^{2}},\quad\xi'=\xi-\frac{p^{2}}{2\kappa_{P}}.
    \end{equation} 
Equation \eqref{2ndordersaddle} is again a Gaussian in $\sigma$ followed by another Gaussian in $\rho$. Finally, the expression for the density of states for the primaries $D(\Delta,\xi, j,p)$ is given by
\begin{align}
D(\Delta,\xi, j,p)=\left(\frac{c_{M}}{2\xi^{\prime}}\right)\exp(2\pi\left(\Delta^{\prime}\sqrt{\frac{c_{M}}{2\xi^{\prime}}}+\tilde{c}_{L}\sqrt{\frac{\xi^{\prime}}{2c_{M}}}\right)).
\end{align}
%
\subsubsection*{Exact Analysis}\label{sec:ExactPartFunction} 
%
Taking into account the relations between modular and Hamiltonian parameters \eqref{eq:ModularParametersHamiltonian} one can see that the inversions involve two inverse Fourier transformations ($\beta$ and $\mu_p$) and two inverse Laplace transformations ($\Omega$ and $\mu_J$). In the following, we will first perform the inverse Fourier transformations followed by the inverse Laplace transformations.\\
In the limit $-2\pi i \sigma = \Omega\to0$ the integral \eqref{eq:DOSIntegralInversion} will be dominated by large $\Delta$. Inserting the low-twist expansion of the partition function \eqref{eq:Zhigh-T} into the previous expression and setting $\tilde{D}(0,0,0,0)=1$ one has
    \begin{equation}\label{eq:SpectralDensityEntropyHighT}
        D(\Delta,\xi,j,p)=c_M\kappa_p\int_{-\infty}^\infty\extd\beta\int_{-\infty}^\infty\extd\mu_P\int_{a-i\infty}^{a+i\infty}\extd\Omega\int_{b-i\infty}^{b+i\infty}\extd\mu_J\frac{e^{-2\pi iF+L}}{\Omega^2}\left(1-e^{-\frac{4\pi^2}{ \Omega}}\right)^2,
    \end{equation}
with
    \begin{subequations}\label{eq:Fshorthand}
        \begin{align}
        F & =\beta\left(\frac{2\xi-c_M}{4\pi}-\frac{\pi\left(\kappa_P\mu_J^2+c_M\right)}{\Omega^2}\right)+\mu_P\left(p+\frac{2\pi\kappa_P\mu_J}{\Omega}\right),\\
        L & = \Omega\left(\Delta-\frac{\tilde{c}_L}{2}+\frac{2\pi^2}{\Omega^2}\left(\kappa_J\mu_J^2+\tilde{c}_L\right)\right)+2\pi\mu_Jj,
        \end{align}
    \end{subequations}
and where we have used the property of the Dedekind eta function $\eta(-1/\sigma)=\sqrt{-i\sigma}\,\eta(\sigma)$. The real numbers $a$ and $b$ are chosen such that the contour of integration of the Bromwich integrals encloses all the poles of the integrand. As it turns out, this integral can be solved \emph{exactly}.\\
Performing first the integral with respect to $\mu_p$ yields a delta distribution
    \begin{equation}
            \int_{-\infty}^\infty\extd\mu_P\,e^{-2\pi i\mu_P\left(p+\frac{2\pi\kappa_P\mu_J}{\Omega}\right)} = \frac{|\Omega|}{2\pi\kappa_P}\delta\left(\mu_J+\frac{\Omega p}{2\pi\kappa_P}\right).
    \end{equation}
Performing the $\beta$ integration yields another delta distribution
    \begin{equation}
        \int_{-\infty}^\infty\extd\beta\, e^{-i\beta\left(\xi'-\frac{c_M}{2}-\frac{2\pi^2c_M}{\Omega^2}\right)}=2\pi\delta\left(\xi'-\frac{c_M}{2}-\frac{2\pi^2c_M}{\Omega^2}\right)=\sum_{\star=+,-}\frac{|\Omega^3_\star|}{2\pi c_M}\delta\left(\Omega-\Omega_\star\right),
    \end{equation}
which simplifies the remaining integrals to
    \begin{equation}
        D(\Delta,\xi,j,p)=\sum_{\star=+,-}\frac{\Omega^2_\star}{4\pi^2}e^{\Omega_\star\left(\Delta'-\frac{\tilde{c}_L}{2}\left(1-\frac{4\pi^2}{\Omega_\star^2}\right)\right)}\oint\extd\Omega\oint\extd\mu_J\,\delta\left(\Omega-\Omega_\star\right)\delta\left(\mu_J+\frac{\Omega_\star p}{2\pi\kappa_P}\right),
    \end{equation}
where we used again the spectral flow invariant quantities \eqref{eq:SpectralFlowInvariants} and $\Omega_\pm$ are the solutions to $\xi'-\frac{c_M}{2}\left(1-\frac{1}{\sigma^2}\right)=0$ which are given by
    \begin{equation}
        \Omega_\pm=\pm\frac{2\pi}{\sqrt{\frac{2\xi'}{c_M}-1}}.
    \end{equation}
Since we started with the assumption that $\Omega$ is real, we also have to demand that $\frac{2\xi'}{c_M}>1$. Furthermore in the limit $\Omega\to0$ one also has $\frac{2\xi'}{c_M}\gg1$.\\
The Bromwich integral used to invert the Laplace transform is a contour integral in the complex plane. Thus, one has to convert the delta distribution to its complex analog to perform the actual integration, which amounts to a formal replacement of\footnote{See e.g., Section~4.2 in \cite{Bagchi:2020rwb} for more details.}
    \begin{equation}
        \delta(x-x_0)\to\frac{1}{2\pi i}\frac{1}{z-z_0}.
    \end{equation}
Then, after closing the Bromwich contour with a semi-circle that encloses all poles and using the residue theorem, one obtains
    \begin{equation}
        D(\Delta,\xi,j,p) =\frac{8c_M}{\left(2\xi'-c_M\right)}\sinh^2\left(\pi\sqrt{\tfrac{2\xi'}{c_M}-1}\right)\cosh\left(2\pi\frac{\Delta'+1-\tilde{c}_L+\tfrac{\xi'}{c_M}\left(\tilde{c}_L-2\right)}{\sqrt{\tfrac{2\xi'}{c_M}-1}}\right).
    \end{equation}
Taking the limit of large $\Delta'$ and $\xi'$ of this expression yields
    \begin{equation}
        D(\Delta,\xi,j,p) = \frac{c_M}{2\xi'}e^{2\pi\left(\sqrt{\frac{\xi'}{c_M}}\tilde{c}_L+\Delta'\sqrt{\frac{c_M}{\xi'}}\right)}.
    \end{equation}
The thermal entropy is then simply given by the natural logarithm of this expression
    \begin{equation}\label{PrimeCardy}
        S_{\textrm{primary}}=\log D(\Delta,\xi,j,p)= 2\pi\left(\sqrt{\frac{\xi'}{c_M}}\left(c_L-\frac{1}{3}\right)+\Delta'\sqrt{\frac{c_M}{\xi'}}\right) - \log\left(\frac{2\xi'}{c_M}\right).
    \end{equation}
    We note here that the expression for the entropy is shifted compared to the vanilla BMS case (reviewed in Appendix A), which had a $(c_L- 1/6)$ factor instead of the $(c_L- 1/3)$ here, apart from the Sugawara shifted weights{\footnote{The shift of the central charge $c_L$ (and no shift in $c_M$) was also noticed in \cite{Merbis:2019wgk} a different context.}}. 

\subsection{Spectral Density for Primaries and All Descendants}
\label{Density_All}
In the previous subsection, we derived the density of primaries. However, for later purposes, the spectral densities of primaries and their descendants $d(\Delta,\xi,j,p)$ originally derived in \cite{Basu:2017aqn} will also be necessary. We exploit the modular properties of BMS$\oplus\mathfrak{u}(1)$ partition function  $Z(\sigma,\rho,\alpha,\gamma)$ to derive the expression of $d(\Delta,\xi,j,p)$.
\begin{eqnarray}
Z(\sigma,\rho,\alpha,\gamma)&=& \Tr e^{2\pi i\left(\sigma \left(L_{0}-\frac{c_{L}}{2}\right)+\rho\left(M_{0}-\frac{c_{M}}{2}\right)+\alpha J_{0}+\gamma P_{0} \right)}=\sum d(\Delta,\xi,j,p) e^{2\pi i\left(\sigma\left(\Delta-\frac{c_{L}}{2}\right)+\rho\left(\xi-\frac{c_{M}}{2}\right)+\alpha j+\gamma p\right)}.\nonumber\\
\end{eqnarray}
Replacing the summation with integration and inverting it, we get
\begin{eqnarray}
d(\Delta,\xi,j,p)&=&\int\extd^{4}\Omega  e^{-2\pi i\left(\sigma\left(\Delta-\frac{c_{L}}{2}\right)+\rho\left(\xi-\frac{c_{M}}{2}\right)+\alpha j+\gamma p\right)}e^{-\pi i \left(\frac{\alpha^{2}\kappa_{J}}{\sigma}+\frac{2\alpha\gamma \kappa_{P}}{\sigma}-\frac{\alpha^{2}\rho\kappa_{P}}{\sigma^{2}}\right)}Z\left(-\frac{1}{\sigma},\frac{\rho}{\sigma^{2}},\frac{\alpha}{\sigma},\frac{\gamma}{\sigma}-\frac{\alpha\rho}{\sigma^{2}}\right)\nonumber\\
&=&\int\extd^{4}\Omega  e^{-2\pi i\left(\sigma\left(\Delta-\frac{c_{L}}{2}\right)+\rho\left(\xi-\frac{c_{M}}{2}\right)+\alpha j+\gamma p\right)}e^{-\pi i \left(\frac{\alpha^{2}\kappa_{J}}{\sigma}+\frac{2\alpha\gamma \kappa_{P}}{\sigma}-\frac{\alpha^{2}\rho\kappa_{P}}{\sigma^{2}}\right)}e^{2\pi i\left(\frac{c_{L}}{2\sigma}-\frac{\rho c_{M}}{2\sigma^{2}}\right)}.
\end{eqnarray} 
We have used the modular properties of the partition function to express $Z(\sigma,\rho,\alpha,\gamma)$ in terms of $Z\left(-\frac{1}{\sigma},\frac{\rho}{\sigma^{2}},\frac{\alpha}{\sigma},\frac{\gamma}{\sigma}-\frac{\alpha\rho}{\sigma^{2}}\right)$.\\
For high twist value the partition function $Z\left(-\frac{1}{\sigma},\frac{\rho}{\sigma^{2}},\frac{\alpha}{\sigma},\frac{\gamma}{\sigma}-\frac{\alpha\rho}{\sigma^{2}}\right)$ is approximated by contributions only from vacuum states. Then, performing similar computations as before one obtains for $d(\Delta,\xi,j,p)$: 
\begin{equation}\label{eq:DOSAllStates}
    d(\Delta,\xi,j,p)=\left(\frac{c_{M}}{2\xi^{\prime}}\right)^{2}
    \exp(2\pi\left(\Delta^{\prime}\sqrt{\frac{c_{M}}{2\xi^{\prime}}}+c_{L}\sqrt{\frac{\xi^{\prime}}{2c_{M}}}\right)).
\end{equation}
%
\section{Asymptotic Structure Constants}\label{sec:ThrePointCoefficients}
%
In the previous sections, we computed the characters of the BMS$\oplus\mathfrak{u}(1)$ field theory and constructed the partition function of the theory. We then used the modular properties of the partition function to calculate the spectral density of primary states and the corresponding entropy. As per expectations and previous similar analyses, the entropy of primaries contributes the maximal part of the total thermal entropy. We will now move beyond the zero point function on the torus, i.e., the partition function, and consider the one-point function on the torus and its modular properties to deduce an asymptotic formula for the structure constants of the theory.
The coefficients of the three-point functions or the structure constants form one of the three sets of data that determines the BMS$\oplus\mathfrak{u}(1)$ field theory, the other two being the spectrum of primaries and the central charges. These are, however, not fixed by symmetries. Below, we exploit the modular properties of one point function on the torus to find an asymptotic expression of the structure constants in a heavy-light-heavy\footnote{Heavy (Light) in this context means that the BMS weights $\xi$ and $\Delta$ are large (small) in comparison to the central charges $c_M$ and $c_L$.} limit, similar to the BMS-Cardy analysis for the entropy. In contrast to the previous section, we will not restrict ourselves to primary states anymore but will consider the full spectrum of operators in the theory. 
%
\subsection{Modular Transformations of One-Point Functions}
%
The one-point function of an operator $\mathcal{O}$ in a BMS$\oplus\mathfrak{u}(1)$ theory on the torus is given by
    \begin{equation}\label{eq:OnePointTrace}
        \langle\mathcal{O}(u,\phi)\rangle=\Tr[\mathcal{O}(u,\phi)e^{2\pi i\sigma(L_{0}-c_{L}/2)}e^{2\pi i\rho(M_{0}-c_{M}/2)}e^{2\pi i\alpha J_{0}}e^{2\pi i\gamma P_{0}}].
    \end{equation}
Taking into account the transformation behavior of both the partition function \eqref{eq:PartitionFunctionModularTransformations} and one-point functions in pure BMS field theories without additional $\mathfrak{u}(1)$ symmetries (see e.g. \cite{Bagchi:2020rwb}) one finds that under the modular transformations \eqref{eq:ModularTransformations} (with $a=0, b=-1, c=1, d=0$) the one-point function \eqref{eq:OnePointTrace} transforms as
    \begin{equation}\label{eq:OnePointModularTransformations}
       \langle\mathcal{O}(u^{\prime}, \phi^{\prime})\rangle=\sigma^{\Delta_{\mathcal{O}}}e^{-\frac{\xi_{\mathcal{O}}\rho}{\sigma}}e^{2\pi i \left(\frac{\alpha^2\kappa_J}{2\sigma}+\left(\frac{\gamma}{\sigma}-\frac{\alpha\rho}{2\sigma^2}\right)\alpha\kappa_P\right)}\langle\mathcal{O}(u,\phi)\rangle,
    \end{equation}
where $u^{\prime}=\frac{u}{\sigma}-\frac{\phi\rho}{\sigma^{2}}$ and $\phi^{\prime}=\frac{\phi}{\sigma}$. The trace in \eqref{eq:OnePointTrace} can be replaced by a sum over eigenstates as
    \begin{equation}\label{eq:OnePointFunctionEigenstateSum}
        \langle\mathcal{O}(u,\phi)\rangle=\sum_n\langle\psi_n|\mathcal{O}(u,\phi)|\psi_n\rangle e^{2\pi i\sigma(\Delta_n-c_{L}/2)}e^{2\pi i\rho(\xi_n-c_{M}/2)}e^{2\pi i\alpha j_n }e^{2\pi i\gamma p_n},
    \end{equation}
where the eigenstates $\psi_n$ with eigenvalues $(\Delta_n, \xi_n, j_n, p_n)$ are defined as \eqref{eq:Eigenvalues} and $\langle\psi_n|\mathcal{O}(u,\phi)|\psi_n\rangle=D(\Delta_n,\xi_n,j_n,p_n)c_{\psi_n^\dagger\mathcal{O}\psi_n}$ with $c_{\psi_n^\dagger\mathcal{O}\psi_n}$ being the three-point function coefficient. At high twist $\Omega=-2\pi i \sigma\to\infty$ this sum can be written as
    \begin{equation}
        \langle\mathcal{O}\rangle_{\textrm{high-}\Omega}\backsimeq\langle\chi|\mathcal{O}|\chi\rangle e^{2\pi i\sigma(\Delta_\chi-c_{L}/2)}e^{2\pi i\rho(\xi_\chi-c_{M}/2)}e^{2\pi i\alpha j_\chi }e^{2\pi i\gamma p_\chi}+\ldots,
    \end{equation}
where $\chi$ labels the lowest primary state with a non-vanishing three-point function coefficient $c_{\chi^\dagger\mathcal{O}\chi}$ and $\langle\chi|\mathcal{O}|\chi\rangle=D(\Delta_\chi,\xi_\chi,j_\chi,p_\chi)c_{\chi^\dagger\mathcal{O}\chi}$. Using \eqref{eq:OnePointModularTransformations} one can relate this expression to the low-twist regime and obtain
    \begin{align}\label{eq:OnePointFunctionHighT}
        \langle\mathcal{O}\rangle_{\textrm{low-}\Omega}\backsimeq\langle\chi|\mathcal{O}|\chi\rangle&\sigma^{-\Delta_{\mathcal{O}}}e^{\frac{\rho}{\sigma}\xi_{\mathcal{O}}}e^{-2\pi i \left(\frac{\alpha^2\kappa_J}{2\sigma}+\left(\frac{\gamma}{\sigma}-\frac{\alpha\rho}{2\sigma^2}\right)\alpha\kappa_P\right)}\nonumber\\
        &\times e^{2\pi i \left(-\frac{1}{\sigma}\left(\Delta_{\chi}-c_L/2\right)+\frac{\rho}{\sigma^{2}}\left(\xi_{\chi}-c_M/2\right)+\frac{\alpha}{\sigma}j_{\chi}+\left(\frac{\gamma}{\sigma}-\frac{\rho\alpha}{\sigma^{2}}\right)p_{\chi}\right)}.
    \end{align}
Let us now we write the sum over eigenstates \eqref{eq:OnePointFunctionEigenstateSum} as an integral
    \begin{equation}
        \langle\mathcal{O}\rangle=\int\extd\Delta\left(\frac{\extd\xi}{c_M}\right)\extd j\left(\frac{\extd\gamma}{\kappa_P}\right) d(\Delta,\xi,j,p)c_{\psi^\dagger\mathcal{O}\psi}e^{2\pi i\sigma(\Delta-c_{L}/2)}e^{2\pi i\rho(\xi-c_{M}/2)}e^{2\pi i\alpha j}e^{2\pi i\gamma p}.
    \end{equation}
Inverting this relation gives the weighted spectral density
    \begin{equation}\label{eq:WeightedSpectralDensity}
        d(\Delta,\xi,j,p)c_{\psi^\dagger\mathcal{O}\psi}=\int\extd^4\Omega\,\langle\mathcal{O}\rangle e^{-2\pi i\sigma(\Delta-c_{L}/2)}e^{-2\pi i\rho(\xi-c_{M}/2)}e^{-2\pi i\alpha j}e^{-2\pi i\gamma p},
    \end{equation}
where $\extd^4\Omega$ is defined as in \eqref{weight}. One can again see that in the low-twist limit $\Omega=-2\pi i \sigma\to0$, this integral will be dominated by large $\Delta$. Since we are interested in three-point coefficients of the type heavy-light-heavy one should use the low-twist expansion \eqref{eq:OnePointFunctionHighT} of the one-point function $\langle\mathcal{O\rangle}$ in \eqref{eq:WeightedSpectralDensity}. Doing so yields
    \begin{equation}\label{eq:WeightedSpectralDensityStart}
        d(\Delta,\xi,j,p)c_{\psi^\dagger\mathcal{O}\psi}=\int\extd^4\Omega\, D(\Delta_\chi,\xi_\chi,j_\chi,p_\chi)c_{\chi^\dagger\mathcal{O}\chi}\,\sigma^{-\Delta_\mathcal{O}}e^{\frac{\rho}{\sigma}\xi_\mathcal{O}}e^{-2\pi i G},
    \end{equation}
with
    \begin{align}
        G =& \frac{\alpha^2\kappa_J}{2\sigma}+\left(\frac{\gamma}{\sigma}-\frac{\alpha\rho}{2\sigma^2}\right)\alpha\kappa_P+\frac{1}{\sigma}\left(\Delta_{\chi}-\frac{c_L}{2}\right)-\frac{\rho}{\sigma^{2}}\left(\xi_{\chi}-\frac{c_M}{2}\right)-\frac{\alpha}{\sigma}j_{\chi}-\left(\frac{\gamma}{\sigma}-\frac{\rho\alpha}{\sigma^{2}}\right)p_{\chi}\nonumber\\
        &+\sigma(\Delta-\frac{c_L}{2})+\rho(\xi-\frac{c_M}{2})+\alpha j+\gamma p.
    \end{align}
Similarly to the previous section, we will first perform a saddle-point approximation to evaluate this integral and then present an \emph{exact} way to solve it.
%
\subsection{Saddle-Point Approximation}
%
The saddle point approximation to determine the structure constants follows closely the saddle point approximation done in \ref{saddle_approx}. We first carry out the two Gaussian integrations in $\alpha$ and $\gamma$. For the remaining integrals, we do a saddle point approximation for large $\Delta$ and $\xi$:
\begin{eqnarray}
\label{saddle_3point}
d(\Delta,\xi,j,p) c_{\psi^{\dagger}\mathcal{O}\psi}&=& D(\Delta_{\chi},\xi_{\chi},j_{\chi},p_{\chi})c_{\chi^{\dagger}\mathcal{O}\chi} c_{M}\exp(2\pi i\left(\frac{p p_{\chi}\kappa_{J}}{\kappa_{P}^{2}}-\frac{2p j_{\chi}}{\kappa_{P}}-\frac{2p_{\chi}j}{\kappa_{P}}\right))\nonumber\\
&&\int\extd\sigma\extd\rho\,\sigma e^{-2\pi i\left(\sigma\left(\Delta^{\prime}-\frac{c_{L}}{2}\right)+\rho\left(\xi^{\prime}-\frac{c_{M}}{2}\right)+\frac{1}{\sigma}\left(\Delta_{\chi}^{\prime}-\frac{c_{L}}{2}\right)-\frac{\rho}{\sigma^{2}}\left(\xi_{\chi}^{\prime}-\frac{c_{M}}{2}\right)\right)} , 
\end{eqnarray} 
where we have used the spectral flow invariant parameters in as defined in \eqref{eq:SpectralFlowInvariants}. Likewise $\Delta_{\chi}^{\prime}$ and $\xi^{\prime}_{\chi}$ are defined as,
\begin{align}
  \Delta_\chi'=\Delta_\chi-\frac{j_\chi p_\chi}{\kappa_{P}}+\frac{p_\chi^{2}\kappa_{J}}{2\kappa_{P}^{2}},\quad\xi_\chi'=\xi_\chi-\frac{p_\chi^{2}}{2\kappa_{P}}.   
\end{align}
To be consistent with the analysis of the density of all states in Section~\ref{Density_All} we carry out the saddle point approximation in \eqref{saddle_3point} up to second order in $\sigma$ and $\rho$.    
In the holographic limit ( $\xi_{\chi}^{\prime}\ll c_{M}$ and $\Delta_{\chi}^{\prime}\ll c_{L} $ ) the saddle points are at,
\begin{align}
    \sigma_{c}=i\sqrt{\frac{c_{M}}{2\xi^{\prime}}},\hspace{3mm} \rho_{c}=i\frac{\xi^{\prime}c_{L}-\Delta^{\prime} c_{M}}{(2\xi^{\prime})^{3/2}\sqrt{c_{M}}}.
\end{align} 
We now get the following expression for $c_{\psi^{\dagger}\mathcal{O}\psi}$
\begin{eqnarray}
\label{3pt_saddle}
    c_{\psi^{\dagger}\mathcal{O}\psi}&=&\frac{D(\Delta_{\chi},\xi_{\chi},j_{\chi},p_{\chi})}{d(\Delta,\xi,j,p)}c_{\chi^{\dagger}\mathcal{O}\chi}e^{2\pi i\left(\frac{p p_{\chi}\kappa_{J}}{\kappa_{P}^{2}}-\frac{2p j_{\chi}}{\kappa_{P}}-\frac{2p_{\chi}j}{\kappa_{P}}\right)}i^{\Delta_{\mathcal{O}}}\left(\frac{2\xi^{\prime}}{c_{M}}\right)^{-\frac{\Delta_{\mathcal{O}}}{2}+2}e^{2\pi\left(\frac{c_{L}}{2}\sqrt{\frac{2\xi^{\prime}}{c_{M}}}+\Delta\sqrt{\frac{c_{M}}{2\xi}}\right)}\nonumber\\
    &&e^{\frac{\xi_{\mathcal{O}}}{2}\left(-\frac{\Delta^{\prime}}{\xi^{\prime}}+\frac{c_{L}}{c_{M}}\right)}e^{-2\pi\sqrt{\frac{\xi^{\prime}}{2c_{M}}}\left(\Delta_{\chi}^{\prime}+\xi_{\chi}^{\prime}\left(\frac{\Delta^{\prime}}{\xi^{\prime}}-\frac{c_{L}}{c_{M}}\right)\right)}.
\end{eqnarray} 
Inserting the expression of $d(\Delta,\xi,j,p)$ from \eqref{eq:DOSAllStates} in \eqref{3pt_saddle} one obatins,
\begin{eqnarray}
  c_{\psi^{\dagger}\mathcal{O}\psi}&=&d(\Delta_{\chi},\xi_{\chi},j_{\chi},p_{\chi})c_{\chi^{\dagger}\mathcal{O}\chi}i^{\Delta_{\mathcal{O}}}\left(\frac{c_{M}}{2\xi^{\prime}}\right)^{\frac{\Delta_{\mathcal{O}}}{2}}e^{2\pi i\left(\frac{p p_{\chi}\kappa_{J}}{\kappa_{P}^{2}}-\frac{2p j_{\chi}}{\kappa_{P}}-\frac{2p_{\chi}j}{\kappa_{P}}\right)}\nonumber\\
    &&e^{\frac{\xi_{\mathcal{O}}}{2}\left(-\frac{\Delta^{\prime}}{\xi^{\prime}}+\frac{c_{L}}{c_{M}}\right)}e^{-2\pi\sqrt{\frac{\xi^{\prime}}{2c_{M}}}\left(\Delta_{\chi}^{\prime}+\xi_{\chi}^{\prime}\left(\frac{\Delta^{\prime}}{\xi^{\prime}}-\frac{c_{L}}{c_{M}}\right)\right)}.   
\end{eqnarray}
%
\subsection{Exact Analysis}
%
The integral \eqref{eq:WeightedSpectralDensityStart} can again be \emph{exactly} computed by following similar steps as in Section~\ref{sec:ExactPartFunction}. Inserting the relations \eqref{eq:ModularParametersHamiltonian} into \eqref{eq:WeightedSpectralDensityStart}, shifting $\mu_J\to\mu_J-i\frac{p_\chi}{\kappa_P}$, and evaluating the $\mu_P$ integral yields again formally a delta distribution 
    \begin{equation}
            \int_{-\infty}^\infty\extd\mu_P\,e^{-2\pi i\mu_P\left(p+\frac{2\pi\kappa_P\mu_J}{\Omega}\right)} = \frac{|\Omega|}{2\pi\kappa_P}\delta\left(\mu_J+\frac{\Omega p}{2\pi\kappa_P}\right).
    \end{equation}
Making again use of the spectral flow invariant expressions \eqref{eq:SpectralFlowInvariants} in addition to
    \begin{equation}
        \Delta_\chi'=\Delta_\chi-\frac{j_\chi p_\chi}{\kappa_{P}}+\frac{p_\chi^{2}\kappa_{J}}{2\kappa_{P}^{2}},\quad\xi_\chi'=\xi_\chi-\frac{p_\chi^{2}}{2\kappa_{P}},
    \end{equation}
and evaluating the $\beta$ integration gives another delta distribution
    \begin{align}
        \int_{-\infty}^\infty\extd\beta\, e^{-i\beta\left(\xi'-\frac{c_M}{2}+\frac{4\pi^2}{\Omega^2}\left(\xi_\chi'-\frac{c_M}{2}\right)+\frac{\xi_\mathcal{O}}{\Omega}\right)}&= 2\pi\delta\left(\xi'-\frac{c_M}{2}+\frac{4\pi^2}{\Omega^2}\left(\xi_\chi'-\frac{c_M}{2}\right)+\frac{\xi_\mathcal{O}}{\Omega}\right)\nonumber\\
        &=\sum_{\star=+,-}\frac{|\Omega^3_\star|}{ \left|2\pi\left(2\xi_\chi'-c_M\right)+\xi_\mathcal{O}\frac{\Omega_\star}{2\pi}\right|}\delta\left(\Omega-\Omega_\star\right),
    \end{align}
where $\Omega_\pm$ are the solutions to $\xi'-\frac{c_M}{2}+\frac{4\pi^2}{\Omega^2}\left(\xi_\chi'-\frac{c_M}{2}\right)+\frac{\xi_\mathcal{O}}{\Omega}=0$ which are given by
    \begin{equation}
        \Omega_\pm=\frac{-\xi_\mathcal{O}\pm\sqrt{\xi_\mathcal{O}^2+4\pi^2\left(2\xi'-c_M\right)\left(c_M-2\xi_\chi'\right)}}{2\xi'-c_M}.
    \end{equation}
In a heavy-light-heavy expansion, $\xi_\mathcal{O}$ is very small. Hence, for $\Omega_\pm$ to be real and to be consistent with our starting assumptions, one has to obey the following inequalities
    \begin{equation}
        \frac{2\xi'}{c_M}>1,\qquad\frac{2\xi_\chi'}{c_M}<1.
    \end{equation}
Using the Cauchy representation of the delta distributions and evaluating the remaining Bromwich integrals, like in Section~\ref{sec:ExactPartFunction}, one obtains
    \begin{equation}
        \frac{D(\Delta,\xi,j,p)c_{\psi^\dagger\mathcal{O}\psi}}{D(\Delta_\chi,\xi_\chi,j_\chi,p_\chi)c_{\chi^\dagger\mathcal{O}\chi}} = \frac{c_M}{4\pi^2}\sum_{\star=\pm}\frac{|\Omega_\star^4|}{|4\pi^2(2\xi_\chi'-c_M)+\xi_\mathcal{O}\Omega_\star|}\left(\frac{i\Omega_\star}{2\pi}\right)^{\Delta_\mathcal{O}}e^{\Omega_\star\left(\Delta'-\frac{c_L}{2}-\frac{4\pi^2}{\Omega_\star^2}\left(\xi_\chi'-\frac{c_L}{2}\right)\right)}.
    \end{equation}
The density of states $D(\Delta,\xi,j,p)$ can be obtained by performing the above computation again, but now with the operator $\mathcal{O}$ being the identity. Effectively this means computing the BMS-Cardy formula \eqref{eq:DOSAllStates} \cite{Bagchi:2012xr,Bagchi:2013qva,Basu:2017aqn,Bagchi:2019unf} by setting $\Delta_\chi'=\xi_\chi'=\Delta_\mathcal{O}=\xi_\mathcal{O}=j_\chi=p_\chi=0$ and using the normalization convention $c_{\mathbb{1}\mathbb{1}\mathbb{1}} = 1$. Assuming that the lowest-lying primary is non-degenerate, i.e., $D(\Delta_\chi,\xi_\chi,j_\chi,p_\chi)=1$ and taking the limit of large $\Delta'$ and $\xi'$ we obtain one of the central results of this work
    \begin{equation}
        \boxed{
            c_{\psi^\dagger\mathcal{O}\psi}=\frac{i^{\Delta_\mathcal{O}}}{c_M}(2\xi')^{\frac{\Delta_\mathcal{O}}{2}}(c_M-2\xi_\chi')^{1-\frac{\Delta_\mathcal{O}}{2}}e^\mathfrak{A}e^\mathfrak{B}
        },
    \end{equation}
with
    \begin{subequations}
        \begin{align}
            \mathfrak{A}&=-\frac{2\pi i}{\kappa_{P}}\left(-\frac{\kappa_{J}}{\kappa_{P}}p p_{\chi}+p j_{\chi}+j p_{\chi}\right),\\
            \mathfrak{B}&=\frac{\xi_\mathcal{O}}{2}\left(\frac{c_L-2\Delta'_\chi}{c_M-2\xi'_\chi}-\frac{\Delta'}{\xi'}\right)-\frac{2\pi\left(c_M\Delta'+c_L\xi'\right)}{\sqrt{2\xi'c_M}}+\frac{2\pi\left(c_M\Delta'+c_L\xi'-2\left(\Delta_\chi'\xi'+\Delta'\xi_\chi'\right)\right)}{\sqrt{2\xi'\left(c_M-2\xi_\chi'\right)}}.
        \end{align}
    \end{subequations}
The quadratic shifts by the $\mathfrak{u}(1)$ charges are consistent with what one would expect from the underlying spectral flow automorphism, which forces the physical observables to be invariant under this symmetry. Furthermore, the additional phase $\mathfrak{A}$ present is reminiscent of known AdS/CFT results \cite{Das:2017vej}.\\
In the holographic limit $\xi_\chi'\ll c_M$ and $\Delta_\chi'\ll c_L$ the above expression further simplifies to
    \begin{equation}\label{eq:HolographicThreePoint}
        \boxed{
            c_{\psi^\dagger\mathcal{O}\psi}=i^{\Delta_\mathcal{O}}\left(\frac{2\xi'}{c_M}\right)^{\frac{\Delta_\mathcal{O}}{2}}e^{-\frac{2\pi i}{\kappa_{P}}\left(-\frac{\kappa_{J}}{\kappa_{P}}p p_{\chi}+p j_{\chi}+j p_{\chi}\right)}e^{\frac{\xi_\mathcal{O}}{2}\left(\frac{c_L}{c_M}-\frac{\Delta'}{\xi'}\right)-2\pi \sqrt{\frac{\xi^{\prime}}{2c_{M}}} \left(2\Delta_{\chi}^{\prime}+\left(\frac{\Delta^{\prime}}{\xi^{\prime}}-\frac{c_{L}}{c_{M}}\right)\xi_{\chi}^{\prime}\right)}
        }.
    \end{equation}
We will reproduce this with a bulk calculation in the next section.     
%
\section{Bulk Computations}\label{sec:BulkComputations}
%
This section is focused on bulk computations that reproduce the field theory results that we computed previously. We first present suitable dual bulk gravity models to a BMSFT supplemented with two $\mathfrak{u}(1)$ currents. We then give a lightning review of the dual bulk computation first presented in \cite{Basu:2017aqn} reproducing the Cardy formula \eqref{eq:DOSAllStates} in the limit of large central charges $c_L$ and $c_M$. Finally, we show how to reproduce the three-point coefficient \eqref{eq:HolographicThreePoint} in the limit of large central charge $c_M$ using holographic methods.
%
\subsection{Einstein Gravity with Additional $\mathfrak{u}(1)$ Currents}
%
The holographic dual to the thermal state of a BMS field theory at a non-zero chemical potential is that of a flat space cosmology (FSC), with additional $\mathfrak{u}(1)$ hair. Thus, we will first consider Einstein gravity with a vanishing cosmological constant in three spacetime dimensions supplemented by two $\mathfrak{u}(1)$ Chern-Simons gauge fields that are dual to the spin-1 conserved currents in the BMS field theory. We will also show how a "reloaded" theory of gravity can be used to recover the general field theory results with $c_L\neq0$ and $\kappa_J\neq0$.\\
Let us first consider Einstein gravity with vanishing cosmological constant in 3d supplemented by Chern-Simons gauge fields $\mathcal{C}$ with coupling $\frac{\kappa_P}{2}$ \cite{Basu:2017aqn} i.e.
    \begin{equation}\label{eq:EH+u1}
        I_{\textrm{EH}+\textrm{CS}}=\frac{1}{16\pi G}\int R+\frac{\kappa_P}{8\pi}\int\left\langle \mathcal{C}\wedge\extd \mathcal{C}\right\rangle,
    \end{equation}
where $R$ is the Ricci scalar, $G$ is Newton's constant in 3d, $\mathcal{C}$ a $\mathfrak{u}(1)\oplus\mathfrak{u}(1)$ valued gauge field, and $\left\langle\ldots\right\rangle$ denotes a suitable invariant bilinear form on the gauge algebra of $\mathcal{C}$.\\
We use a basis for $\mathfrak{u}(1)\oplus\mathfrak{u}(1)$ with generators $\Jt$ and $\Pt$, i.e. $\mathcal{C}=C^J \Jt + C^P \Pt$ that have the following invariant bilinear form: $\langle \Jt\Jt\rangle=\langle \Pt\Pt\rangle=0$ as well as $\langle \Jt\Pt\rangle=2$.\\
Let us assume that the topology of the manifold we are working on is that of a solid cylinder. In addition, we choose coordinates such that there is a radial direction $0\leq r<\infty$ and the boundary of the cylinder is parameterized by a retarded time coordinate $-\infty<u<\infty$ as well as an angular coordinate $\varphi\sim\varphi+2\pi$.\\
The flat connections of the $\mathfrak{u}(1)$ Chern-Simons gauge field have the following non-vanishing components:
    \begin{equation}\label{eq:u1CSConnection}
        C^J = \frac{P}{\kappa_P} \extd\varphi, \qquad C^P = \frac{J}{\kappa_P}\extd\varphi + \frac{P}{\kappa_P}\extd u,
    \end{equation}
where $J$ and $P$ are the canonical charges associated with the $\mathfrak{u}(1)\oplus\mathfrak{u}(1)$ symmetry. The metric of a charged FSC in these coordinates is given by
    \begin{equation}\label{eq:ChargedFSCMetricBondi}
        \extd s^2 = \mathcal{M}\extd u^2 +\mathcal{N}\extd u \extd\varphi -2 \extd u\extd r+r^2 \extd\varphi^2,
    \end{equation}
where in general, one has
    \begin{align}\label{eq:FSCZeroModes}
        \frac{\mathcal{M}}{8G} & = M - \frac{P^2}{2\kappa_P} - \frac{c_M}{2} = \xi - \frac{p^2}{2\kappa_P} - \frac{c_M}{2} = \xi' - \frac{c_M}{2},\\
        \frac{\mathcal{N}}{8G} & = L - \frac{J P}{\kappa_P} + \frac{\kappa_J}{2 \kappa_P^2}P^2 - \frac{c_L}{2} = \Delta - \frac{j p}{\kappa_P} + \frac{\kappa_J}{2 \kappa_P^2}p^2 - \frac{c_L}{2} = \Delta' - \frac{c_L}{2}.
    \end{align}
Here $M$ and $L$ denote the shifted zero modes of the holographic mass and angular momentum operators of the FSC, respectively, and we used $\xi'$ and $\Delta'$ previously defined in \eqref{eq:SpectralFlowInvariants}. For the model described by \eqref{eq:EH+u1} one has $c_M=\frac{1}{4G}$, $c_L = 0$, $\kappa_P\neq0$ and $\kappa_J = 0$.\\
For later purposes, it is convenient to make a change of coordinates as
    \begin{equation}
        \extd\varphi = \extd \phi - \frac{r_0}{r_+\left(r^2-r_0^2\right)}\extd r ,\qquad\extd u = \extd t + \frac{r^2}{r_+^2\left(r^2-r_0^2\right)}\extd r,
    \end{equation}
where
    \begin{equation}
        r_0 = \frac{\mathcal{N}}{2\sqrt{\mathcal{M}}},\qquad r_+ = \sqrt{\mathcal{M}}.
    \end{equation}
This yields the following metric
    \begin{equation}\label{eq:FSCBHCoordinates}
        \extd s ^2 = r_+^2 \extd t^2 - \frac{r^2}{r_+^2\left(r^2-r_0^2\right)}\extd r^2 + 2 r_0 r_+ \extd\phi\extd t + r^2 \extd\phi^2.
    \end{equation}
%
\subsection*{Einstein Gravity with Additional $\mathfrak{u}(1)$ Currents  ``Reloaded"}
%
Up to boundary terms the action \eqref{eq:EH+u1} can be equivalently formulated in terms of an $\mathfrak{isl}(2,\mathbb{R})\oplus\mathfrak{u}(1)\oplus\mathfrak{u}(1)$ Chern-Simons theory with the following action \cite{Witten:1988hc,Basu:2017aqn}
\begin{equation}\label{eq:CSFS}
		I_{\rm CS}[\mathcal{A}] = \frac{1}{16\pi G} \int \langle\mathcal{A} \wedge \extd \mathcal{A} +\frac23 \mathcal{A} \wedge \mathcal{A} \wedge \mathcal{A}\rangle\,+\frac{\kappa_P}{8\pi}\int\left\langle \mathcal{C}\wedge\extd \mathcal{C}\right\rangle,
	\end{equation}
where $\mathcal{A}\in\mathfrak{isl}(2,\mathbb{R})$ and $\mathcal{C}\in\mathfrak{u}(1)\oplus\mathfrak{u}(1)$\footnote{For the $\mathfrak{u}(1)\oplus\mathfrak{u}(1)$ part we use the same conventions as in the previous subsection.}.	
We use a basis for $\mathfrak{isl}(2,\mathbb{R})$ with generators $\Lt_n,\,\Mt_n$ with $n=0,\pm1$ that have the following non-vanishing Lie brackets:
	\begin{subequations}\label{eq:isl2RBasis}
	\begin{align}
		[\Lt_n,\Lt_m]&=(n-m)\Lt_{n+m},\\
		[\Lt_n,\Mt_n]&=(n-m)\Mt_{n+m}.
	\end{align}
	\end{subequations}
The corresponding invariant bilinear form is given by $\langle \Lt_n\Lt_m\rangle=\langle \Mt_n\Mt_m\rangle=0$ as well as
	\begin{subequations}\label{eq:ISLInvBilForm}
	\begin{align}
		\langle \Lt_n\Mt_m\rangle & =-2\left(
			\begin{array}{c|ccc}
				  &\Mt_1&\Mt_0&\Mt_{-1}\\
				\hline
				\Lt_1&0&0&1\\
				\Lt_0&0&-\frac{1}{2}&0\\
				\Lt_{-1}&1&0&0
			\end{array}\right).
	\end{align}		
	\end{subequations}
With an appropriate choice of boundary conditions for the gauge fields $\mathcal{A}$ and $\mathcal{C}$ such as the ones first presented in \cite{Basu:2017aqn} one can precisely reproduce the metric \eqref{eq:ChargedFSCMetricBondi}.\\
Among many other advantages, one particularly nice feature of the Chern-Simons formulation of gravity is that it is straightforward to also generate a model of gravity with $c_L\neq0$, $c_M\neq0$, $\kappa_P\neq0$, and $\kappa_J\neq0$. This is done by changing the invariant bilinear form used in the Chern-Simons action \eqref{eq:CSFS} such as in, e.g. \cite{Giacomini:2006dr,Barnich:2014cwa} in order to obtain ``reloaded'' versions of Einstein gravity. Consider for example \eqref{eq:CSFS} with the bilinear form \eqref{eq:ISLInvBilForm} and in addition
    \begin{equation}\label{eq:InvBilFormModEinstein}
        \langle \Lt_n\Lt_m\rangle = \mu \langle \Lt_n\Mt_m\rangle,\qquad \langle \Jt_n\Jt_m\rangle = \nu \langle \Jt_n\Pt_m\rangle.
    \end{equation}
This change in bilinear form changes some of the canonical boundary charges of the theory and, thus also the central charges of the boundary BMSFT. More precisely, the angular momentum $L$ and $\mathfrak{u}(1)$ charge $J$ are shifted as
    \begin{equation}\label{eq:ReloadedEinsteinCharges}
        L\Rightarrow L+\mu M-\frac{P^2}{\kappa_P}(\mu-\nu),\qquad
        J\Rightarrow J+\nu P.
    \end{equation}
The resulting asymptotic algebra generated by the modified canonical boundary charges satisfies \eqref{bmsu1} with
    \begin{equation}\label{eq:ReloadedCentralCharges}
        c_L=\mu c_M,\qquad c_M=\frac{1}{4G},\qquad \kappa_J=\nu \kappa_P.
    \end{equation}
%
\subsection{Thermal Entropy and Cardy Formula}
%
In this subsection, we give a lightning review on how to compute the thermal entropy of FSCs and its relation to the general BMS Cardy formula that counts \emph{all} states, not just the primary ones. We will first start with Einstein gravity in the presence of two $\mathfrak{u}(1)$ fields described by \eqref{eq:EH+u1} and then extend our considerations to the ``reloaded" version of that theory described by \eqref{eq:CSFS}.\\ 
Similarly to black hole solutions such as e.g., the BTZ black hole in AdS$_3$ \cite{Banados:1992wn,Banados:1992gq}, the thermal entropy of an FSC is given by its horizon area $\textrm{A}_\textrm{H}$ divided by $4G$:
    \begin{equation}
        S = \frac{\textrm{A}_\textrm{H}}{4G}.
    \end{equation}
To compute the area of the cosmological horizon, one first has to find its location by looking at the point where the determinant of the induced metric on slices of constant radius vanishes, that is
    \begin{equation}
        g_{uu}\,g_{\varphi\varphi}-\left(g_{u\varphi}\right)^2=0.
    \end{equation}
For the metric \eqref{eq:FSCBHCoordinates} this happens at
    \begin{equation}\label{eq:Horizon}
        r = r_0 = \frac{\mathcal{N}}{2\sqrt{\mathcal{M}}}.
    \end{equation}
One can now compute the area of the horizon by
    \begin{equation}\label{eq:HorizonArea}
        \textrm{A}_\textrm{H}=\int\limits_0^{2\pi}\extd\varphi\sqrt{|g_{\varphi\varphi}|}\,\Big|_{r=r_0}=\int\limits_0^{2\pi}\extd\varphi\frac{\mathcal{N}}{2\sqrt{\mathcal{M}}}.
    \end{equation}
This can be trivially integrated to yield the area
    \begin{equation}
        {\rm A}_{\rm H}^0=\pi\frac{\mathcal{N}}{\sqrt{\mathcal{M}}},
    \end{equation}
and the corresponding thermal entropy \cite{Bagchi:2012xr,Basu:2017aqn}
    \begin{equation}\label{eq:ThermalEntropyArea}
        S_{\rm Th}=\frac{{\rm A}_{\rm H}^0}{4G_N}=\frac{\pi}{4G}\frac{\mathcal{N}}{\sqrt{\mathcal{M}}}.
    \end{equation}
Assuming that the mass and angular momentum of the FSC are very large, the thermal entropy reads
    \begin{equation}\label{eq:ThermalEntropyAreaBMSFT}
        S_{\rm Th}=\frac{{\rm A}_{\rm H}^0}{4G_N}=\frac{\pi}{4G}\frac{\Delta'}{\sqrt{\xi'}} = 2\pi\sqrt{\frac{c_M}{2\xi'}}\Delta'.
    \end{equation}
This expression precisely coincides with the BMS Cardy formula, counting \emph{all} states in the presence of two additional currents \cite{Basu:2017aqn}
    \begin{equation}\label{eq:CardyAllStates}
        S = \pi \sqrt{2c_M\xi'}\left(\frac{\Delta'}{\xi'}+\frac{c_L}{c_M}\right),
    \end{equation}
for $c_L=0$ and $\kappa_J = 0$. At large $c_M$ the expression \eqref{PrimeCardy} -- at leading order in $c_M$ -- is indistinguishable from \eqref{eq:CardyAllStates}.\\

\smallskip

Repeating the same computation in the "reloaded" theory and taking into account the shifts \eqref{eq:ReloadedEinsteinCharges} and \eqref{eq:ReloadedCentralCharges} the thermal entropy changes accordingly as
    \begin{equation}\label{eq:ThermalEntropyAreaReloaded}
        S_{\rm Th}=\frac{{\rm A}_{\rm H}^0}{4G_N}=\pi \sqrt{2c_M\xi'}\left(\frac{\Delta'}{\xi'}+\frac{c_L}{c_M}\right),
    \end{equation}
which is precisely \eqref{eq:CardyAllStates}, as advertised in the beginning of this section.
%
\subsection{Holographic Three-Point Coefficients}
\begin{figure}
\centering

\tikzset{every picture/.style={line width=0.75pt}} 

\begin{tikzpicture}[x=0.75pt,y=0.75pt,yscale=-1,xscale=1]

\draw   (297.58,140.04) .. controls (297.8,126.11) and (309.27,114.99) .. (323.2,115.2) .. controls (337.13,115.41) and (348.26,126.88) .. (348.04,140.82) .. controls (347.83,154.75) and (336.36,165.88) .. (322.42,165.66) .. controls (308.49,165.45) and (297.37,153.98) .. (297.58,140.04) -- cycle ;
\draw  [color={rgb, 255:red, 0; green, 0; blue, 0 }  ,draw opacity=1 ][fill={rgb, 255:red, 255; green, 255; blue, 255 }  ,fill opacity=1 ][dash pattern={on 4.5pt off 4.5pt}] (193.48,140.66) .. controls (193.48,69.45) and (251.21,11.71) .. (322.42,11.71) .. controls (393.64,11.71) and (451.37,69.45) .. (451.37,140.66) .. controls (451.37,211.88) and (393.64,269.61) .. (322.42,269.61) .. controls (251.21,269.61) and (193.48,211.88) .. (193.48,140.66) -- cycle ;
\draw [color={rgb, 255:red, 74; green, 144; blue, 226 }  ,draw opacity=1 ][line width=1.5]    (226.2,55.6) -- (301.2,122.6) ;
\draw  [color={rgb, 255:red, 74; green, 144; blue, 226 }  ,draw opacity=1 ][line width=0.75]  (297.42,140.66) .. controls (297.42,126.85) and (308.62,115.66) .. (322.42,115.66) .. controls (336.23,115.66) and (347.42,126.85) .. (347.42,140.66) .. controls (347.42,154.47) and (336.23,165.66) .. (322.42,165.66) .. controls (308.62,165.66) and (297.42,154.47) .. (297.42,140.66) -- cycle ;
\draw  [color={rgb, 255:red, 255; green, 255; blue, 255 }  ,draw opacity=1 ][line width=3] [line join = round][line cap = round] (323.2,140.6) .. controls (323.2,134.63) and (316.01,139.22) .. (314.2,135.6) .. controls (312.95,133.1) and (308.52,125.6) .. (307.2,125.6) ;
\draw  [color={rgb, 255:red, 255; green, 255; blue, 255 }  ,draw opacity=1 ][line width=3] [line join = round][line cap = round] (327.2,138.6) .. controls (320.5,138.6) and (316.2,139.59) .. (316.2,132.6) ;
\draw  [color={rgb, 255:red, 255; green, 255; blue, 255 }  ,draw opacity=1 ][line width=3] [line join = round][line cap = round] (235.2,84.6) .. controls (230.79,82.4) and (226.8,76.6) .. (221.2,76.6) ;
\draw  [color={rgb, 255:red, 255; green, 255; blue, 255 }  ,draw opacity=1 ][line width=3] [line join = round][line cap = round] (227.2,92.6) .. controls (227.2,88.88) and (220.12,86.6) .. (216.2,86.6) ;
\draw  [color={rgb, 255:red, 255; green, 255; blue, 255 }  ,draw opacity=1 ][line width=3] [line join = round][line cap = round] (223.2,100.6) .. controls (219.53,100.6) and (215.87,100.6) .. (212.2,100.6) ;
\draw  [color={rgb, 255:red, 74; green, 144; blue, 226 }  ,draw opacity=1 ][dash pattern={on 3.75pt off 3pt on 7.5pt off 1.5pt}] (294.04,140.66) .. controls (294.04,124.98) and (306.75,112.27) .. (322.42,112.27) .. controls (338.1,112.27) and (350.81,124.98) .. (350.81,140.66) .. controls (350.81,156.34) and (338.1,169.05) .. (322.42,169.05) .. controls (306.75,169.05) and (294.04,156.34) .. (294.04,140.66) -- cycle ;

\draw (271,75) node [anchor=north west][inner sep=0.75pt]  [color={rgb, 255:red, 0; green, 0; blue, 0 }  ,opacity=1 ] [align=left] {$\displaystyle \phi _{0}$};
\draw (339,159) node [anchor=north west][inner sep=0.75pt]   [align=left] {$\displaystyle \phi _{\chi }$};
\draw (122,153) node [anchor=north west][inner sep=0.75pt]   [align=left] {};
\draw (298.58,137.04) node [anchor=north west][inner sep=0.75pt]  [font=\footnotesize] [align=left] {r = r$\displaystyle _{0}$};

\end{tikzpicture}

\caption{The field $\phi_\mathcal{O}$ comes from the asymptotic boundary at null infinity and splits into a pair of fields $\phi_\chi$ ($\phi_\chi^\dagger$) that wrap around the cosmological horizon at $r_0$.}
\label{fig:GeodesicConfiguration}
\end{figure}
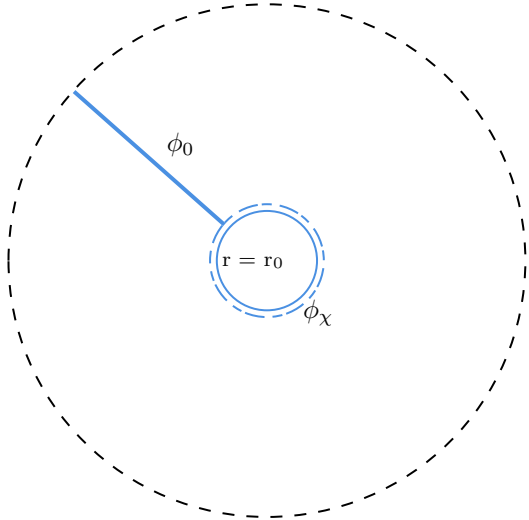

%
We are interested in computing $\langle E|O|E\rangle$ for high energy $E$ using holographic methods. On the bulk side, this means we have to calculate the one-point
function of a light operator\footnote{We are working in the probe limit, and hence we can neglect the backreaction of $\mathcal{O}$ on the FSC geometry.} $\mathcal{O}$, in the background of a heavy state $|E\rangle$, which is given by an FSC solution. This is pictorially depicted in Figure~\ref{fig:GeodesicConfiguration}.
We are working in a regime where $\xi_\mathcal{O}, \xi_\chi\gg1$ but the fields are light in the sense that $\xi_\mathcal{O}, \xi_\chi\ll c_M$. We are also assuming $\xi - \frac{p^2}{2\kappa_P}\gg c_M$ for the FSC weights.\\
In this limit, one can use the geodesic approximation and compute the two-point function for a bulk scalar field with mass $m$ using the (regularized) length $\gamma_L$ of a geodesic connecting the two spacetime points. The corresponding two-point function is then given by $e^{-m\gamma_L}$ \cite{Hijano:2017eii}. We will denote the bulk-neutral scalar field dual to the BMSFT probe $\mathcal{O}$ as $\phi_\mathcal{O}$ and it's mass (which is taken to be large) as $m_\mathcal{O}$. Similarly, the charged scalar with mass $m_\chi$ (also assumed to be large) dual to the operator $\chi$ will be denoted by $\phi_\chi$. In the large mass limit, one has
    \begin{equation}
        m_\mathcal{O} \approx \xi_\mathcal{O}, \qquad
        m_\chi \approx \xi_\chi - \frac{p_\chi^2}{2\kappa_P}.
    \end{equation}
For the bulk to boundary scalar $\phi_\mathcal{O}$ one has to calculate the geodesic length from the location of the FSC horizon at $r = r_0$ to some cutoff surface $r = \Lambda$ very close to the asymptotic boundary $r = \infty$. The length of this geodesic is given by
    \begin{equation}
        \gamma_L = \int_{r_0}^\Lambda \frac{r\extd r}{r_+\sqrt{r^2-r_0^2}} = \frac{\sqrt{\Lambda^2-r_0^2}}{r_+^2}.
    \end{equation}
In the limit $\Lambda\rightarrow\infty$ this expression diverges, so we have to regularize it. Taking the logarithm on both sides of the equation, one obtains
    \begin{equation}
        \log \gamma_L = \frac{1}{2}\log\left(\frac{\Lambda^2}{r_0^2}-1\right) - \log\left(\frac{r_+}{r_0}\right).
    \end{equation}
Removing the first term yields a finite expression; thus, the regulated geodesic length is given by $\gamma_L=\frac{r_0}{r_+}$. Therefore the contribution to the one-point function is
    \begin{equation}
        e^{-m_\mathcal{O} \gamma_L} \approx e^{-\xi_\mathcal{O}\frac{r_0}{r_+}} = 
        e^{-\xi_\mathcal{O}\frac{\Delta'}{2\xi'}}.
    \end{equation}
The charge neutral scalar field $\phi_\mathcal{O}$ can be thought of as a result of the fusion of two oppositely charged scalar fields $\phi_\chi$ and $\phi_\chi^\dagger$ that wrap around the horizon whose vertex is proportional to $\langle\chi|\mathcal{O}|\chi\rangle$. Since the scalar $\phi_\chi$ with mass $m_\chi$ has additional $\mathfrak{u}(1)$ charges $(j_\chi,p_\chi)$ there are two contributions to the amplitude coming from the relevant classical Euclidean action of a charged scalar field in the presence of a gravitational field.
    \begin{equation}
        S^{(\chi)}_\textrm{geodesic} = S^{(\chi)}_\textrm{gravity} + S^{(\chi)}_{\mathfrak{u}(1)},
    \end{equation}
with
    \begin{equation}
        S^{(\chi)}_\textrm{gravity} = m_\chi\int\extd\lambda\sqrt{g_{\mu\nu}\dot{x}^\mu\dot{x}^\nu},\qquad S^{(\chi)}_{\mathfrak{u}(1)} = i j_\chi\oint C^J_\mu \extd x^\mu + i p_\chi\oint C^P_\mu \extd x^\mu.
    \end{equation}
The gravitational on-shell action simply corresponds to the horizon area of the FSC
    \begin{equation}
        S^{(\chi)}_\textrm{gravity}\big|_{\textrm{on-shell}} = 2\pi r_0 m\chi.
    \end{equation}
The $\mathfrak{u}(1)$ on-shell action can be explicitly evaluated using the boundary conditions \eqref{eq:u1CSConnection} and yields
    \begin{equation}
        S^{(\chi)}_{\mathfrak{u}(1)}\big|_{\textrm{on-shell}} = i j_\chi\int_0^{2\pi} C^J_\phi \extd \phi + i p_\chi\int_0^{2\pi} C^P_\phi \extd \phi = \frac{2\pi i}{\kappa_P} \left(j_\chi p + p_\chi j\right).
    \end{equation}
Combining all these contributions, the full amplitude for this particular geodesic configuration is
    \begin{equation}
        \langle\chi|\mathcal{O}|\chi\rangle e^{-m_\mathcal{O}\frac{r_0}{r_+}}e^{-2\pi r_0 m_\chi}e^{-\frac{2\pi i}{\kappa_P} \left(j_\chi p + p_\chi j\right)},
    \end{equation}
or more explicitly
    \begin{equation}\label{eq:HolOnePointCompact}
        \langle\chi|\mathcal{O}|\chi\rangle e^{-\xi_\mathcal{O}\frac{\Delta'}{2\xi'}}e^{-2\pi \left(\xi_\chi - \frac{p_\chi^2}{2\kappa_P}\right)\frac{\Delta'}{\sqrt{2c_M \xi'}}}e^{-\frac{2\pi i}{\kappa_P} \left(j_\chi p + p_\chi j\right)}.
    \end{equation}
which has precisely the same form as the previously derived expression in \eqref{eq:HolographicThreePoint} with $c_M = \frac{1}{4G}$, $c_L = 0$, $\kappa_P\neq0$, $\kappa_J = 0$, and $\Delta_\mathcal{O} = \Delta_\chi = 0$.\\
To obtain the general expression \eqref{eq:HolographicThreePoint} from a gravity computation, we now consider the ``reloaded" version of the model \eqref{eq:EH+u1} that we presented previously. Introducing the bilinear form \eqref{eq:InvBilFormModEinstein} effectively only changes the zero modes of the cosmological solution and thus, in turn, also its BMS weights as
    \begin{equation}
        \Delta'\Rightarrow \Delta'+\frac{c_L}{c_M} \xi-\frac{p^2}{\kappa_P}\left(\frac{c_L}{c_M}-\frac{\kappa_J}{\kappa_P}\right),\qquad
        j\Rightarrow j + \frac{\kappa_J}{\kappa_P}p.
    \end{equation}
Replacing $\Delta'$ and $j$ in \eqref{eq:HolOnePointCompact} one immediately reproduces \eqref{eq:HolographicThreePoint}.

\section{Conclusions}\label{sec:Conclusions}
\subsection*{Summary}
In this paper, we investigated the modular properties of 2d field theories with an underlying BMS$_3$ symmetry augmented with $\mathfrak{u}(1)$ currents. We constructed the characters of the highest weight representations and used these to construct the partition function of the theory. Using first a saddle-point approximation and then an exact method, we found the density of primary states and a corresponding BMS-Cardy formula for these theories. The formula for entropy exhibits the usual BMS-Cardy form with Sugawara-shifted weights along with a shift of the central term $c_L \to c_L -1/3$. A shift in $c_L \to c_L -1/6$ was also observed in the usual BMS theory without the $\mathfrak{u}(1)$'s \cite{Bagchi:2020rwb}. Including the $\mathfrak{u}(1)$ currents introduces an additional shift of $-1/6$. We then mcomputed the asymptotic structure constants from the torus one-point function. Similar to previous results in AdS$_3$/CFT$_2$, an Aharonov-Bohm like phase appeared in the expression for the asymptotic structure constants. A bulk analysis of a scalar probe in the background of a charged FSC solution reproduced the boundary analysis, providing further support for our calculations and a holographic interpretation of the result. 

\subsection*{Future directions}
There are several interesting topics that require further attention. In this work, we have considered the zero-point (partition function) and one-point functions here with additional $\mathfrak{u}(1)$ symmetry. The obvious next step is the construction of torus-two point functions with and without the additional $\mathfrak{u}(1)$'s. The calculation of the two-point function would have several interesting consequences. For example, understanding whether or not the Eigenstate Thermalization Hypothesis \cite{Srednicki:1995pt} holds beyond usual Lorentz invariant CFTs \cite{Brehm:2018ipf,Romero-Bermudez:2018dim,Hikida:2018khg}, more specifically in the case of 2d Galilean and Carrollian CFTs. 

The torus two-point function in a 2d CFT also carries information of the dual bulk theory. The poles of the two-point function give the spectrum of quasi-normal modes (QNM) of the BTZ black hole in the bulk. It would be intriguing to investigate whether such a connection remains valid for BMSFTs and the dual FSCs. QNMs for Flat Space Cosmologies have not been worked out as of yet, partly due to a lack of understanding a physical interpretation of QNMs for cosmological horizons and what the correct boundary conditions are for this problem. This computation of the torus two-point function would provide valuable insights into this critical question. Generalizing to computations with additional $\mathfrak{u}(1)$ symmetry would also lead to a better general understanding of holography in asymptotically flat spacetimes, as we have mentioned in the introduction. 

Apart from these applications to flat holography, again as we have stressed in the introduction, BMS field theories with additional charges turned on are important for a variety of problems encompassing generic charged black holes to condensed matter systems where flat bands appear. The study of modular properties that we have developed for these theories in our paper would have an important role to play in the study for all of these problems.

\subsection*{Acknowledgements}

AB would like to acknowledge the warm hospitality of the Erwin Schr\"odinger Institute (University of Vienna), Technische Universit\"at Wien, and Ecole Polytechnique Paris during the final stages of completion of this project. 

The research of AB is partially supported by a Swarnajayanti Fellowship (SB/SJF/2019-20/08) of the Science and Engineering Research Board (SERB), a visiting professorship at Ecole Polytechnique Paris, and by the following grants: CRG/2020/002035 (SERB), Research-in-groups grant (Erwin Schrodinger Institute).   
The research of MR is supported by the European Union’s Horizon 2020 research and innovation program under the Marie Skłodowska-Curie grant agreement No.~832542.
The research of RC is supported by the CSIR grant File No: 09/092(0991)/2018-EMR-I.

\section*{Appendix}
\appendix

\section{Review of 2d BMSFT}\label{ApA} 
The generators of 2d BMSFT can be obtained as the In{\"o}n{\"u}-Wigner contraction of the linear combinations of two copies of the generators of the Virasoro algebra. 
\begin{equation}
\label{vir_gen}
    L_{n}=\mathcal{L}_{n}-\bar{\mathcal{L}}_{-n},\quad M_{n}=\lim_{\epsilon\to 0}\epsilon(\mathcal{L}_{n}+\bar{\mathcal{L}}_{-n}).
\end{equation} 
This is also called ultra-relativistic contraction. 
The Virasoro algebra is given by 
\begin{equation}
[\mathcal{L}_{n},\mathcal{L}_{m}]=(n-m) \mathcal{L}_{n+m}+\frac{c}{12}(n^{3}-n)\delta_{n+m,0},\hspace{3mm} 
[\bar{\mathcal{L}}_{n},\bar{\mathcal{L}}_{m}]=(n-m) \bar{\mathcal{L}}_{n+m}+\frac{\bar{c}}{12}(n^{3}-n)\delta_{n+m,0}.
\end{equation}
Using the commutation relations of the Virasoro generators $\mathcal{L}_{n}$ and $\bar{\mathcal{L}}_{n}$, one obtains the following algebra for 2d BMSFT. 
\begin{align}
    [L_{n},L_{m}]&=(n-m)L_{n+m}+c_{L}(n^{3}-n)\delta_{n+m,0}, \\
    [L_{n},M_{m}]&=(n-m)M_{m+n}+c_{M}(n^{3}-n)\delta_{m+n,0},\\
    [M_{n},M_{m}]&=0,
\end{align}
where $c_L$ and $c_M$ are the central charges of the algebra. They are related to the central charges of CFT as $c_{L}=\frac{c-\bar{c}}{12}$ and $c_{M}=\frac{\epsilon(c+\bar{c})}{12}$.  

\subsection*{Coordinate Representation of BMS Algebra}
2d BMSFT is defined on a spacetime having the following metric
\begin{align}
ds^2=0 \times du^2+ d\theta^{2}.
\end{align}
The topology of the spacetime is that of a cylinder, which is $\mathbb{R}_{u}\times S^{1}$. The ultra-relativistic nature of the contraction becomes evident considering the coordinate representation of  $L_{n}$ and $M_{n}$ by taking the limit $u\to\epsilon u$, $\theta\to\theta$ with $\epsilon\to 0$ on the coordinate representation of Virasoro generators $\mathcal{L}_{n}$ and $\bar{\mathcal{L}}_{n}$  
\begin{equation}
    \mathcal{L}_{n}=e^{inw}\partial_{w},\quad\bar{\mathcal{L}}_{n}=e^{-in\bar{w}}\partial_{\bar{w}},\quad \text{where}\quad w,\bar{w}=u\pm\theta.
\end{equation}
Using \eqref{vir_gen}, the coordinate representation of the generators $L_{n}$ and $M_{n}$ one finds
\begin{align}
    L_{n}=e^{in\theta}(\partial_{\theta}+inu\partial_{u}),\quad M_{n}=e^{in\theta}\partial_{u}.
\end{align}

\subsection*{2d BMS Highest Weight Representation}
 The center of the BMS algebra is $\{L_{0},M_{0}\}$. The states of the BMSFT are labeled by the simultaneous eigenstates of $L_{0}$ and $M_{0}$.
\begin{align}
    L_{0}\ket{\Delta,\xi}=\Delta\ket{\Delta,\xi},\hspace{5mm}M_{0}\ket{\Delta,\xi}=\xi\ket{\Delta,\xi}.
\end{align}
Similar to CFT, we define the BMS primary states as follows:
\begin{align}
   L_{n}\ket{\Delta,\xi}=M_{n}\ket{\Delta,\xi}=0,\hspace{5mm}\forall n>0. 
\end{align}
The BMS algebra dictates that the action of $L_{n}$ and $M_{n}$ on $\ket{\Delta,\xi}$ is
\begin{align}
    L_{0}L_{n}\ket{\Delta,\xi}=(\Delta-n)L_{n}\ket{\Delta,\xi},\hspace{5mm}L_{0}M_{n}\ket{\Delta,\xi}=(\Delta-n)M_{n}\ket{\Delta,\xi}.
\end{align}
Hence, $\Delta$ can be increased by $n$ when $L_{-n}$ and $M_{-n}$ acts on a primary state $\ket{\Delta,\xi}$. States derived from the primary states in this way serve as a basis of BMS module $\mathcal{B}(\Delta,\xi,c_{L},c_{M})$. These basis states are given by
\begin{align}
    L^{n_1}_{-1}L^{n_2}_{-2}....L_{-a}^{n_a}M^{m_1}_{-1}M^{m_2}_{-2}....M^{m_b}_{-b}\ket{\Delta,\xi}=L_{\Vec{n}}M_{\Vec{m}}\ket{\Delta,\xi}.
\end{align}
In the above $\vec{n}=(n_1,n_2,....n_{a})$ and $\vec{m}=(m_1,m_2....m_b)$. The $L_{0}$ eigenvalue of this state is
\begin{align}
    L_{0}L_{\Vec{n}}M_{\Vec{m}}\ket{\Delta,\xi}=(N+\Delta)L_{\Vec{n}}M_{\Vec{m}}\ket{\Delta,\xi},\quad N=\sum_{i}ip_{i}+\sum_{j}jq_{j}.
\end{align}
Here $N$ is called the level of the state. The Hilbert space of BMSFT hence is the direct sum of the BMS modules of all BMS primaries
\begin{align}
    \mathcal{H}(c_L,c_M)=\oplus_{(\Delta,\xi)}\mathcal{B}(c_{L},c_{M},\Delta,\xi).
\end{align}
\subsection*{Partition Function, Modular Invariance, and Cardy Formula}
The partition function of BMSFT is defined as
\begin{align}
    Z(\sigma,\rho)=tr(e^{2\pi i\sigma(L_{0}-c_L/2)}e^{2\pi i\rho(M_{0}-c_M/2)})=\sum_{\Delta,\xi}d(\Delta,\xi)e^{2\pi i\Delta}e^{2\pi i\xi}.
\end{align}
where $d(\Delta,\xi)$ is the density of states having weights ($\Delta,\xi$). From the limit of the parent CFT, one can derive the modular transformation for BMSFT, which is \cite{Bagchi:2013qva}
\begin{align}
    \sigma\to\frac{a\sigma+b}{c\sigma+d}\hspace{2mm},\quad\rho\to\frac{\rho}{(c\sigma+d)^2}.
\end{align}
The invariance of the partition function under S-modular transformation gives the BMS-Cardy formula for entropy \cite{Bagchi:2013qva} which is 
\begin{align}
    S^{(0)}=\log{d(\Delta,\xi)}=2\pi\Big(c_L\sqrt{\frac{\xi}{2c_M}}+\Delta\sqrt{\frac{c_M}{2\xi}}\Big).
\end{align}
\subsection*{Highest Weight Character and BMS Cardy Formula for Primaries}
The character of the BMS module is defined as
\begin{align}
    \chi_{(\Delta,\xi,c_L,c_M)}(\sigma,\rho)=tr_{\Delta,\xi}(e^{2\pi i\sigma(L_{0}-c_L/2)}e^{2\pi i\rho(M_{0}-c_M/2)}),
\end{align}
where the trace is taken over all the descendant states of $\ket{\Delta,\xi}$ in the module $\mathcal{B}(\Delta,\xi,c_{L},c_{M})$. One can rewrite the partition function in terms of characters as follows
\begin{align}
    Z(\sigma,\rho)=\sum_{\Delta,\xi}D(\Delta,\xi) \chi_{(\Delta,\xi,c_L,c_M)}(\sigma,\rho),
\end{align}
where $D(\Delta,\xi)$ is the density of primaries. The character for BMS module $\mathcal{B}(\Delta,\xi,c_{L},c_{M})$ for non-zero $\Delta$ and $\xi$ is given by \cite{Bagchi:2019unf}
\begin{align}
    \chi_{(c_L,c_M,\Delta,\xi)}(\sigma,\rho)=\frac{e^{\frac{2\pi i\sigma}{12}}e^{-2\pi i\big(\sigma\frac{c_L}{2}+\rho\frac{c_M}{2}\big)}e^{2\pi i(\sigma\Delta+\rho\xi)}}{\eta(\sigma)^2}.
\end{align}
For the vacuum ($\Delta=\xi=0$), the character is given by \cite{Bagchi:2019unf}
\begin{align}
    \chi_{(c_L,c_M,0,0)}(\sigma,\rho)=\frac{e^{\frac{2\pi i\sigma}{12}}e^{-2\pi i\big(\sigma\frac{c_L}{2}+\rho\frac{c_M}{2}\big)}}{\eta(\sigma)^2}(1-e^{2\pi i\sigma})^2.
\end{align}
Hence, the partition function can be written in terms of the density of primary states as \cite{Bagchi:2019unf}
\begin{align}
    Z(\sigma,\rho)=\frac{e^{\frac{2\pi i\sigma}{12}}e^{-2\pi i\big(\sigma\frac{c_L}{2}+\rho\frac{c_M}{2}\big)}}{\eta(\sigma)^2}\sum_{\Delta,\xi}D(\Delta,\xi)e^{2\pi i(\sigma\Delta+\rho\xi)}.
\end{align}
Here one can again use the invariance of the partition function under the S-modular transformation to find the Cardy formula for primary states \cite{Bagchi:2019unf}
\begin{align}
    S_{primary}=\log{D(\Delta,\xi)}=2\pi\Bigg(\sqrt{\frac{\xi}{2c_M}}\Big(c_L-\frac{1}{6}\Big)+\Delta\sqrt{\frac{c_M}{2\xi}}\Bigg).
\end{align}
Hence one can see that the Cardy formula for primary has a shift from the Cardy formula for all states given by
\begin{align}
    c_L\to c_L-\frac{1}{6}.
\end{align}

\section{Review of Virasoro $\oplus$ $\mathfrak{u}(1)$ Theories}\label{ApB}
For a CFT$_2$ with $\mathfrak{u}(1)$ symmetry, the algebra gets enhanced due to the presence of the generators of $\mathfrak{u}(1)$. The algebra of two copies of Virasoro generators $\mathcal{L}_{n}$, $\bar{\mathcal{L}}_{n}$ and $\mathfrak{u}(1)$ generators $j_{n}$ and $\bar{j}_{n}$ is given by  
\begin{equation}
  [\mathcal{L}_{n},\mathcal{L}_{m}]=(n-m)\mathcal{L}_{n+m}+\frac{c}{12}(n^{3}-n)\delta_{n+m,0},\hspace{3mm}[\mathcal{L}_{n}, j_{m}]= -m j_{n+m} ,\hspace{3mm}[j_{n},j_{m}]=kn\delta_{n+m,0},
\end{equation}
\begin{equation}
  [\mathcal{\bar{L}}_{n},\mathcal{\bar{L}}_{m}]=(n-m)\mathcal{\bar{L}}_{n+m}+\frac{\bar{c}}{12}(n^{3}-n)\delta_{n+m,0},\hspace{3mm}[\mathcal{\bar{L}}_{n}, \bar{j}_{m}]= -m \bar{j}_{n+m},\hspace{3mm}[\bar{j}_{n},\bar{j}_{m}]=\bar{k}n\delta_{n+m,0}. 
\end{equation}
Here $c, \bar{c}$ are the central charges, $k, \bar{k}$ are the levels of the current algebras.\\
We work with a highest weight representation of the full algebra, denoted by highest weight states $\ket{\psi}$ satisfying
\be 
    \begin{split}
        \mathcal{L}_0\ket{\psi} = h\ket{\psi}, \ & \ \bar{\mathcal{L}}_0\ket{\psi} = \bar{h}\ket{\psi} \ , \
        j_0\ket{\psi} = q\ket{\psi} \ , \ \bar{j}_0\ket{\psi} = \bar{q}\ket{\psi},\\
        \mathcal{L}_n\ket{\psi} = \bar{\mathcal{L}}_n\ket{\psi} &= j_n\ket{\psi} = \bar{j}_n\ket{\psi} = 0, \ \ \ \forall n>0.
    \end{split}
\ee 
The total Hilbert space of the theory is made up of the set of highest weight states $\{\ket{\psi_i}\}$ and their descendants. 

\subsubsection*{Character and Partition Function}
For a 2D CFT with $\mathfrak{u}(1)$ symmetry, the grand canonical partition function on the torus with non-vanishing chemical potential $\nu$ is given by \cite{Das:2017vej}
\begin{equation}
    Z(\tau,\bar{\tau},\nu,\bar{\nu})=\Tr\left[e^{2\pi i\tau(\mathcal{L}_{0}-c/2)}e^{-2\pi i\bar{\tau}(\bar{\mathcal{L}}_{0}-\bar{c}/2}e^{2\pi i\nu J_{0}}e^{-2\pi i\nu\bar{J}_{0}}\right]  .
\end{equation} 
Here, $\tau$ is the modular parameter of a torus with a spatial circle of circumference $L$ and temporal circle of circumference $\beta$ and is given by $\tau=i\beta/L$. Under a modular transformation, which takes $\tau$ to $\frac{a\tau+b}{c\tau+d}$ and $\nu$ to $\frac{\nu}{c\tau+d}$, the partition function transforms as \cite{Benjamin:2016fhe}
\begin{equation}
\label{cftp}
\mathcal{Z}\left(\frac{a\tau+b}{c\tau+d},\frac{\nu}{c\tau+d}\right)=\exp{\left(\frac{i k c\pi \nu^{2}}{c\tau+d}-\frac{i \bar{k} c\pi \bar{\nu}^{2}}{c\bar{\tau}+d}\right)}\mathcal{Z}(\tau,\nu).
\end{equation}
The character is defined as a trace over states of a module $\mathcal{B}(h,q,c,k)$. The character is then given by
\be 
\chi_{\{h, \bar{h}, \mathcal{Q}, \bar{\mathcal{Q}}\}} = e^{2\pi i\tau(h-c/24)}e^{2\pi i\tau(\bar{h}-\bar{c}/24)} e^{2\pi i\nu \mathcal{Q}} e^{2\pi i\bar{\nu} \bar{\mathcal{Q}}} \frac{e^{\frac{\pi i(\tau + \bar{\tau})}{6}}}{\eta^{2}(\tau)\eta^{2}(\bar{\tau})} .
\ee

\subsubsection*{Cardy Formula}
Starting with the torus partition function, and using its modular transformation properties (\ref{cftp}) one obtains the density of states $\rho(h,\bar{h},q,\bar{q})$ as
\begin{align}
    \rho(h, \bar{h},q, \bar{q})\approx \tilde{\rho}(0,0) \exp\bigg\{2\pi\sqrt{\frac{c'}{6}\left(h'-\frac{c'}{24}\right)}+ \sqrt{\frac{\bar{c}'}{6}\left(\bar{h}'-\frac{\bar{c}'}{24}\right)}\bigg\},
\end{align}
which means that the entropy of primaries is
\begin{align}
    S=2\pi\bigg(\sqrt{\frac{c'}{6}\left(h'-\frac{c'}{24}\right)}+\sqrt{\frac{\bar{c}'}{6}\left(\bar{h}'-\frac{\bar{c}'}{24}\right)}\bigg).
\end{align}
Here $h'=h-\frac{q^2}{2k}$, and $c'=c-2$ and similar relations hold true for the anti-holomorphic counterparts.

\subsubsection*{Torus 1-Point Function}
A primary operator $\mathcal{O}(w,\bar{w})$ with weights $h,\bar{h}$ transforms under a modular transformation in the following manner
\begin{equation}
    \mathcal{O}(w^{\prime},\bar{w}^{\prime})\vert_{\frac{a\tau+b}{c\tau+d}}=\left(\frac{\partial w^{\prime}}{\partial w}\right)^{-h}\left(\frac{\partial \bar{w}^{\prime}}{\partial \bar{w}}\right)^{-\bar{h}}\mathcal{O}(w,\bar{w})\vert_{\tau}.
\end{equation} 
Here, $w$ is the elliptic coordinate on torus and the modular-transformed coordinate $w^{\prime}$ is related to $w$ by 
\begin{equation}
    w^{\prime}=\frac{w}{c\tau+d}.
\end{equation}
The one-point function of an operator $\mathcal{O}(w,\bar{w})$ on a torus with modular parameter $\tau$ is given by 
\begin{equation}
    \langle\mathcal{O}(w,\bar{w})\rangle_{\tau}=\Tr\left[\mathcal{O}(w,\bar{w})e^{2\pi i\tau(\mathcal{L}_{0}-c/2)}e^{2\pi i\nu J_{0}}\right].
\end{equation}
Under a modular transformation the one-point function transforms as
\begin{equation}
\begin{split}
    \langle\mathcal{O}(w^{\prime},\bar{w}^{\prime})\rangle_{\left(\frac{a\tau+b}{c\tau+d},\frac{\nu}{c\tau+d}\right)}&= \exp(\frac{i k c\pi \nu^{2}}{c\tau+d}-\frac{i \bar{k} c\pi \bar{\nu}^{2}}{c\bar{\tau}+d})\left(\frac{\partial w^{\prime}}{\partial w}\right)^{-h}\left(\frac{\partial \bar{w}^{\prime}}{\partial \bar{w}}\right)^{-\bar{h}}\langle\mathcal{O}(w,\bar{w})\rangle_{\tau,\nu}\\
    &=\exp(\frac{i k c\pi \nu^{2}}{c\tau+d}-\frac{i \bar{k} c\pi \bar{\nu}^{2}}{c\bar{\tau}+d})(c\tau+d)^{h}(c\bar{\tau}+d)^{\bar{h}}\langle\mathcal{O}(w,\bar{w})\rangle_{\tau,\nu}.
\end{split}
\end{equation}

\section{BMS$\oplus\mathfrak{u}(1)$ Hilbert Space, Gram Matrix, and Character}\label{ApC}
In this section we generalize the calculation of the trace of the operator
\be{}
\hat{\mathcal{O}}=e^{2\pi i\sigma(L_{0}-c_{L}/2)}e^{2\pi i\rho(M_{0}-c_{M}/2)}e^{2\pi i\alpha J_{0}}e^{2\pi i\gamma P_{0}},
\ee  
for any given level $N$. The basis states are the $N^\textrm{th}$ level descendant states of a highest weight primary 
$$\ket{A}=\vert\Delta,\xi,j,p\rangle. $$ 
In the following we will show that 
\be{}
(\Tr\hat{\mathcal{O}})_{N}=dim_{N}e^{2\pi i\sigma(\Delta-c_{L}/2)}e^{2\pi i\rho(\xi-c_{M}/2)}e^{2\pi i\alpha j}e^{2\pi i\gamma p},
\ee 
as claimed in (\ref{tr1}).      
\subsection*{Equivalent States} 
Before determining the contribution to the character, we need to define a notion of ``equivalent'' operators or states. Any descendant state is created by applying a string of $L_{-n}, J_{-n}, M_{-n}, P_{-n}$ operators on a highest weight state. Among these operators, $L_{-n}$ and $J_{-n}$ are `similar' and the same is true for $M_{-n}$ and $P_{-n}$. We will make this notion of `similarity' clear in the next section. First let us define a general descendant state as 
\begin{equation}\label{desc1}
    \begin{split}
    \ket{\psi^{(N)}}&= L_{-l_{1}}^{p_1} \dots L_{-l_{a}}^{p_a} M_{-m_{1}}^{q_1}\dots M_{-m_{b}}^{q_b} J_{-l^{\prime}_{1}}^{r_1}\dots J_{-l^{\prime}_{c}}^{r_c} P_{-m^{\prime}_{1}}^{s_1}\dots P_{-m^{\prime}_{d}}^{s_d}\ket{A},
    \end{split}
\end{equation}
where $1\leq l_{1}<l_{2}<...<l_{a}$ , $1\leq m_{1}<m_{2}<...<m_{b}$ , $1\leq l^{\prime}_{1}<l^{\prime}_{2}<...<l^{\prime}_{c}$ , $1\leq m_{1}^{\prime}<m^{\prime}_{2}<...<m^{\prime}_{d}$. Furthermote, $(\sum_{i=1}^{a}p_{i}l_{i}+\sum_{i=1}^{b}m_{i}q_{i}+\sum_{i=1}^{c}r_{i}l^{\prime}_{i}+\sum_{i=1}^{d}s_{i}m^{\prime}_{i})=N$ denotes the level of the descendant state. Based on our notion of similarity, we define the set of equivalent states of $\ket{\psi^{(N)}}$ to be the set of states that one can obtain from $\ket{\psi^{(N)}}$ by swapping one or more $L_{-n} \rightarrow J_{-n}$ or $M_{-n} \rightarrow P_{-n}$ operators, and vice versa. One can represent the equivalent class of $\ket{\psi^{(N)}}$ as $\{\ket{\psi^{(N)}}\}$ or with any of its members. For the sake of simplicity, we will denote the equivalence sets with the elements constructed from only the BMS generators $L_{i},M_{i}$. For example,
\begin{equation}
    \begin{split}
        \{L_{-1}M_{-1}\ket{A}\} &\equiv \{ L_{-1}M_{-1}\ket{A}, L_{-1}P_{-1}\ket{A}, J_{-1}M_{-1}\ket{A}, J_{-1}P_{-1}\ket{A}\}, \\
        \{L_{-1}^2\ket{A}\} &\equiv \{L_{-1}^2\ket{A}, L_{-1}J_{-1}\ket{A}, J_{-1}^2\ket{A}\}.
    \end{split}
\end{equation}
\subsection*{Nonzero Inner Product Rule} 
In BMS field theory the inner product of two states $L_{-l_{1}}^{p_{1}}\dots L_{-l_{a}}^{p_{a}}M_{-m_{1}}^{q_{1}}\dots M_{-m_{b}}^{q_{b}}\ket{\Delta,\xi}$ and \\$L_{-l^{\prime}_{1}}^{p^{\prime}_{1}}\dots L_{-l^{\prime}_{a^{\prime}}}^{p^{\prime}_{a^{\prime}}}\ket{\Delta,\xi}$ has non-zero inner-product only if they are at the same level and for every $M_{m^{\prime}_{i}}$ there should be strings of $L$ operators that add up to $L_{-m^{\prime}_{i}}$. Similarly, for every $M_{-m_{i}}$ there should be combinations of $L$ adding up to $L_{m_{i}}$. For a detailed discussion see \cite{Bagchi:2020rwb}. This is how we motivate our notion of 'similarity' between $L, J$ and $M, P$ that was mentioned in the previous section.     
Looking at the commutation rules of the generators one has schematically
\begin{equation}
    \begin{split}
        [L_m, L_n] \sim L_{n+m}\ + f(n)\delta_{m+n,0}  \ &, \ [L_m, J_n] \sim J_{m+n},\\
        [L_m, M_n] \sim M_{n+m}\  + f(n)\delta_{m+n,0}  \,, \, [L_m, P_n]& \sim P_{m+n}\,, \, [M_m, J_n] \sim P_{m+n}.
    \end{split}
\end{equation}
Alternatively, one could also write (up to additive terms like $f(n)\delta_{m+n,0}$)
\begin{equation}
    [L_m, \{L_n\}] \sim \{L_{m+n}\} \ , \ [M_m, \{L_n\}] \sim \{M_{m+n}\}.
\end{equation}
Because of this similarity in commutators, the nonzero inner product rules of the BMSFT Hilbert space also apply to the BMS$\oplus\mathfrak{u}(1)$ Hilbert space. Due to the commutation relations of generators in an BMS$\oplus\mathfrak{u}(1)$ theory, an analogue of ``L majority rule'' is also here. For descendant states that only contain $J$ and $P$ operators, the total number of $J_{i}$'s for a given level $i$ should be more than the total number of $P_{-i}$'s inside an inner product to get a non-zero value. Another way to state this rule is that for two states consisting of all operators $L, M, J, P$, one should  first obtain the equivalent representative element of the states, that is, convert all $J$ to $L$ and $P$ to $M$. Then, one has to compare the number of $M$ and $L$ for each level. If the number of $M$ surpasses the number of $L$, the inner product will be zero. With this condition satisfied, the sum of $L$ indices should be greater that that of $M$ indices for a non-zero inner product.   

\subsection*{Ordering Rule for BMS$\oplus\mathfrak{u}(1)$ States}
Based on the inner product properties, one can order the states in an efficient fashion to obtain a particularly simple Gram matrix. Before going into the details of the ordering rule we first define conjugate states. Two states that can be related to each other by swapping $L\leftrightarrow M$ and $J\leftrightarrow P$ are conjugate to each other. The ordering rules are similar to those of BMS descendant states, and are given by
\begin{enumerate}
    \item For any level N, collect `similar' states into a set of equivalent states represented by a state created purely of $L$ and $M$. These sets are to be ordered following the BMS ordering rule as explained in \cite{Bagchi:2020rwb}. 
    \item Within each set, the ordering is not important, but the ordering of the conjugate states should follow a certain order. Suppose the number of states in the equivalence class $\{\ket{\psi^{(N)}}\}$ is $D$. Now assume that a state $\ket{\psi_{i}^{N}}$ is placed at $d$th position. Then one has to make sure that its conjugate state $\ket{\psi^{N,c}_{i}}$ (obtained by swapping all $L \leftrightarrow M$, $J \leftrightarrow P$ of $\ket{\psi_{i}^{N}}$) is placed at $(D-d+1)$ th position within the conjugate set $\{\ket{\psi^{c}}\}$. For the self-conjugate states, any ordering within an equivalent set would work.
\end{enumerate}
Take as an example for this ordering the second level descendant states. Here, the equivalent states are given by
\begin{equation}\label{l2sets}
    \begin{split}
        &\{L_{-1}^2\} = \{L_{-1}^2, J_{-1}^2, L_{-1}J_{-1}, \},\\
        &\{L_{-2}\} = \{L_{-2}, J_{-2}\},\\
        &\{L_{-1}M_{-1}\} = \{L_{-1}M_{-1}, J_{-1}M_{-1}, L_{-1}P_{-1}, J_{-1}P_{-1}\},\\
        &\{M_{-2}\} = \{M_{-2}, P_{-2} \},\\
        &\{M_{-1}^2\} = \{ M_{-1}P_{-1}, P_{-1}^2, M_{-1}^2\}.
    \end{split}
\end{equation}
 As stated, the sets will be ordered as $\{\{L_{-1}^2\}, \{L_{-2}\}, \{L_{-1}M_{-1}\}, \{M_{-2}\} , \{M_{-1}^2\}\}$. To maintain symmetry among the conjugate states, we choose the following ordering
\be\label{l2order}
\scriptstyle \{L_{-1}^2, L_{-1}J_{-1}, J_{-1}^2\}, \{L_{-2}, J_{-2}\}, \{L_{-1}M_{-1}, J_{-1}M_{-1}, L_{-1}P_{-1}, J_{-1}P_{-1}\}, \{P_{-2} , M_{-2}\}, \{P_{-1}^2, M_{-1}P_{-1}, M_{-1}^2\}.
\ee
Note that the set $\{L_{-1}M_{-1}\}$ is a self conjugate set, so we have just chosen one possible ordering.

\subsection*{Gram Matrix in Terms of Equivalent State Blocks}
The ordering of states ensures that the BMS Gram matrix is upper-anti-triangular for odd level. With the elaborate ordering rule that we just laid out we argue that if an element of the BMS Gram matrix is zero because of the ``M majority rule'', then in the BMS$\oplus \mathfrak{u(1)}$ Hilbert space, the block of the Gram matrix created from the equivalent sets of the previously mentioned BMS states will be a zero block. One can see that the Gram Matrix of the BMS$\oplus\mathfrak{u}(1)$ Hilbert space will be block-wise upper-anti-triangular, where each block will have variable size, depending on the number of states in the sets $\{\ket{\psi^{N}}\}$
\be
K^{(N=\text{odd})}_{\text{BMS$\oplus\mathfrak{u}(1)$}} = \left(\begin{array}{ccccc}
\boxed{K_{1,1}} &  & \cdots &  & \boxed{A_{\tilde{D}}}\\
 &  &  & \iddots & 0\\
\vdots &  & \iddots &  & \vdots\\
 & \boxed{A_{2}}\\
\boxed{A_{1}} & 0 & \cdots &  & 0
\end{array}\right) .
\label{oddmatrix}
\ee
Similarly for even level, the Gram matrix will have the same structure as a BMS even level Gram matrix, but in blocks
\be
K^{(N=\text{even})}_{\text{BMS$\oplus\mathfrak{u}(1)$}}=\left(\begin{array}{cccccccccc}
\boxed{K_{1,1}} & \cdots & \cdots &  & \cdots &  & \cdots &  & \cdots & \boxed{A_{\tilde{D}}}\\
\vdots & \ddots &  &  &  &  &  &  & \iddots & 0\\
 &  & \ddots &  &  &  &  & \boxed{A_{k+l}} &  & \vdots\\
 &  &  & \boxed{B_{1}} & 0 & \cdots & 0\\
\vdots &  &  & 0 & \boxed{B_{2}} & \cdots & 0 &  &  & \vdots\\
\vdots &  &  & \vdots & \vdots & \ddots & \vdots &  &  & \vdots\\
 &  &  & 0 & 0 & \cdots & \boxed{B_{l}}\\
 &  & \boxed{A_{k}} &  &  &  &  & 0\\
\vdots & \iddots &  &  &  &  &  &  & \ddots & \vdots\\
\boxed{A_{1}} & 0 & \cdots &  & \cdots &  & \cdots &  & \cdots & 0
\end{array}\right).
\label{evenmatrix}
\ee
Here, $\tilde{D}$ is the number of equivalent sets at level $N$. One can see that this is actually equal to the number of BMS states at level $N$, denoted by $\widetilde{\rm {dim}_N}$. Note that this is different from the number of BMS$\oplus\mathfrak{u}(1)$ descendants at level $N$, ${dim_N}$. 

The anti-diagonal blocks $A_k$ will be square, as they are the inner product between conjugate states. Due to the nature of the commutation relations between the generators, one has (for odd levels)
\be\label{oddmatrix2}
\boxed{A_{(i)}} = \boxed{A_{(D-i+1)}^T}.
\ee
One can verify this in the case of the level 2 Gram matrix explicitly. Here, the equivalent sets of states are ordered as (\ref{l2order}) and the corresponding Gram matrix is
\begin{equation}
    \begin{psmallmatrix}
    8\Delta^2+4\Delta &&\cdots&&\cdots&&\cdots&&\cdots&&\cdots& 4\xi p & 2p^{2} & 8\xi^{2} \\
     &\ddots&&&\cdots&&\cdots&&\cdots&&\cdots& 2\kappa_{P}p & 2\kappa_{P}^{2} & 2p^{2}\\
     &&\ddots&&&&\cdots&&\cdots&&\cdots& 2\kappa_{P}\xi+p^{2} & 2\kappa_{P}p & 4\xi p\\ 
     \vdots&\vdots&&\ddots&&&\cdots&&\cdots& 2p & 4\xi &0&0& 0\\
     &&&&\ddots&&\cdots&&\cdots& 2\kappa_{P}&2p &0&0& 0\\
     \vdots&\vdots&\vdots&\vdots&& 4\xi^{2} & 2\xi p & 2\xi p & p^{2} & 0&0&0&0&0\\
     &&&&&  4\xi p & p^{2} & 2\xi\kappa_{P} & \kappa_{P}p & 0&0&0&0&0\\
     &&&&&  2\xi p & 2\xi\kappa_{P} & p^{2} & \kappa_{P}p & 0&0&0&0&0\\
    &&&&&  p^{2} & \kappa_{P}p & \kappa_{P}p & p^{2} & 0&0&0&0&0\\
    \vdots&\vdots&\vdots& 2Q_K & 2\mathcal{K}_1 &0 &0&0&0&0&0&0&0&0\\
    &&& 4\xi & 2Q_K &0 &0&0&0&0&0&0&0&0\\
    4\xi Q_K & 2\mathcal{K}_1Q_K & 2\mathcal{K}_1\xi+Q_K^2&0&0&0&0&0&0&0&0&0&0&0\\ 
    2Q_K^2 & 2\mathcal{K}_1^2& 2\mathcal{K}_1Q_K &0&0&0&0&0&0&0&0&0&0&0\\
    8\xi^2 & 2Q_K^2 & 4\xi Q_K &0&0&0&0&0&0&0&0&0&0&0  
    \end{psmallmatrix},
\end{equation}
which has 4 anti-diagonal blocks $A_i$ and one diagonal block $B_1$ in the middle (because of self conjugate set $\{L_{-1}M_{-1}\}$).

\subsection*{Inverse Gram Matrix}
Because of the block-wise upper-anti-triangular structure of the odd level Gram matrix, the inverse Gram matrix will be lower-anti-triangular, with the elements on the anti diagonal being the inverse of the anti diagonal element of the Gram matrix. So the inverse Gram matrix will look like
\begin{equation}
\label{igram1}
\bigl(K^{(N=\text{odd})}_{\text{BMS$\oplus\mathfrak{u}(1)$}}\bigr)^{-1} \equiv K_{(N=\text{odd})} = \left(\begin{array}{ccccc}
$0$ & \cdots & $0$ & \cdots & \fbox{$A_{1}^{-1}$}\\
$0$ &  &  & \fbox{$A_{2}^{-1}$} & \vdots\\
\vdots &  & \iddots &  & \vdots\\
 & \fbox{$A_{D-1}^{-1}$} &&&\\
\fbox{$A_{D}^{-1}$} & \iddots & \cdots & \cdots & \fbox{$K^{D,D}$}
\end{array}\right) .
\end{equation}
Similarly for even levels, the inverse Gram matrix looks like
\be\label{igram2}
K_{(N=\text{even})}=\left(\begin{array}{cccccccccc}
0 & \cdots & \cdots &  & \cdots &  & 0 &  & \cdots & \fbox{$A_{1}^{-1}$}\\
\vdots & \ddots &  &  &  &  &  &  & \iddots & \\
 &  & \ddots &  &  &  &  & \fbox{$A_{k}^{-1}$} &  & \vdots\\
 &  &  & \fbox{$B_{1}^{-1}$} & 0 & \cdots & 0\\
\vdots &  &  & 0 & \fbox{$B_2^{-1}$} & \cdots & 0 &  &  & \vdots\\
0 &  &  & \vdots & \vdots & \ddots & \vdots &  &  & \vdots\\
 &  &  & 0 & 0 & \cdots & \fbox{$B_{l}^{-1}$}\\
 &  & \fbox{$A_{k+l}^{-1}$} &  &  &  &  & \ddots \\
\vdots & \iddots &  &  &  &  &  &  & \ddots & \vdots\\
\fbox{$A_{D}^{-1}$} &  & \cdots &  & \cdots &  & \cdots &  & \cdots & \fbox{$K^{D,D}$}
\end{array}\right).
\ee

\subsection*{Character Computation}
Due to the non-orthogonal nature of the Hilbert space, we define the character of a highest weight state $\ket{A}= \ket{\Delta_A, \xi_A, \mathcal{Q}_{J,A}, \mathcal{Q}_{K, A}}$ as 
\begin{equation}
    \chi = \sum_{N,i,j} ({K^{(N)}}^{-1})_{ij}\bra{\psi^{(N)}_i}\hat{O}\ket{\psi^{(N)}_j} = \sum_{N,i,j} ({K_{(N)}})_{ij} \hat{O}_{ij},
\end{equation}
where ${K^{(N)}}^{-1} \equiv K_{(N)}$ is the Gram matrix of the level N descendants of $\ket{A}$, and the operator\\
$\hat{O} = e^{2\pi i \sigma(L_0-\frac{c_L}{2})}e^{2\pi i \rho(M_0-\frac{c_M}{2})}e^{2\pi i \gamma J_0}e^{2\pi i \delta P_0}$.\\
Take the general descendant state as defined in (\ref{desc1}). One can simplify the character expression by noting the following properties of the generators
\begin{equation}
    L_0\ket{\psi^{(N)}} = (\Delta_A + N)\ket{\psi^{(N)}} \ , \ J_0\ket{\psi^{(N)}}= \mathcal{Q}_{J,A}\ket{\psi^{(N)}} \ , \ P_0\ket{\psi^{(N)}} = \mathcal{Q}_{K,A}\ket{\psi^{(N)}}.
\end{equation}
Using these expressions one obtains 
\begin{equation}
    \hat{O}\ket{\psi^{(N)}} = \bigl(q_L^{(\Delta_A + N -\frac{c_L}{2})}q_M^{(-\frac{c_M}{2})}e^{2\pi i \gamma \mathcal{Q}_{J,A}}e^{2\pi i \delta \mathcal{Q}_{K,A}}\bigr)q_M^{M_0}\ket{\psi^{(N)}},
\end{equation}
where $q_M = e^{2\pi i \rho} , q_L = e^{2\pi i \sigma}$. Thus, one can write the character of the highest weight state $\ket{A}$ as
\begin{equation}\label{charac1}
    \chi_A = \bigl(q_L^{(\Delta_A  -\frac{c_L}{2})}q_M^{(-\frac{c_M}{2})}e^{2\pi i \gamma Q_{J,A}}e^{2\pi i \delta Q_{K,A}}\bigr)\sum_{N,i,j} q_L^N {K^{(N)}}^{-1}_{ij} \bigl( q_M^{M_0}\bigr)_{ij}.
\end{equation}
This means that the remaining computation boils down to computing the matrix elements of the operator $q_M^{M_0}$. In particular, we will see that the matrix of $q_M^{M_0}$ will be upper-anti-triangular, just like the Gram matrix. As a result, only the anti diagonal matrix elements of the matrices ${K^{(N)}}^{-1}$ and $q_M^{M_0}$ will contribute to the sum in (\ref{charac1}).

\subsection*{Action of $q_{M}^{M_0}$ on Equivalent State Blocks}
For the following calculation, we will assume $N$ to be odd for simplicity. The fact that $N$ = even gives the same answer for the character contribution can be shown in a similar fashion.

We first note that 
\be
[M_0, L_{-i}] \sim M_{-i} \ ; \ [M_0, J_{-i}] \sim P_{-i},
\ee
that is, $M_0$ converts $L_{-i} \rightarrow M_{-i}, J_{-i}\rightarrow P_{-i}$ and destroys the operators $M_{-i}, P_{-i}$. So its effect on a generic state can be seen as 
\begin{equation}
    \begin{split}
        &M_0\ket{A,N;\vec{l}; \vec{m}; \vec{j}; \vec{k}} = M_0\ket{\psi}\\
        &=  \xi_A\ket{\psi} + \sum_i \bigl(C_i\ket{A,N; \{.. l_i-1, ..\}; \{.. , m_i+1..\};\vec{j}; \vec{k}} \\
        &+ D_i\ket{A,N; \vec{l}; \vec{m}; \{.., j_i-1,..\}; \{ .., P_i+1, ..\}},
    \end{split} 
\end{equation}
where, in the sum the $M_0$ have converted one $L_{-i} \to M_{-i}$ or one $J_{-i} \to P_{-i}$. Since the states inside the sum on the right-hand-side have more $M$($P$) operator content, they are not equivalent states of $\ket{\psi}$. Rather they belong to an equivalent set that is placed to the `right ' of the set $\{\ket{\psi}\}$ according to the ordering rule. So we can write this in terms of equivalent sets as
\begin{equation}
    q_{M}^{M_0}\{\ket{\psi}\} = q_{M}^{\xi}\{\ket{\psi}\} + \{\ket{{\psi}_R}\}.
\end{equation}
That is, $q^{M_0}$ acting on some state will give that state, plus a linear combination of other states which belong to the equivalence sets placed to the right of $\ket{\psi}$.

For example, let us take the state $J_{-1}P_{-1}\ket{a}$. Then one has
\begin{equation}
    q_{M}^{M_0}J_{-1}P_{-1}\ket{A} = q_{M}^{\xi_A} J_{-1}P_{-1}\ket{A} + \log(q_{M}) q_{M}^{\xi_A} P_{-1}^2\ket{A}.
\end{equation}
Here, the state $P_{-1}^2\ket{A}$ is placed to the right of $J_{-1}P_{-1}\ket{A}$ as per our ordering rule because it has more $M$($P$) operator content.

\subsection*{Matrix Structure of ${q_M^{M_0}}$}
We first look at the the action of $q^{M_0}$ on the anti-diagonal blocks (for odd $N$). We take states $\ket{\psi},\ket{\phi}$ from the blocks denoted by $\{\ket{\psi}\}, \{\ket{\phi}\}$, such that $\bra{\phi}\ket{\psi}$ is placed in one of the antidiagonal blocks in the Gram matrix. One then has
\begin{equation}
    \begin{split}
        \bra{\phi}q_{M}^{M_0}\ket{\psi} &= q_{M}^{\xi_A}\bra{\phi}\ket{\psi} + \bra{\phi}\ket{\psi_R}\\
        &= q_{M}^{\xi_A} K_{ij} ,
    \end{split}
\end{equation}
where $K_{ij}= \bra{\phi}\ket{\psi}$, which is an element of the Gram matrix. Here, we used the fact that the $\bra{\phi}\ket{\psi}$ is an element of one anti-diagonal block, and $\bra{\phi}\ket{\psi_R}$ consists of matrix elements that are below the anti-diagonal, which are identically zero as one can see from the Gram matrix (\ref{oddmatrix}).\\
So for the $d^\textrm{th}$ anti-diagonal block elements, denoted by $\{q_M^{M_0}(d)\}_{ij}$
\begin{equation}\label{block}
    \{q_M^{M_0}(d)\}_{ij} = q_{M}^{\xi_A}\{K^{(N)}(d)\}_{ij} = q_{M}^{\xi_A}\{A_d\}_{ij}.
\end{equation}
Here, ($K^{(N)}(d) $ is the $d^\textrm{th}$ anti-diagonal block/submatrix of the Gram matrix, which is denoted as $A_d$ according to (\ref{oddmatrix}).)

Next let us look at the lower anti-triangular elements of $q_M^{M_0}$ (elements below the anti diagonal). We choose $\ket{\phi}, \ket{\psi}$ such that $\braket{\phi}{\psi}$ is below the anti-diagonal. Just like before one has
\begin{equation}
    \begin{split}
        \bra{\phi}q_{M}^{M_0}\ket{\psi} &= q_{M}^{\xi_A}\bra{\phi}\ket{\psi} + \bra{\phi}\ket{\psi_R}.
    \end{split}
\end{equation}
However, now, both of the terms fall below the anti-diagonal in the Gram matrix, so both $\bra{\phi}\ket{\psi} = \bra{\phi}\ket{\psi_R} =0$ (see also \eqref{oddmatrix}).

So we see that the matrix of $q_M^{M_0}$ is block-wise upper-anti-triangular, just like the Gram matrix, and its anti-diagonal blocks are proportional to those of the Gram matrix. So in the expression $(K_{(N)})_{ij} (q_M^{M_0})_{ij}$, only the anti-diagonal blocks will contribute.

\subsection*{Contribution to Character}
From \eqref{charac1}, we consider the nontrivial summation, at a given level $N$ (writing ${K^{(N)}}^{-1}$ as $K_{(N)}$ and $q_M^{M_0}$ as $\hat{O}$ as shorthand, also taking $N$ to be odd for simplicity)
\begin{equation}
    \begin{split}
        \sum_{i,j}(K_{(N)})_{ij}(\hat{O})_{ij} =\sum_{d}\sum_{i', j'} (K_{(N)}(d))_{i'j'} (\hat{O}(d))_{i'j'},
    \end{split}
\end{equation}
where we used the fact that the sum had support only from the diagonal blocks. Here $\sum_d$ runs over the diagonal blocks at level $N$, and $K_{(N)}(d)$ and $\hat{O}_d$ are the $d^\textrm{th}$ diagonal block/submatrix of the corresponding matrices. Now we use \eqref{block} to replace $\hat{O}_d$
\begin{equation}
    \begin{split}
        \sum_{i,j}(K_{(N)})_{ij}(\hat{O})_{ij}=&\sum_{d}\sum_{i', j'} (K_{(N)}(d))_{i'j'} (\hat{O}(d))_{i'j'}\\
        =&\sum_{d}\sum_{i', j'} q_M^{\xi_A}(A_{(D-d+1)}^{-1})_{i'j'} (A_d)_{i'j'}.
    \end{split}
\end{equation}
Here, we have used$(K_{(N)}(d)) = A_{(D-d+1)}^{-1} = ({A_d^T})^{-1}$ from \eqref{igram2} and \eqref{oddmatrix2}.

Now one can insert this into the sum, and obtain
\begin{equation}\label{fr11}
    \begin{split}
        \sum_{d}\sum_{i', j'} q_M^{\xi_A}(K_{(N)}(d))_{i'j'} (A_d)_{i'j'} = &\sum_{d}\sum_{i', j'} q_M^{\xi_A}({A_d}^T)^{-1}_{i'j'} (A_d)_{i'j'}\\
        =&\sum_{d}\sum_{i', j'} q_M^{\xi_A}({A_d})^{-1}_{j'i'} (A_d)_{i'j'}\\
        =&\sum_{d}\sum_{j'}q_M^{\xi_A} \delta_{j'j'} =q_M^{\xi_A}\sum_{d}dim(d)=q_M^{\xi_A} {dim_N}.
    \end{split}
\end{equation}
Using this and \eqref{charac1}, the character of a highest weight state $\ket{A}$ becomes
\begin{equation}
\label{ChR1}
    \begin{split}
        \chi_A = \bigl(q_L^{(\Delta_A  -\frac{c_L}{2})}q_M^{(\xi_A-\frac{c_M}{2})}e^{2\pi i \gamma \mathcal{Q}_{J,A}}e^{2\pi i \delta \mathcal{Q}_{K,A}}\bigr)\sum_{N} q_L^N {dim_N},
    \end{split}
\end{equation}
as claimed in \eqref{tr2}. 

\section{Character for BMS Doublet States} \label{chmlt}
In this appendix we are calculating the character for BMS doublet (multiplet for rank 2). Then, we perform the same calculation for the BMS$\oplus\mathfrak{u}(1)$ doublet.
\subsection*{Character for BMS doublet}
In case of rank two we need to start from two primary operators $\mathcal{O}_1$ and $\mathcal{O}_2$. The corresponding primary states will satisfy the following
\begin{align}\label{chh10}
   [L_0,\mathcal{O}_i]&=\Delta\mathcal{O}_i,\hspace{5mm}[M_0,\mathcal{O}_i]=(\hat{\xi}\mathcal{O})_{i}, \nonumber
\end{align}
where
\begin{equation}\label{chh10'}
    \hat{\xi}=\left(\begin{array}{cc}
\xi & 0\\
1 & \xi\\
\end{array}\right).
\end{equation}
In case of singlets, the lowest level of the Gram matrix was 1, since we had only one primary to begin with. In this case, however, we are beginning from two-component primaries resulting in two primary states and hence there will be a level 0 $2\times 2$ Gramm matrix. As shown in \cite{Hao:2021urq}, the inner product of two components of a doublet is
\begin{align}\label{chh11}
    \bra{\mathcal{O}_{1}}\ket{\mathcal{O}_{1}}=\bra{\mathcal{O}_{2}}\ket{\mathcal{O}_{2}}=0 \ , \  \bra{\mathcal{O}_{1}}\ket{\mathcal{O}_{2}}=1.
\end{align}
Hence, at level 0, the Gram matrix will be
\begin{align}{\label{chh12}}
    \mathcal{K}_{0}=\left(\begin{array}{cc} 
0 & 1\\
1 & 0\\
\end{array}\right).
\end{align}
Now, in case of usual BMS singlets elements of the Gram matrices, for generic level, are given by 
\begin{align}\label{chh13}
    &\bra{\psi}\Big(\prod_{i}L_{i}^{n_{i}}\prod_{j}M_{j}^{m_{j}}\Big)\Big(\prod_{i^{\prime}}M_{-i^{\prime}}^{m^{\prime}_{i^{\prime}}}\prod_{j^{\prime}}L_{-j^{\prime}}^{n^{\prime}_{j^{\prime}}}\Big)\ket{\psi}\nonumber\\
    =&\bra{\psi}\Big[\prod_{i}L_{i}^{n_{i}}\prod_{j}M_{j}^{m_{j}},\prod_{i^{\prime}}M_{-i^{\prime}}^{m^{\prime}_{i^{\prime}}}\prod_{j^{\prime}}L_{-j^{\prime}}^{n^{\prime}_{j^{\prime}}}\Big]\ket{\psi}\nonumber\\
    =& F(\Delta,\xi)\bra{\psi}\ket{\psi}=F(\Delta,\xi),
\end{align}
where $\ket{\psi}$ is the primary or highest weight state. In case of a doublet, however, the Gram matrix elements for generic level change to
\begin{align}\label{chh14}
     &\bra{\mathcal{O}_{a}}\Big(\prod_{i}L_{i}^{n_{i}}\prod_{j}M_{j}^{m_{j}}\Big)\Big(\prod_{i^{\prime}}M_{-i^{\prime}}^{m^{\prime}_{i^{\prime}}}\prod_{j^{\prime}}L_{-j^{\prime}}^{n^{\prime}_{j^{\prime}}}\Big)\ket{\mathcal{O}_{b}}\nonumber\\
    =&\bra{\mathcal{O}_{a}}\Big[\prod_{i}L_{i}^{n_{i}}\prod_{j}M_{j}^{m_{j}},\prod_{i^{\prime}}M_{-i^{\prime}}^{m^{\prime}_{i^{\prime}}}\prod_{j^{\prime}}L_{-j^{\prime}}^{n^{\prime}_{j^{\prime}}}\Big]\ket{\mathcal{O}_{b}}\nonumber\\
    =& F(\Delta,\xi)\bra{\mathcal{O}_{a}}\ket{\mathcal{O}_{b}}=F(\Delta,\xi)(\mathcal{K}_{0})_{ab},
\end{align}
where $\mathcal{K}_{0}$ is given by \eqref{chh12}. Hence, each element of the BMS Gram matrix for the singlet is replaced by a $2\times 2$ block.

Let us suppose that we have the following BMS Gram matrix for the singlet at level $N$ (the upper-anti-triangular structure was discussed in \cite{Bagchi:2019unf})
\begin{align}\label{chh15}
    \mathcal{K}^{s}_{N}=\left(\begin{array}{ccccccc}
\cdots & \cdots & \cdots & C_{1,j} & \cdots & \cdots & C_{1,M}\\
\cdots & \cdots & \cdots & \cdots & \cdots &C_{2,M-1} &  0\\
\vdots &  &  & & \iddots  & 0 & 0 \\
\vdots & & & \iddots & & \vdots & \vdots\\
\vdots & & \iddots & & & \vdots & \vdots\\
\vdots & C_{M-1,2} & 0 & & & \vdots & \vdots  \\
C_{M,1} & 0 & 0 & \cdots & \cdots & \cdots & 0\\
\end{array}\right),
\end{align}
where $M = \widetilde{dim_N}$ is the number of independent BMS states at level $N$.

Then, as per our discussion above, the Gram matrix of the doublet case will be
\begin{align}\label{chh16}
    (\mathcal{K}^{d}_{N})=\left(\begin{array}{ccccccc}
\cdots & \cdots & \cdots & \fbox{$C_{1,j}$} & \cdots & \cdots & \fbox{$C_{1,M}$}\\
\cdots & \cdots & \cdots & \cdots & \cdots & \fbox{$C_{2,M-1}$} &  0\\
\vdots &  &  & & \iddots  & 0 & 0 \\
\vdots & & & \iddots & & \vdots & \vdots\\
\vdots & & \iddots & & & \vdots & \vdots\\
\vdots & \fbox{$C_{M-1,2}$} & 0 & & & \vdots & \vdots  \\
\fbox{$C_{M,1}$} & 0 & 0 & \cdots & \cdots & \cdots & 0\\
\end{array}\right),
\end{align}
where
\begin{align}\label{chh17}
    \fbox{$C_{i,j}$}=C_{i,j}\mathcal{K}_{0}=C_{i,j}\left(\begin{array}{cc}
0 & 1\\
1 & 0\\
\end{array}\right).
\end{align}
Using the fact that $\mathcal{K}_{0}$ is its own inverse, we find that the inverse of Gramm matrix is
\begin{align}\label{chh18}
 (\mathcal{K}^{d}_{N})^{-1}= \left(\begin{array}{ccccccc}
0 & \cdots & \cdots & 0 & \cdots & \cdots & \fbox{$C^{-1}_{1,M}$}\\
0 & \cdots & \cdots & \cdots & 0 & \fbox{$C^{-1}_{2,M-1}$} & \vdots \\
0 &  &  & & \iddots  & \vdots  & \vdots \\
\vdots & & & \iddots & & \vdots & \vdots\\
\vdots & & \iddots & & & \vdots & \vdots\\
0 & \fbox{$C^{-1}_{M-1,2}$} &  & & & \vdots & \vdots  \\
\fbox{$C^{-1}_{M,1}$} & \cdots & \cdots & \fbox{$C_{M,j}$} & \cdots & \cdots & \cdots\\
\end{array}\right) , 
\end{align}
where,
\begin{align}\label{chh19}
    (C_{i,j})^{-1}=\frac{1}{C_{i,j}}\left(\begin{array}{cc}
0 & 1\\
1 & 0\\
\end{array}\right).
\end{align}
Now, we want to determine the trace of the operator $\widehat{O}=e^{2\pi i\sigma (L_{0}-c_{L}/12)}e^{2\pi i\rho (M_{0}-c_{M}/12)}$ over a primary doublet and its descendants. In order to do so, one needs to know the action of the operator $e^{2\pi i\rho M_{0}}\equiv q^{M_{0}}_{M}$ on the primary doublet which is given by
\begin{align}\label{chh20}
   q^{M_{0}}_{M}\left(\begin{array}{c}
\ket{\mathcal{O}_{1}}\\
\ket{\mathcal{O}_{2}}\\
\end{array}\right)=q_{M}^{\xi}\left(\begin{array}{cc}  
1 & 0\\
\log{q_{M}} & 1\\
\end{array}\right)\left(\begin{array}{c}
\ket{\mathcal{O}_{1}}\\
\ket{\mathcal{O}_{2}}\\
\end{array}\right).
\end{align} 
On the right-hand-side of the above expression, $\hat{\xi}$ has been exponentiated as given in \eqref{chh10'}. For a BMS singlet, the operator $q_{M}^{M_{0}}$ acts on a generic descendant as
\begin{align*}
    q_{M}^{M_{0}}\ket{\psi}=q_{M}^{\xi}\ket{\psi}+\ket{\psi_{R}}.
\end{align*} 
Here too, one has 
\begin{align}{\label{chh21}}
     q_{M}^{M_{0}}\ket{\psi_{a}}=q_{M}^{\xi}\ket{\psi_{a}}+\ket{\psi_{R,a}},
\end{align}
where $\ket{\psi_{a}}$ is descendant of the doublet $\ket{\mathcal{O}_{a}}$ 
\begin{align}\label{chh22}
    \ket{\psi_{1}}&=\Big(\prod_{i}L^{n_{i}}_{-i}\prod_{j}M^{m_{j}}_{-j}\Big)\ket{\mathcal{O}_{1}}, \nonumber\\
     \ket{\psi_{2}}&=\Big(\prod_{i^{\prime}}L^{n_{i^{\prime}}}_{-i^{\prime}}\prod_{j^{\prime}} M^{m_{j^{\prime}}}_{-j^{\prime}}\Big)\ket{\mathcal{O}_{2}}.
\end{align}
Then, $q_{M}^{M_{0}}$ acts on such descendant states in the following manner
\begin{align}{\label{chh23}}
   q_{M}^{M_{0}}\ket{\psi_{1}}&=q_{M}^{\xi}\ket{\psi_{1}}+\ket{\psi_{R,1}},\nonumber\\
   q_{M}^{M_{0}}\ket{\psi_{2}}&=\log{q_{M}}\hspace{.5mm}q_{M}^{\xi}\ket{\psi_{1}}+q_{M}^{\xi}\ket{\psi_{2}}+\ket{\psi_{R,2}}.
\end{align} 
Similarly to the singlet case one has anti-diagonal blocks in the gram matrix (the only difference here is that each anti-diagonal element of the singlet Gram matrix will be replaced by $2\times 2$ blocks as in \eqref{chh17}). Let us define two descendants states $\ket{\psi_{a}}$ and $\ket{\phi_{b}}$ in such a way that $\bra{\phi_{b}}\ket{\psi_{a}}$ belongs to the anti-diagonal block of the Gram matrix. As before, one has
\begin{align}{\label{chh24}}
  \bra{\phi_{b}}q_{M}^{M_{0}}\ket{\psi_{a}}=\bra{\phi_{b}}q_{M}^{\hat{\xi}}\ket{\psi_{a}}.
\end{align}
In the above expression, we have used the fact that $\bra{\phi_{b}}\ket{\psi_{R,a}}=0$, since they are below the anti-diagonal blocks of the Gram matrices. Writing \eqref{chh24} component-wise, one obtains
\begin{align}\label{chh25}
    \bra{\phi_{1}}q_{M}^{M_{0}}\ket{\psi_{1}}&=q_{M}^{\xi} \bra{\phi_{1}}\ket{\psi_{1}}=q_{M}^{\xi}(C_{\phi,\psi})_{1,1}\, ,\nonumber\\
    \bra{\phi_{1}}q_{M}^{M_{0}}\ket{\psi_{2}}&=\log{q_{M}}\hspace{1mm} q_{M}^{\xi}(C_{\phi,\psi})_{1,1}+q_{M}^{\xi}(C_{\phi,\psi})_{1,2}\, ,\\
     \bra{\phi_{2}}q_{M}^{M_{0}}\ket{\psi_{1}}&=q_{M}^{\xi}(C_{\phi,\psi})_{2,1}\, ,\nonumber\\
   \bra{\phi_{2}}q_{M}^{M_{0}}\ket{\psi_{2}}&=\log{q_{M}}\hspace{1mm} q_{M}^{\xi}(C_{\phi,\psi})_{2,1}+q_{M}^{\xi}(C_{\phi,\psi})_{2,2}\, .\nonumber
\end{align}
Following \eqref{chh17} one has
\begin{align}\label{chh26}
    (C_{\phi,\psi})_{1,1}&=(C_{\phi,\psi})_{2,2}=0\ , \  (C_{\phi,\psi})_{1,2}=(C_{\phi,\psi})_{2,1}=1.
\end{align}
Hence, the entries of $q_{M}^{M_{0}}$ of the singlet picture, for any level, are being replaced by $2\times 2$ blocks as given below
\begin{align}\label{chh27}
    (q_{M}^{M_{0}})_{\phi,\psi}=q_{M}^{\xi}\bra{\phi}\ket{\psi}\left(\begin{array}{cc}
       0  &  1 \\
       1  & \log{q_{M}}
    \end{array}\right).
\end{align}
The matrix form of $q_{M}^{M_{0}}$ looks like
\begin{align}\label{chh28}
    q_{M}^{M_{0}}=q_{M}^{\xi}\left(\begin{array}{ccccccc}
\cdots & \cdots & \cdots & \fbox{$\widehat{C}_{1\hspace{.5mm}j}$} & \cdots & \cdots & \fbox{$\widehat{C}_{1\hspace{.5mm}M}$}\\
\cdots & \cdots & \cdots & \cdots & \cdots & \fbox{$\widehat{C}_{2\hspace{.5mm}(M-1)}$} &  0\\
\vdots &  &  & & \iddots  & 0 & 0 \\
\vdots & & & \iddots & & \vdots & \vdots\\
\vdots & & \iddots & & & \vdots & \vdots\\
\vdots & \fbox{$\widehat{C}_{(M-1)\hspace{.5mm}2}$} & 0 & & & \vdots & \vdots  \\
\fbox{$\widehat{C}_{M\hspace{.5mm}1}$} & 0 & 0 & \cdots & \cdots & \cdots & 0\\
\end{array}\right),
\end{align}
where
\begin{align}\label{chh29}
    \widehat{C}_{ij}=C_{ij}\left(\begin{array}{cc}
       0 & 1 \\
       1 & \log{q_{M}}
    \end{array}\right).
\end{align}
The character of the highest weight doublet, is 
\begin{align}\label{chh30}
    \chi=\sum_{N,i,j}(\mathcal{K}^{d}_{N})^{-1}_{ij}\bra{\phi^{N}_{i}}\widehat{O}\ket{\phi^{N}_{j}}.
\end{align}
As argued previously, one can see that only the anti-diagonal blocks of both $(\mathcal{K}_{N})^{-1}$ and $q_{M}^{M_{0}}$ contribute. Using \eqref{chh18}, \eqref{chh19}, \eqref{chh27},\eqref{chh28} and \eqref{chh29}, the character of a primary doublet $\{\mathcal{O}_1, \mathcal{O}_2\}$ with weights ($\Delta_{A},\xi_{A}$) turns out the be
\begin{align}\label{chh31}
   \chi^{\{\mathcal{O}_1, \mathcal{O}_2\}}_{A} &=\sum_N q_{M}^{\Delta_{A}-\frac{c_{L}}{2}}q_{M}^{-\frac{c_{M}}{2}}\sum_{a,i,j}q_{M}^{\xi_{A}}(C_{a,M-a})\left(\begin{array}{cc}
       0 & 1 \\
       1 & \log{q_{M}}
    \end{array}\right)_{ij}\frac{1}{(C_{a,M-a})}\left(\begin{array}{cc}
       0 & 1 \\
       1 & 0
    \end{array}\right)_{ij}\nonumber\\
    &= q_{M}^{\Delta_{A}-\frac{c_{L}}{12}}q_{M}^{\xi_{A}-\frac{c_{M}}{12}} \sum_N q^{N}_{L}\hspace{.5mm}(2\widetilde{dim_N}),
\end{align}
where $\widetilde{dim_N}$ represents the number of independent BMS descendant states at level $N$. Hence, the character contribution at any level of the doublet primary states, is twice the contribution of the singlet states. The factor of 2 represents the fact that we have taken a doublet primary state.
\subsection*{Character for BMS$\oplus\mathfrak{u}(1)$ Doublet}
In the case of pure BMSFT it was found that $L_{0}$ and $M_{0}$ cannot be simultaneously diagonalized. Hence, one had to start from a basis where $L_{0}$ was diagonalized but $M_{0}$ had a Jordan cell structure. As elucidated previously, for BMS$\oplus$ $\mathfrak{u}(1)$, we work in a basis where $L_{0}$, $P_{0}$ and $J_{0}$ are simultaneously diagonalized but $M_{0}$ has a Jordan structure (see \eqref{chh7}, \eqref{chh8}, \eqref{chh9}). Hence, the algorithm for computing the character of a BMS$\oplus\mathfrak{u}(1)$ doublet primary is very similar to that of BMS doublet primary. The only difference here will be due to the presence of the charges $P_{0}$ and $J_{0}$. As a result, \eqref{chh22} will be replaced by
\begin{align}
    \ket{\psi_{1}}&=\Big(\prod_{p}L^{n_{p}}_{-p}\prod_{q}M^{m_{q}}_{-q}\prod_{r}P^{k_{r}}_{-r}\prod_{s}J^{l_{s}}_{-s}\Big)\ket{\mathcal{O}_{1}}, \nonumber\\
     \ket{\psi_{2}}&=\Big(\prod_{p^{\prime}}L^{n'_{p^{\prime}}}_{-p^{\prime}}\prod_{q^{\prime}} M^{m'_{q^{\prime}}}_{-q^{\prime}}\prod_{r^{\prime}}P^{k'_{r^{\prime}}}_{-r^{\prime}}\prod_{s^{\prime}}J^{l'_{s^{\prime}}}_{-s^{\prime}}\Big)\ket{\mathcal{O}_{2}}.
\end{align}
Hence, each element of the BMS$\oplus\mathfrak{u}(1)$ Gram matrices is given by the following expression
\begin{align}
    &\bra{\mathcal{O}_{a}}\Big(\prod_{s}J_{s}^{l_{s}}\prod_{r}P_{r}^{k_{r}}\prod_{q}M^{m_{q}}_{q}\prod_{p}L^{n_{p}}_{p}\Big)\Big(\prod_{p^{\prime}}L_{-p^{\prime}}^{n^{\prime}_{p^{\prime}}}\prod_{q^{\prime}}M_{-q^{\prime}}^{m^{\prime}_{q^{\prime}}}\prod_{r^{\prime}}P^{k^{\prime}_{r^{\prime}}}_{-r^{\prime}}\prod_{s^{\prime}}J^{l^{\prime}_{s^{\prime}}}_{-s^{\prime}}\Big)\ket{\mathcal{O}_{b}}\nonumber\\
    =&\bra{\mathcal{O}_{a}}\Big[\prod_{s}J_{s}^{l_{s}}\prod_{r}P_{r}^{k_{r}}\prod_{q}M^{m_{q}}_{q}\prod_{p}L^{n_{p}}_{p},\prod_{p^{\prime}}L_{-p^{\prime}}^{n^{\prime}_{p^{\prime}}}\prod_{q^{\prime}}M_{-q^{\prime}}^{m^{\prime}_{q^{\prime}}}\prod_{r^{\prime}}P^{k^{\prime}_{r^{\prime}}}_{-r^{\prime}}\prod_{s^{\prime}}J^{l^{\prime}_{s^{\prime}}}_{-s^{\prime}}\Big]\ket{\mathcal{O}_{b}}\nonumber\\
    =& F(\Delta,\xi,p,j)\bra{\mathcal{O}_{a}}\ket{\mathcal{O}_{b}}=F(\Delta,\xi,p,j)(\mathcal{K}_{0})_{ab}.
\end{align}
This means again that each element of the Gram matrix is getting replaced by 2 $\times$ 2 block. The action of $q_{M}^{M_{0}}$ on the primary and descendant states, however, will be same as \eqref{chh20} and \eqref{chh23}, respectively. Hence, the matrix form of $q_{M}^{M_{0}}$ will have a structure similar to \eqref{chh28} and \eqref{chh29}. The operator whose trace we are going to calculate this time will be
\begin{align}
    \widehat{O}=e^{2\pi i\sigma(L_{0}-c_{L}/12)}e^{2\pi i\rho(M_{0}-c_{M}/12)}e^{2\pi i\alpha P_{0}}e^{2\pi i\gamma J_{0}}.
\end{align}
Hence, following the same algorithm for BMS$\oplus\mathfrak{u}(1)$ one ends up with the following expression for the character of a primary doublet labeled by ($\Delta_{A},\xi_{A},p_A,j_A$),
\begin{align}
\label{ChR2}
    \chi^{\{\mathcal{O}_1, \mathcal{O}_2\}}_{A}=2q_{L}^{\Delta_{A}-\frac{c_{L}}{12}}q_{M}^{\xi_{A}-\frac{c_{M}}{12}}q_{K}^{p_A}q_{J}^{j_A}\sum_{N}q^{N}_{L}\hspace{.5mm}{dim_N}.
\end{align}
Here $dim_N$ is the number of independent BMS$\oplus\mathfrak{u}(1)$ descendants at level $N$ (which is different from $\widetilde{dim_N}$). Hence, again, the character contribution of the doublet primary states for any level turns out to be twice than the one of the singlet.

\section{Character for $r$-Multiplets}\label{ApE}
As already mentioned previously, in the space of multiplets the action of $L_0, J_0, P_0$ are diagonal, but the action of $M_0$ is given by the $r \times r$ dimensional Jordan matrix. The operator relations for the primary fields $\{\mathcal{O}_i\}$ of the $r$-multiplet is given by 
\begin{align}\label{chh97}
[L_0,\mathcal{O}_i]=\Delta\mathcal{O}_i,\hspace{5mm}[M_0,\mathcal{O}_i]=(\hat{\xi}_r\mathcal{O})_{i},\hspace{5mm}[P_{0},\mathcal{O}_i]=p,\mathcal{O}_i\hspace{5mm}[J_{0},\mathcal{O}_i]=j\mathcal{O}_i,
\end{align}
where highest weight conditions are 
\begin{align}\label{chh98}
    [L_{n},\mathcal{O}_i]=0,\hspace{5mm}[M_{n},\mathcal{O}_i]=0,\hspace{5mm}[P_{n},\mathcal{O}_i]=0,\hspace{5mm}[J_{n},\mathcal{O}_i]=0,\hspace{5mm} \forall n>0.
\end{align}
Here, $\hat{\xi}_r$ is a Jordan cell of rank $r$
\begin{align}\label{chh38}
\hat{\xi}_r = \left(\begin{array}{ccccc}
\xi & 0 & \cdots & & 0\\
1 & \xi & 0 & \cdots \vdots\\
0 & 1 &  \xi &\ddots  & \vdots   \\
\vdots & \vdots& &\ddots  & 0\\
0 & \cdots & & 1 & \xi\\
\end{array}\right)_{r\times r}.
\end{align}
The highest weight states of the multiplet follow the inner product relation \cite{Hao:2021urq}
\be 
\mathcal{K}^{r}_{0}=\left(\begin{array}{cccc} 
0 & \dots & \dots & 1\\
0 & \dots & 1 & 0\\
\vdots & \iddots & \dots &0\\
1 & \dots &\dots & 0\\ \end{array}\right).
\ee 
One can also obtain the action of $q_M^{M_0}$ within a multiplet as
\be 
(q_M^{M_0})_{multiplet} = (q_M)^{\hat{\xi}_r} = \left(\begin{array}{ccccc}
(q_M)^\xi & 0 & \cdots & \cdots & 0\\
\log(q_M)(q_M)^\xi & (q_M)^\xi & 0 & \cdots & 0 \\
\frac{1}{2!}(\log(q_M))^2 (q_M)^\xi & \log(q_M)(q_M)^\xi &  (q_M)^\xi  & 0 & \vdots   \\
\vdots & \vdots &\vdots &\ddots  & 0\\
\frac{1}{(r-1)!}(\log(q_M))^r (q_M)^\xi & \frac{1}{(r-2)!}(\log(q_M))^{r-1} (q_M)^\xi & \dots & \log(q_M)(q_M)^\xi & (q_M)^\xi\\
\end{array}\right).\\
\ee 
Alternatively, one can write 
\be 
(q_M^{M_0})_{multiplet} = (q_M)^{\hat{\xi}_r} = (q_M)^{\xi} \mathcal{Q}^r_0 .
\ee 
Here,
\be 
\mathcal{Q}^r_0 = \left(\begin{array}{ccccc}
1 & 0 & \cdots & \cdots & 0\\
\log(q_M) & 1 & 0 & \cdots & 0 \\
\frac{1}{2!}(\log(q_M))^2  & \log(q_M) &  1  & 0 & \vdots   \\
\vdots & \vdots &\vdots &\ddots  & 0\\
\frac{1}{(r-1)!}(\log(q_M))^r  & \frac{1}{(r-2)!}(\log(q_M))^{r-1}  & \dots & \log(q_M) & 1\\
\end{array}\right).
\ee 
Using the Gram matrix of descendants of a singlet BMS primary (denoted by ${K^{(1)}}_{(N)}$, where $1$ denotes singlet), and the inner product matrix of the multiplets $\mathcal{K}^r_0$ (or the zero level Gram matrix), one can write the $N$-level Gram matrix  of the multiplet primaries as a simple tensor product of the two matrices,
\be 
{K^{(r)}}_{(N)} = {K^{(1)}}_{(N)} \otimes \mathcal{K}^r_0.
\ee 
This will be an ($rN \times rN$)-dimensional matrix, and each element of the singlet Gram matrix will be replaced by the multiplet inner product matrix $\mathcal{K}_{0}$,
\be
(K^{(1)}_{(N)})_{ij} \to \fbox{$\bigl(K^{(r)}_{(N)}\bigr)_{ij}$}= (K^{(1)}_{(N)})_{ij}\times \mathcal{K}^{r}_{0}.
\ee

The action of $q_{M}^{M_0}$ can also be written as 
\be 
q_{M (r)}^{M_{0}} = q_{M (1)}^{M_{0}} \otimes \bigl( \mathcal{K}^r_0\mathcal{Q}^r_0 \bigr).
\ee 
Here, 1 and $r$ in the subscript mean singlet and r-multiplet respectively.

Every element of the previously calculated singlet matrix of $(q_M)^{M_0}$ will be replaced by the nontrivial matrix to take into account the multiplet structure of the Hilbert space. One can verify this result by looking at the case of $r=2$.

Writing the new matrices in terms of tensor products makes computing the character very straightforward, as we will show below. One has 
\be 
    ({K^{(r)}}_{(N)})^{-1} = ({K^{(1)}}_{(N)})^{-1} \otimes \mathcal[{K}^r_0]^{-1} = ({K^{(1)}}_{(N)})^{-1} \otimes \mathcal{K}^r_0,
\ee 
where we have used $\mathcal[{K}^r_0]^{-1} = \mathcal{K}^r_0$.

Now, let us go back to the nontrivial part of the character contribution which is given in terms of the following expression 
\be 
    \begin{split}
        \sum_{i,j}\bigl({K^{(r)}}_{(N)}\bigr)^{-1}_{ij} ((q_M)^{M_0}_{r})_{ij}&= \sum_{i,j} \Bigl( ({K^{(1)}}_{(N)})^{-1} \otimes \mathcal{K}^r_0\Bigr)_{ij}\Bigl( q_{M (1)}^{M_0} \otimes \bigl( \mathcal{K}^r_0\mathcal{Q}^r_0 \bigr)\Bigr)_{ij}\\
        &= \Bigl[\sum_{i',j'}\bigl({K^{(1)}}_{(N)}\bigr)^{-1}_{i'j'}((q_M)^{M_0}_{(1)})_{i'j'}\Bigr] \times \Bigl[\sum_{k,l} \bigl((\mathcal{K}^r_0)_{kl} (\mathcal{K}^r_0\mathcal{Q}^r_0)_{kl}\bigr)\Bigr]\\
        &= q_{M}^\xi (dim_N) \times \Bigl( Tr\bigl((\mathcal{K}^r_0)^2\mathcal{Q}^r_0\bigr)\Bigr)= q_{M}^\xi (dim_N)\times \Bigl( Tr\bigl(\mathcal{Q}^r_0\bigr)\Bigr)\\
        &= q_{M}^{\xi} (dim_N) \times r .
    \end{split}
\ee 
In the third line, the first expression is the usual expression for the singlet case, which we have derived in \eqref{fr11}. Thus, we have replaced it by $q_{M}^\xi dim_N$. Also in the calculation, we used that $\mathcal{K}^r_0$ is symmetric, and $(\mathcal{K}^r_0)^2 = I$.

Since the other operators are diagonal in the multiplet Hilbert space, the character formula is given by 
\be 
\chi_{(r)}^{\{\mathcal{O}_i\}} =  r\bigl(q_L^{(\Delta  -\frac{c_L}{2})}q_M^{(\xi-\frac{c_M}{2})}e^{2\pi i \gamma j}e^{2\pi i \alpha p}\bigr)\sum_{N} q_L^N \times ( {dim_N}).
\ee 
This expression is same as \eqref{ChR1}, except for a factor of $r$. For $r=2$  one again obtains \eqref{ChR2}.

\newpage
\bibliographystyle{JHEP}
\bibliography{ref-flat_v2}	
\end{document}